\documentclass[%
nofootinbib,
 amsmath,amssymb,
onecolumn
]{revtex4-2}

\usepackage{graphicx}
\usepackage{dcolumn}
\usepackage{bm}
\usepackage{dsfont}

\usepackage[utf8]{inputenc}
\usepackage{kbordermatrix}
\usepackage{listings}
\usepackage{xcolor}

\definecolor{codegreen}{rgb}{0,0.6,0}
\definecolor{codegray}{rgb}{0.5,0.5,0.5}
\definecolor{codepurple}{rgb}{0.58,0,0.82}
\definecolor{backcolour}{rgb}{0.95,0.95,0.92}
\definecolor{orangered}{RGB}{239,134,64}
\definecolor{darkblue}{rgb}{0.0,0.0,0.6}
\definecolor{realblue}{rgb}{0.0,0.0,255}
\definecolor{realred}{rgb}{255,0,0}

\lstdefinestyle{mystyle}
{
    backgroundcolor=\color{backcolour},   
    commentstyle=\color{codegreen},
    numberstyle=\tiny\color{codegray},
    stringstyle=\color{codepurple},
    basicstyle=\ttfamily\footnotesize,
    breakatwhitespace=false,         
    captionpos=b,                    
    keepspaces=true,                 
    numbersep=5pt,                  
    showspaces=false,                
    showstringspaces=false,
    showtabs=false,                  
    tabsize=2,
	language = Python,
	keywordstyle=  [1]{\color{magenta}},
	keywordstyle = [2]{\color{realblue}},
	otherkeywords = {+,-,*,/,(,),[,],=},
    morekeywords = [2]{+,-,*,/,(,),[,],=},
}

\begin{document}

\preprint{APS/123-QED}

\title{Bayesian Quantum State Tomography with Python's PyMC}

\author{Daniel J. Lum}
 \altaffiliation{The MITRE Corporation's Quantum Technologies Group; Princeton, NJ 08540}
\email{dlum[at]mitre.org}
\author{Yaakov Weinstein}
 \affiliation{The MITRE Corporation's Quantum Technologies Group; Princeton, NJ 08540}


\date{\today}

\begin{abstract}
Quantum state tomography (QST) is typically performed from a frequentist viewpoint using maximum likelihood estimation (MLE) which seeks to find the best plausible state consistent with the data by maximizing a likelihood function / distribution. The likelihood function holds an implicit assumption that there is suitable data to infer frequency. In data-starved experiments, this may or may not be a feasible assumption. Moreover, MLE returns no error estimates on the final solution and users are forced to rely on alternative approaches involving either additional measurements or simulated data. Alternatively, Bayesian methods can return a solution with error estimates consistent with the data's uncertainty, but at the expense of a difficult integration over the likelihood distribution. The integration usually requires computational methods with appropriately chosen step sizes in a somewhat complicated problem formulation. This additional complexity serves as a strong deterrent from using Bayesian methods despite the advantages. Probabilistic programming is becoming a common alternative with growing computational power and the development of robust automated integration techniques such as Markov-Chain Monte Carlo (MCMC). Here, we show how to use Python-3's open source PyMC probabilistic programming package to transform an otherwise complicated QST optimization problem into a simple form that can be quickly optimized with efficient under-the-hood MCMC samplers.\footnote{\bf{Approved for Public Release; Distribution Unlimited. Public Release Case Number 21-3375. \newline \copyright 2022 The MITRE Corporation. ALL RIGHTS RESERVED. \newline }}
\end{abstract}

\maketitle

\section{Introduction}

Quantum state tomography (QST) is a method to estimate the form of an unknown quantum state by taking appropriate projective measurements (either orthogonal or non-orthogonal \cite{PhysRevA.66.012303}) from an ensemble of identically prepared states \cite{PhysRevA.64.052312}. Unfortunately, QST is an experimentally expensive and difficult problem. For a system with $n$-qudits (where each qudit spans a $d$-dimensional Hilbert space), the minimum number of measurements grows as $d^{2n}-1$ \cite{PhysRevA.66.012303}. The difficulty in performing a full tomography in the presence of measurement uncertainties and noise has led some researchers to incorporate sparsity assumptions into the tomography reconstruction model in a manner similar to compressed sensing \cite{PhysRevLett.105.150401}. Additional complexity requires one to constrain the set of viable solutions to positive-semidefinite matrices (i.e., those with positive, real eigenvalues) to be consistent with physical measurements. 

One of the most universally used methods for this task is maximum likelihood estimation (MLE) \cite{PhysRevA.55.R1561}. MLE seeks to solve for the parameters of a likelihood distribution that best explains the data (by maximizing the likelihood distribution). The method is robust and can incorporate the error propagation of photon statistics, waveplate angular settings, and values derived from the state estimate as presented in \cite{PhysRevA.66.012303, ALTEPETER2005105} (for polarization-based tomography). However, MLE can introduce errors itself and only yields a singular point estimate for the quantum state estimate. Inferring a spread in values for uncertainty estimation is typically accomplished by either running the experiment numerous times or by simulating additional data consistent with the first solution (sometimes by adding a little noise). Thus, propagating simple sources of measurement errors and uncertainties through MLE is a non-trivial task. For certain applications, such as quantum state discrimination \cite{Chefles2000Quantum}, having a reasonable estimate in the error of the estimated quantum state is invaluable. We propose the Bayesian framework for an easier experimentalist-friendly approach to error propagation.

Over the past 30 years, there has been a growing emphasis on applying Bayesian methods to quantum mechanical systems -- particularly QST. In its early development, a ``quantum Bayes rule'' \cite{PhysRevA.64.014305} was applied to a spin-$1/2$ system in the limit of an infinite number of measurements with a pure-state prior assumption. This Bayesian method was then extended to mixed states by realizing a pure-state prior could be applied if the state of interest was first entangled with a ``reservoir'' state. The reservoir state could then be traced out leaving the posterior distribution for the mixed-state of interest \cite{derka1997quantum,buvzek1998reconstruction}. By applying Bayes rule for conditional probabilities and prior information to a quantum system, we can determine quantum state estimates in the form of a posterior distribution that allows us to extract mean and variance information. This quantum Bayes rule demonstrates that as the number of measurements tends towards infinity, the exact form for the posterior distributions becomes irrelevant. However, it also states that the prior assumptions take more weight with less data. 

More importantly, and what will be emphasized here, is the ability to accurately and honestly estimate a quantum state from a limited number of measurements knowing that errors and uncertainties will propagate to the final solution. Fortunately, the Bayes mean estimate (BME) yields a unique solution to this honest state-estimation problem. In the presence of measurement errors and uncertainty, pure (or rank-deficient) quantum state estimates should be considered as infinitely dishonest estimates \cite{blume2006accurate}. Bayesian analysis allows us to avoid these pure-state pitfalls when applied to experimental data. In the end, the BME outperforms the MLE (on average) with respect to the state-estimate accuracy \cite{schmied2016quantum}. A detailed analysis and a means to utilize Markov-Chain Monte-Carlo (MCMC) sampling to make the problem of finding a Bayesian posterior distribution numerically tractable are presented in \cite{blume2010optimal}. 

Since the Bayesian framework was introduced, there has been significant progress in applying these ideas to experimental problems. One important outcome is the development of adaptive statistical methods. For example, \cite{PhysRevA.85.052120} presents an adaptive method based on Bayesian inference and Shannon's information to pick the next-best measurement in QST data acquisition and shows superior performance over non-adaptive protocols (including mutually unbiased bases) in terms of minimum uncertainty. Since that work, experimental demonstrations have been performed for both a single qubit \cite{PhysRevA.87.062122} and two-qubit system \cite{PhysRevA.93.012103}. Additionally, authors in \cite{PhysRevLett.111.183601} demonstrated that they could adaptively reduce the infidelity between their reconstructed state and the expected form by an order of magnitude with a single additional measurement \cite{PhysRevLett.111.183601}.

Despite the advent of adaptive methods for optimal experimental design using Bayesian analysis, the problem of establishing informative and meaningful error bars in quantum tomography is still an active area of research. In MLE, error estimates can be obtained by approximating the Fisher information matrix \cite{PhysRevA.83.012105} (which yields information about the width of the likelihood function). More commonly, a numerical ``bootstrapping'' method \cite{efron1994introduction} utilizes the initial estimate to generate additional data sets to reconstruct and yield a spread of MLE estimates. Unfortunately, this bootstrapping may either underestimate or overestimate the error and does not take into account errors from the MLE numerical optimization \cite{blume2012robust}. As an alternative to error bars, confidence intervals over which the unknown quantum state may be found with high probability are growing in popularity. Rather than quantifying the probability of success, the confidence interval quantifies our confidence in the quantum state estimate and is independent of any prior assumptions on the quantum state. Confidence intervals can generally be easily constructed using likelihood ratios \cite{blume2012robust} or through a type of adaptive procedure by using the first $n$ measurements to construct a testing procedure to be applied to the remaining $k$ measurements \cite{PhysRevLett.109.120403}.

While confidence intervals are robust, they are not particularly easy to use from an experimentalist point of view knowing that there is uncertainty in both the photon statistics, waveplate-angle settings, fiber losses, beam-splitter imbalance, and detector inefficiencies. How does this error affect the final quantum state estimate? One solution presented in \cite{granade2016practical} uses sequential Monte-Carlo sampling \cite{PhysRevA.85.052120,ferrie2014quantum} to numerically generate a posterior distribution by sampling from the set of all possible positive-semidefinite solutions for both time-dependent and time-independent states. In the generated posterior distribution, they numerically find mean and variance information. While their method finds error bars by assuming a prior distribution of the quantum state, they did not propagate all sources of error. Their final mean and error estimates are still consistent with the data, but they assumed ideal experimental measurements. In a typical experiment, measurement error means the data may not be consistent with the ideal measurement. Thus, error estimates in the final states should capture the variance in the measured data \emph{as well as} uncertainty in the measurements themselves. An alternative approach in \cite{PhysRevLett.117.010404} uses Monte-Carlo sampling via the Metropolis-Hastings algorithm \cite{metropolis1953equation} to numerically explore the state space and generate a posterior distribution. They then fit a function composed of three parameters (for the single-qubit tomography) and those three parameters serve as their ``quantum error bars.'' These two methods are robust in that they can also generate confidence regions for any given confidence interval consistent with the data. However, neither approach propagates all forms of experimental uncertainty and may fail if enough samples are not used.

We propose an alternative method to quantum-state estimation (which can be argued to be an extension of \cite{PhysRevLett.117.010404}) where we generate error bars through Bayesian analysis. We utilize a particularly efficient MCMC sampler called the No-U-turn sampler \cite{hoffman2014no} within Python-3's PyMC library \cite{salvatier2016probabilistic} to brute-force the propagation of experimental uncertainty and measurement error through to the final quantum state estimate. All experimental parameters $\theta_i$, such as photon flux, waveplate-angle settings, detection efficiency, beamsplitter crosstalk, and the unknown quantum-state parameters, take on a numerically calculated joint prior distribution $P(\theta_0,\theta_1,...,\theta_n)$ having a mean and variance (and possibly covariance) stating our knowledge (or lack thereof) about each variable. Rather than provide the state projections directly, our formulation requires an experimentalist to provide the waveplate settings and system uncertainties along with the corresponding photon counts. By modeling a likelihood function $P(X|\theta_0,...,\theta_n)$ as the probability of obtaining data $X$ given an unknown set of parameters $\theta_i$, PyMC generates a posterior distribution $P(\theta_0,...,\theta_n|X)$ as the set of possible variable $\theta_i$ that could arise given the data $X$. Formally, this relationship is expressed as Bayes rule which states
\begin{equation}
P(\theta_0,...,\theta_n|X) = \frac{P(X|\theta_0,...,\theta_n) P(\theta_0,...,\theta_n)}{P(X)}.
\label{eq:BayesRule}
\end{equation}
Typically, the normalizing constant $P(X)$ is unknown -- making the posterior distribution practically impossible to calculate analytically. Fortunately, MCMC sampling methods can estimate the posterior distribution since they only require the posterior distribution to be proportional to the product of the likelihood and posterior; i.e., $P(\theta_0,...,\theta_n|X) \propto P(X|\theta_0,...,\theta_n) P(\theta_0,\theta_1,...,\theta_n).$ Given enough numerical samples, an accurate posterior distribution can be obtained to provide mean and variance information for each unknown variable. As is typical with MCMC methods, the model fit may fail if not enough samples are used. Fortunately, there exist tell-tale signs (e.g., a large autocorrelation of each sampled trace) to indicate that more numerical sampling is needed in post-processing.

Because PyMC generates a trace posterior distribution for all parameters (having traced out all model variables from the posterior distribution except the variable of interest), propagating uncertainty is trivial. Thus, we find the BME density matrix and Stokes parameters with their corresponding highest density intervals for an experimentally measured single qubit QST and simulate a double-qubit QST.

The structure of this article is primarily meant as a useful resource to an experimentalist seeking to perform QST in a laboratory while yielding meaningful error bars. As long as the number of numerical samples is large enough such that the PyMC convergence criteria hold (see PyMC documentation for details), the method is robust to noise, measurement uncertainty, and poor tomographic sampling. Our Python code can be found in the appendix.

\subsection{Introductory Notation}
A brief overview of our notation and QST is provided here, but we encourage the reader to refer to \cite{PhysRevA.64.052312, ALTEPETER2005105} for a more comprehensive overview. To begin, we restrict our QST to discrete variables where our basis vectors (pure states) are those of the Poincar\'e sphere for optical polarization tomography measurements. The basis vectors for horizontal $|H\rangle$ and vertical $|V\rangle$ polarizations are presented as
\begin{align}
|H\rangle &= 
\begin{bmatrix}
1 & 0
\end{bmatrix}^T \nonumber \\
|V\rangle &= 
\begin{bmatrix}
0 & 1
\end{bmatrix}^T,
\end{align}
where $\ast^T$ is the transpose operation. Using the conjugate transpose operator $\ast^\dagger$, such that $|\psi\rangle^\dagger=\langle\psi |$, it is immediately clear that $|H\rangle$ and $|V\rangle$ are orthogonal $\langle H|V\rangle=0$. From the $|H\rangle$ and $|V\rangle$ basis vectors, we can represent linear combinations for diagonal $|D\rangle$, antidiagonal $|A\rangle$, right-circular $|R\rangle$, and left-circular $|L\rangle$ polarizations as
\begin{align}
|D\rangle &= \frac{1}{\sqrt{2}}\left( |H\rangle + |V\rangle \right) \nonumber\\
|A\rangle &= \frac{1}{\sqrt{2}}\left( |H\rangle - |V\rangle \right) \nonumber\\
|L\rangle &= \frac{1}{\sqrt{2}}\left( |H\rangle + i|V\rangle \right) \nonumber\\
|R\rangle &= \frac{1}{\sqrt{2}}\left( |H\rangle - i|V\rangle \right)
\label{eq:polstates}
\end{align}
such that $\langle D|A\rangle = 0$ and $\langle L|R\rangle = 0$. 
Because the QST given here is for discrete-valued states, we limit our discussion of states and density matrices to discrete values. Given a finite dimensional complex-valued Hilbert space, pure states, $|\psi\rangle$, are vectors of unit length that reside within the Hilbert space. Superpositions of pure states can form other pure states $|\Psi\rangle$ as long as they are properly normalized such that
\begin{equation}
|\Psi\rangle = \sum\limits_{i = 1}^n \alpha_i |\psi_i\rangle
\end{equation}
where the weights $\alpha_i$ follow
\begin{equation}
\sum\limits_{i=1}^n \alpha_i^2 = 1.
\end{equation}
However, pure states cannot capture all forms of a physical system's uncertainty as there can exist classical mixtures of pure states. A density matrix $\bm{\rho}$, defined as a positive, semi-definite ($\bm{a}^{\dagger}\cdot\bm{\rho}\cdot\bm{a} \geq 0$ for all column vectors $\bm{a}$) Hermitian operator ($\bm{\rho}^\dagger = \bm{\rho}$) of trace 1 ($\text{Tr}\bm{\rho}=1$), can be represented by a sum of pure states $|\psi\rangle$ as 
\begin{equation}
\bm{\rho} = \sum\limits_{i=1}^n P_i |\psi_i\rangle\langle \psi_i |,
\end{equation}
where $P_i$ is the probability of finding state $|\psi_i\rangle\langle \psi_i |$. 

\section{Single-qubit QST}

The objective in QST is to find a positive-definite unity-trace density matrix $\bm{\rho}$ that best describes our measurement results. The problem formulation to ensure we find a physical density matrix for both a single-qubit and a two-qubit density matrix is briefly introduced here. A detailed review can be found in the appendix.

Any single-qubit density matrix can be written as
\begin{equation}
\bm{\rho} = 
\begin{bmatrix} 
A & B e^{i\phi}\\ 
B e^{-i\phi} & 1-A
\end{bmatrix},
\label{eq:OneQubitDensityMatrix0}
\end{equation}
where $A,B,\phi\in\mathbb{R}$ and $B\leq\sqrt{A(1-A)}$ ensures all eigenvalues are real and $\geq 0$. This shows that any single-qubit state is completely defined by three variables. Hence, a minimum of three properly chosen measurements are needed to form a complete set of equations having a unique solution. In general, any density matrix of $n$ qubits will have $4^n-1$ independent variables.

A useful way to parameterize a single-qubit density matrix using the basis vectors in Eq. \ref{eq:polstates} is via the Pauli matrices. The Pauli matrices can be used to represent any single-qubit density matrix as a linear sum with each Pauli matrix weighted by a Stokes parameter $S_i$:
\begin{equation}
\bm{\rho} = \frac{1}{2}\sum\limits_{i=0}^3 S_i\bm{\sigma_i} = \frac{1}{2}\sum\limits_{i=0}^3 \langle \bm{\sigma_i}\rangle\bm{\sigma_i},
\label{eq:stokesdensitymatrix}
\end{equation}
where $\langle \bm{\sigma_i} \rangle = \text{Tr}\left[\bm{\rho}\cdot\bm{\sigma_i}\right]$. The Stokes parameters are used to visualize the qubit on the Bloch sphere (or Poincar\'e sphere for polarization states) with the Stokes parameters as coordinates $\left[S_1 \, \hat{x}, S_2\, \hat{y}, S_3 \,\hat{z}\right]$ where $S_0 = 1$ by the law of total probability. Thus, pure states will reside along the surface of the Poincar\'e sphere while maximally mixed states will reside at the center.

\begin{figure}
  \includegraphics[width=.4\linewidth]{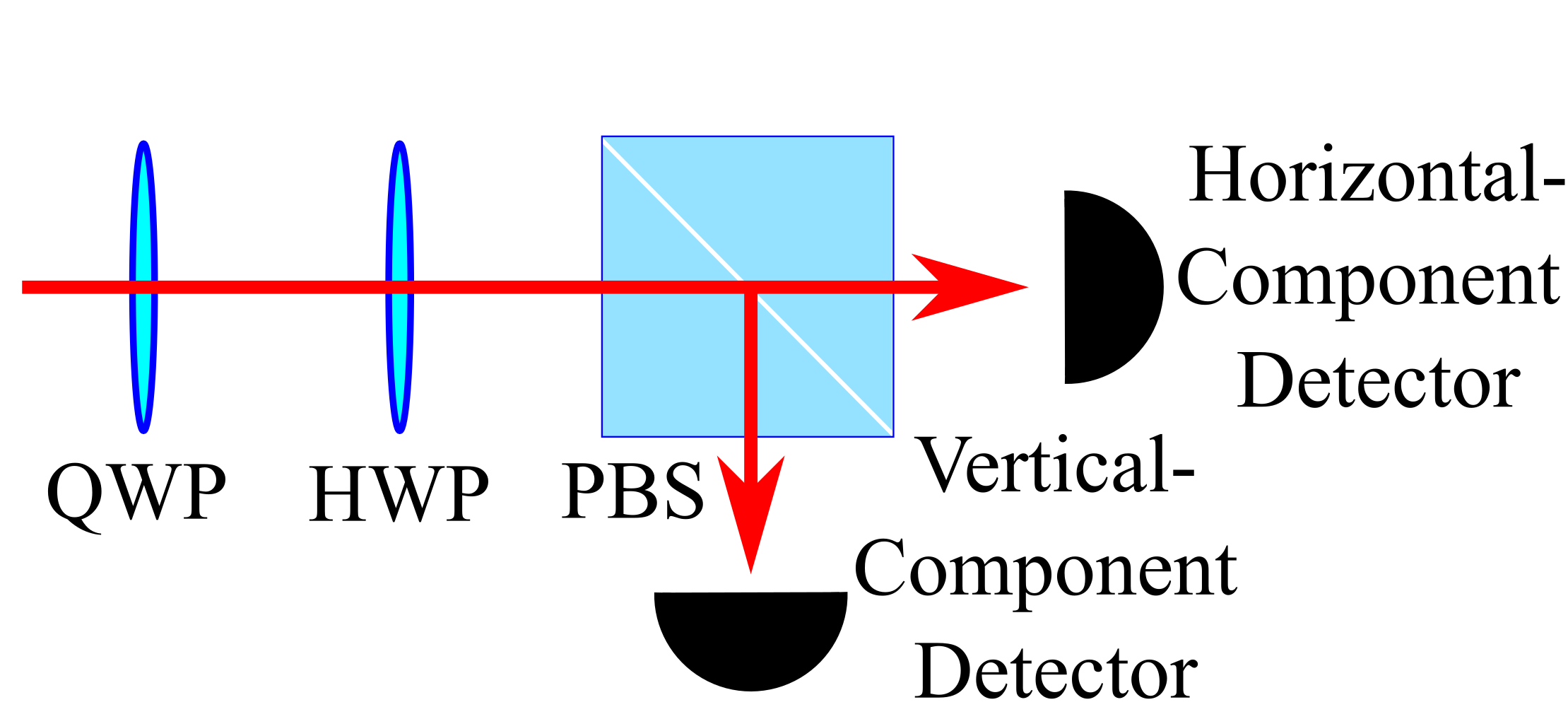}
  \caption{Polarization projections can be made using a half-wave plate (HWP), a quarter-wave plate (QWP), and a polarizing beamsplitter. }
  \label{fig:polprojection}
\end{figure}

In an optical polarization based QST, projection probabilities are experimentally obtained by measuring the two output ports of a polarizing beamsplitter (PBS) where the beamsplitter effectively takes horizontal (transmitted) and vertical (reflected) projections. Given that we cannot take all necessary projections with only a PBS, we use a quarter-wave plate (QWP) and half-wave plate (HWP) to rotate the projection state into the horizontal-vertical basis of the PBS as shown in Fig. \ref{fig:polprojection}. The goal is to take the projections as if $|H\rangle$, $|D\rangle$, and $|L\rangle$ polarization states are incoming and then rotate those states such they are transmitted by the PBS.

With this in mind, we construct the rotation operators $\bm{U}(\theta_Q,\theta_H)$ from QWP and HWP unitary operations ($\bm{Q}$ and $\bm{H}$, respectively) -- each of which have their fast axis rotated $\theta_Q$ and $\theta_H$ radians with respect to the horizontal axis of the PBS. The unitary operators for a QWP and HWP with their fast axes at horizontal are given by
\begin{align}
\bm{Q} &= e^{-i\frac{\pi}{4}}
\begin{bmatrix}
1 & 0 \nonumber\\
0 & i 
\end{bmatrix} \\
\bm{H} &= e^{-i\frac{\pi}{2}}
\begin{bmatrix}
1 & 0 \\
0 & -1 
\end{bmatrix}.
\end{align}
Rotating these operators with a rotation matrix $\bm{R}(\theta)$, where 
\begin{equation}
\bm{R}(\theta) = 
\begin{bmatrix}
\cos(\theta) & -\sin(\theta) \\
\sin(\theta) & \cos(\theta)
\end{bmatrix},
\end{equation}
results in the general polarization rotation operations
\begin{align}
\bm{Q}(\theta_Q) &= \bm{R}(\theta_Q)\cdot\bm{Q}\cdot\bm{R}(\theta_Q)^T \nonumber \\
\bm{H}(\theta_H) &= \bm{R}(\theta_H)\cdot\bm{H}\cdot\bm{R}(\theta_H)^T \nonumber \\
\bm{U}(\theta_Q,\theta_H) &= \bm{H}(\theta_H)\cdot \bm{Q}(\theta_Q).
\label{eq:WPOperations}
\end{align}
Acquiring the necessary projections via a PBS requires the HWP and QWP be placed at the proper angle with respect to the horizontal axis of the PBS. Knowing that $|H\rangle = \bm{U}(\frac{\pi}{4},\frac{\pi}{8})|D\rangle$, $|V\rangle = \bm{U}(\frac{\pi}{4},\frac{\pi}{8})|A\rangle$, $|H\rangle = \bm{U}(\frac{\pi}{4},0)|L\rangle$, and $|V\rangle = \bm{U}(\frac{\pi}{4},0)|R\rangle$, a complete set of observables ($|H\rangle$, $|V\rangle$, $|D\rangle$, $|A\rangle$, $|L\rangle$, and $|R\rangle$) can be measured.

Wile Eq. \ref{eq:WPOperations} presents the unitary operations most commonly used for polarization rotations due to the prevalence QWPs and HWPs, liquid crystal phase retarders are often being used over traditional waveplates for their rapid switching capability. Like traditional waveplates, they have an angular setting $\theta$ defined with respect to their horizontal axis. However, they also have an adjustable phase retardance $\eta$ dependent on the applied voltage to the liquid crystals. These arbitrary waveplates take the form
\begin{equation}
\bm{AWP}(\eta,\theta) = e^{-i\frac{\eta}{2}}
\begin{bmatrix}
\cos^2(\theta)+e^{i\eta}\sin^2(\theta) & 1-e^{i\eta}\cos(\theta)\sin(\theta) \\
1-e^{i\eta}\cos(\theta)\sin(\theta)  & \sin^2(\theta)+e^{i\eta}\cos^2(\theta)
\end{bmatrix},
\label{eq:AWP}
\end{equation}
where
\begin{align}
\bm{AWP}(\eta=\pi,\theta) &= \bm{H}(\theta) \\
\bm{AWP}(\eta=\pi/2,\theta) &= \bm{Q}(\theta).
\end{align}
Because Eq. \ref{eq:AWP} is a general form for the polarization rotations, any result derived with the arbitrary waveplate form is applicable to an experiment using either liquid-crystal wave retarders or QWPs and HWPs. 

Previously, we stated that only three measurements were needed to uniquely identify a qubit state while simultaneously suggesting the need for six projective measurements. Only three measurements would be required if we could measure probabilities directly -- since by measuring $P_H$, we can immediately infer that $P_V = 1 - P_H$ and likewise for the pairs ($P_D$, $P_A$) and ($P_L$, $P_R$). Because we must infer probabilities through counting statistics, we need the total flux and should at least measure $P_H$, $P_D$, $P_L$ under the assumption the flux does not change between measurements. Measuring all six projections will result in an overcomplete set with more accurate results (assuming a steady photon flux between pair measurements).

These six state projections on an ensemble of identically prepared states will enable us to identify the unknown states (see details in appendix). However, the presence of measurement uncertainties and noise does not guarantee that a state constructed with the measured Stokes parameters is positive-definite with unit trace. To overcome this, constrained optimization techniques (such as appropriate formulations of maximum likelihood) are typically used to return a state estimate that is both physical and consistent with the data. Because we desire a probabilistic solution, we show how to formulate the problem similar to that of maximum likelihood estimation but with an emphasis on propagating probability distributions in the next section.

\subsection{Single-Qubit Problem Formulation}

In the problem formulation presented here, we demonstrate how to confine the solution set consistent with the data to positive-definite Hermitian operators while simultaneously incorporating measurement noise / uncertainty from beamsplitter losses, beamsplitter crosstalk, detector efficiencies, fiber coupling, shot noise, etc.

\subsubsection{Model Formulation}

To begin, we construct a measurement model for an arbitrary input state $\bm{\rho}$ in an attempt to predict photon counts on a detector when using arbitrary waveplate-angle settings ($\theta_Q$ and $\theta_H$ for a QWP and a HWP, respectively). While we previously stated that an arbitrary single-qubit density matrix has the form from Eq. \ref{eq:OneQubitDensityMatrix0}, we find better computational performance in reconstructing a density matrix of the form
\begin{equation}
\bm{\rho} = \frac{1}{A+B}
\begin{bmatrix} 
A & \text{Re}(C) - i\text{Im}(C) \\ 
\text{Re}(C) + i\text{Im}(C) & B
\end{bmatrix},
\label{eq:OneQubitDensityMatrix1}
\end{equation}
where we allow for four unknown variables ($A$, $B$, $\text{Re}(C)$, and $\text{Im}(C)$). Equation \ref{eq:OneQubitDensityMatrix1} is constrained to have a unit trace and is Hermitian. While the eigenvalues are real, they are not constrained to be strictly positive. To ensure the density matrix is also positive semi-definite, we follow the design presented in \cite{PhysRevA.64.052312} and \cite{ALTEPETER2005105} using the fact that any positive-definite matrix can be written in the form $\mathbf{T}^\dagger\mathbf{T}$ (for matrix $\mathbf{T}$) because
\begin{equation}
\langle \psi | \mathbf{T}^\dagger\mathbf{T} |\psi\rangle = \langle \psi' | \psi'\rangle \geq 0.
\end{equation}
A lower-diagonal matrix was used for ease of inversion in both \cite{PhysRevA.64.052312} and \cite{ALTEPETER2005105}. We also use the same form for simplicity (but not necessarily for ease of inversion) and prescribe the elements in $\mathbf{T}$ as
\begin{equation}
\mathbf{T} = 
\begin{bmatrix}
t_0 &0 \\
t_1+i t_2 & t_3
\end{bmatrix},
\end{equation}
where $t_0,t_1,t_2,t_3\in\mathbb{R}$. The resulting matrix $\mathbf{T}^\dagger\mathbf{T}$ is
\begin{align}
\mathbf{T}^\dagger \mathbf{T}
&= \begin{bmatrix}
t_0 & t_1-i t_2 \\
0 & t_3
\end{bmatrix} \cdot
\begin{bmatrix}
t_0 & 0 \\
t_1+i t_2 & t_3
\end{bmatrix} \nonumber\\
&= \begin{bmatrix}
t_0^2 + t_1^2 + t_2^2 & t_1 t_3 - i t_2 t_3 \\
t_1 t_3 + i t_2 t_3 & t_3^2
\end{bmatrix}.
\end{align}
To enforce the unity trace condition, we normalize by the trace such that $\bm{\rho} = \mathbf{T}^\dagger\mathbf{T}/\text{Tr}\left(\mathbf{T}^\dagger\mathbf{T}\right)$ as
\begin{equation}
\bm{\rho} = \frac{1}{t_0^2 + t_1^2 + t_2^2 + t_3^2}
\begin{bmatrix}
t_0^2 + t_1^2 + t_2^2 & t_1 t_3 - i t_2 t_3 \\
t_1 t_3 + i t_2 t_3 & t_3^2
\end{bmatrix}.
\label{eq:ProbDensityMatrix0}
\end{equation}
Comparing Eq. \ref{eq:OneQubitDensityMatrix1} and \ref{eq:ProbDensityMatrix0}, we immediately find
\begin{align}
A &= t_0^2 + t_1^2 + t_2^2 \nonumber\\
B &= t_3^2 \nonumber \\
\text{Re}C &= t_1 t_3 \nonumber \\
\text{Im}C &= t_2 t_3
\label{eq:MatrixElements0}
\end{align}
with the Stokes parameters being
\begin{align}
S_1 &= 2\,\text{Re}C \nonumber \\
S_2 &= 2\,\text{Im}C \nonumber \\
S_3 &= A - B 
\end{align}

Again, it should be noted that the above formalism suggests we need to solve for four unknowns. However, $t_3$ is technically constrained such that $t_3^2 = 1 - t_0^2-t_1^2-t_2^2$. This constraint could be incorporated in the model allowing us to do away with the normalization factor. However, we found significantly faster performance in the numerical simulation by leaving it as an independent variable knowing that the problem still admits a globally maximum solution.

To simplify the model formulation moving forward, we restrict the nomenclature to use only $A$, $B$, $\text{Re}(C)$, and $\text{Im}(C)$ knowing Eq. \ref{eq:MatrixElements0} applies.

To construct polarization-measurement probabilities, we make arbitrary polarization projections. The following notation uses bold values to designate vector quantities ($\in \mathbb{R}^n$ for $n$ measurements). Because we use a PBS for our projections, we assume that the state projection $|\bm{\psi}\rangle = \left[\mathbf{h} \,\, \mathbf{v}\right]^T$ is properly normalized with horizontal component $\mathbf{h} \in \{0,1\}$ and vertical component $\mathbf{v}\in \{0,1\}$. As we measure output ports of our PBS independently, $\mathbf{h}$ and $\mathbf{v}$ can only ever have values 0 or 1 and will never both be simultaneously 1. Utilizing the arbitrary waveplate operator for presented in Eq. \ref{eq:AWP}, we define general quarter- and half-waveplate operators as $\bm{AWP}(\eta_Q,\theta_Q)$ and $\bm{AWP}(\eta_H,\theta_H)$, respectively. The resulting probability $\bm{P}_\psi$ of obtaining a pure-state polarization projection $\bm{\psi}$ from an unknown quantum state $\bm{\rho}$ is expressed as

\begin{align}
\bm{P}_\psi = &\left\langle \bm{\psi} \left| \Bigl( \bm{AWP}\left(\bm{\eta_H}, \bm{\theta_H}\right)\cdot \bm{AWP}\left(\bm{\eta_Q}, \bm{\theta_Q}\right)\Bigr)^\dagger \cdot \bm{\rho} \cdot \Bigl( \bm{AWP}\left(\bm{\eta_H}, \bm{\theta_H}\right)\cdot \bm{AWP}\left(\bm{\eta_Q}, \bm{\theta_Q}\right)\Bigr) \right| \bm{\psi} \right\rangle \nonumber\\
= &\;\frac{1}{32} \Biggl\{ 4(5A\mathbf{h} + 3B\mathbf{h} + 3A\mathbf{v}+5B\mathbf{v}) + 2(A-B)(\mathbf{h}-\mathbf{v})\cos(\bm{\eta_H}-\bm{\eta_Q}) \nonumber\\ 
&+(\mathbf{h}-\mathbf{v})\Biggl[2(A-B)\cos(\bm{\eta_H}+\bm{\eta_Q})+4\cos(\bm{\eta_Q})\Bigl(A-B+8\text{Im}C\sin(\bm{\eta_H})\sin(2\bm{\theta_H})\Bigr) \nonumber\\
&\;\;\;\;\;\;\;\;+ 16\cos^2(\frac{\bm{\eta_Q}}{2})\sin^2(\frac{\bm{\eta_H}}{2})\Bigl((A-B)\cos(4\bm{\theta_H})+2\text{Re}C\sin(4\bm{\theta_H})\Bigr) \nonumber\\
&\;\;\;\;\;\;\;\; + 32\cos(2\bm{\theta_Q})\sin(\bm{\eta_Q})\Bigl(\text{Re}C\sin(\bm{\eta_H})\sin(2\bm{\theta_H})-\text{Im}C\sin^2(\frac{\bm{\eta_H}}{2})\sin(4\bm{\theta_H})\Bigr) \nonumber\\
&\;\;\;\;\;\;\;\; + 8\cos(4\bm{\theta_Q})\sin^2(\frac{\bm{\eta_Q}}{2})\Bigl[A-B+2\sin^2(\frac{\bm{\eta_H}}{2}) \Bigl( (A-B)\cos(4\bm{\theta_H})-2\text{Re}C\sin(4\bm{\theta_H}) \Bigr) \Bigr] \nonumber\\
&\;\;\;\;\;\;\;\; - 8\sin(\bm{\eta_Q})\sin(2\bm{\theta_Q})\Bigl(-4\text{Im}C\cos^2(2\bm{\theta_H}) + 2\text{Im}C\cos(\bm{\eta_H})\cos(4\bm{\theta_H})+2(A-B)\sin(\bm{\eta_H})\sin(2\bm{\theta_H})\Bigr) \nonumber\\
&\;\;\;\;\;\;\;\; - 4\sin^2(\frac{\bm{\eta_Q}}{2})\sin(4\bm{\theta_Q})\Bigl[-4\text{Re}C-4\sin^2(\frac{\bm{\eta_H}}{2})\Bigl(2\text{Re}C\cos(4\bm{\theta_H})+(A-B)\sin(4\bm{\theta_H})\Bigr)\Bigr] \Biggr] \nonumber\\
&+ 4(\mathbf{h}-\mathbf{v})\cos(\bm{\eta_H})\biggl[A-B+4\text{Im}C\sin(\bm{\eta_Q})\sin(2\bm{\theta_Q}) \nonumber\\
&\;\;\;\;\;\;\;\;+ 2\sin^2(\frac{\bm{\eta_Q}}{2})\Bigl((A-B)\cos(4\bm{\theta_Q})+2\text{Re}C\sin(4\bm{\theta_Q})  \Bigr) \biggr] \Biggr\}.
\label{eq:SingleQubitProb}
\end{align}

To include the effects that PBS crosstalk and detector efficiencies have on the photon flux measured at the detector, we define several additional variables here. The photon flux is designated as $\bm{\mathcal{F}}$. The transmitted (reflected) detection efficiency (which includes any fiber coupling efficiency) is defined as $\mu$ ($\nu$). The PBS crosstalk for transmitted horizontal polarization (transmitted vertical polarization) probability is defined as $T_H$ ($T_V$) and the reflected vertical polarization (reflected horizontal polarization) probability is defined as $R_V$ ($R_H$).
Thus, a crosstalk matrix (which incorporates the PBS transmission and reflection components together with the fiber coupling and detector efficiencies) affects the anticipated number of photons per polarization projection $\mathcal{N}_{\psi}$ detected at the output as
\begin{equation}
\begin{bmatrix}
\mathcal{N}_{|H\rangle} \\
\mathcal{N}_{|V\rangle}
\end{bmatrix} =
\mathcal{F} \cdot \Bigl(
\begin{bmatrix}
\mu T_H  & \mu T_V \\
\nu R_H & \nu R_V
\end{bmatrix}
\cdot
\begin{bmatrix}
P_{|H\rangle}\\
P_{|V\rangle}
\end{bmatrix}  \Bigr).
\label{eq:OneQubitCrosstlk}
\end{equation}
Likewise, the same form holds for calculating $[\mathcal{N}_{|D\rangle}, \mathcal{N}_{|A\rangle}]^T$ and $[\mathcal{N}_{|L\rangle}, \mathcal{N}_{|R\rangle}]^T$.

We are interested in a non-adaptive tomography with six measurement settings to sample our unknown quantum state. The matrix of measurement settings, where the row designates the polarization-dependent output photon number $\mathcal{N}_\psi$ and the columns designate the required settings for each variable, is dependent on whether the experiment uses liquid crystal phase retarders or a standard QWP and HWP. Settings for experiments using liquid-crystal waveplates (where the angles are fixed and the retardance is changed with voltage) are given in Eq. \ref{eq:SettingsLC} while settings using standard waveplates (where the retardance is fixed and the angles are changed by rotations) are given in Eq. \ref{eq:SettingsQH} as 
\begin{equation}
\text{Settings}_{\left(\text{Liquid Crystal Waveplates}\right)} = 
\kbordermatrix{                & \bm{\eta_H} & \bm{\eta_Q} & \bm{\theta_H} & \bm{\theta_Q} & \mathbf{h} & \mathbf{v} \cr
                \mathcal{N}_H  &       0     &        0       &   \pi/8    &    \pi/4      &      1     &    0    \cr
                \mathcal{N}_V  &       0     &        0       &   \pi/8    &    \pi/4      &      0     &    1    \cr
                \mathcal{N}_D  &      \pi    &        0       &   \pi/8    &    \pi/4      &      1     &    0    \cr
                \mathcal{N}_A  &      \pi    &        0       &   \pi/8    &    \pi/4      &      0     &    1    \cr
				\mathcal{N}_L  &       0     &      \pi/2     &   \pi/8    &    \pi/4      &      1     &    0    \cr
				\mathcal{N}_R  &       0     &      \pi/2     &   \pi/8    &    \pi/4      &      0     &    1    }
\label{eq:SettingsLC}
\end{equation}
and
\begin{equation}
\text{Settings}_{\left(\text{Standard Waveplates}\right)} = 
\kbordermatrix{                & \bm{\eta_H} & \bm{\eta_Q} & \bm{\theta_H} & \bm{\theta_Q} & \mathbf{h} & \mathbf{v} \cr
                \mathcal{N}_H  &      \pi    &      \pi/2     &     0      &      0      &      1     &    0    \cr
                \mathcal{N}_V  &      \pi    &      \pi/2     &     0      &      0      &      0     &    1    \cr
                \mathcal{N}_D  &      \pi    &      \pi/2     &   \pi/8    &    \pi/4    &      1     &    0    \cr
                \mathcal{N}_A  &      \pi    &      \pi/2     &   \pi/8    &    \pi/4    &      0     &    1    \cr
				\mathcal{N}_L  &      \pi    &      \pi/2     &     0      &    \pi/4    &      1     &    0    \cr
				\mathcal{N}_R  &      \pi    &      \pi/2     &     0      &    \pi/4    &      0     &    1    }.
\label{eq:SettingsQH}
\end{equation}

Using the settings defined in either Eq. \ref{eq:SettingsLC} or Eq. \ref{eq:SettingsQH}, we express the vector of all six output photon numbers as 
\begin{equation}
\begin{bmatrix}
\mathcal{N}_{|H\rangle} \\
\mathcal{N}_{|V\rangle} \\
\mathcal{N}_{|D\rangle} \\
\mathcal{N}_{|A\rangle} \\
\mathcal{N}_{|L\rangle} \\
\mathcal{N}_{|R\rangle}
\end{bmatrix} 
= 
\begin{bmatrix}
\mu T_H & \mu T_V &         & \cdots  &         & 0       \\
\nu R_H & \nu R_V &         &         &         &         \\
        &         & \mu T_H & \mu T_V &         & \vdots  \\
\vdots  &         & \nu R_H & \nu R_V &         &         \\
        &         &         &         & \mu T_H & \mu T_V \\
0       &         & \cdots  &         & \nu R_H & \nu R_V
\end{bmatrix}
\cdot
\begin{bmatrix}
P_{|H\rangle} \cdot \mathcal{F}_{HV} \\
P_{|V\rangle} \cdot \mathcal{F}_{HV} \\
P_{|D\rangle} \cdot \mathcal{F}_{DA} \\
P_{|A\rangle} \cdot \mathcal{F}_{DA} \\
P_{|L\rangle} \cdot \mathcal{F}_{LR} \\
P_{|R\rangle} \cdot \mathcal{F}_{LR}
\end{bmatrix},
\label{eq:OneQubitModel}
\end{equation}
where $\mathcal{F}_{HV}$ represents the total flux for the horizontal and vertical polarizations. Likewise, $\mathcal{F}_{DA}$ and $\mathcal{F}_{LR}$ represent the total flux for their respective polarizations. By not assuming a single constant flux for all polarizations, the outputs tend to be more accurate with experimental data as the flux can potentially change between measurements due to laser fluctuations or some other mechanism. 

With the measurement model complete, a likelihood function consisting of a Gaussian distribution $\mathcal{G}(t_i,\mathcal{N},\sigma)$ having the model parameters $t_i$ given in Eq. \ref{eq:MatrixElements0}, a mean photon number $\mathcal{N}$, and a standard deviation $\sigma=\sqrt{\mathcal{N}}$ is used to approximate the Poisson distribution to model single-photon-counting statistics. By approximating the Poisson distribution using a continuous function, the numerical sampler tends to converge more quickly than with discrete distributions. Thus, we define a likelihood function $\mathcal{L}(t_i | \mathcal{N}_j)$ for $j=1,2,...,6$ measurements as 
\begin{equation}
\mathcal{L}(t_i | \mathcal{N}_i) = 
\prod\limits_{j=1}^6\mathcal{G}(t_i,\mathcal{N}_j,\sqrt{\mathcal{N}_j}).
\label{eq:likelihoodFn}
\end{equation}

MLE is often used to minimize the final convex negative log-likelihood of Eq. \ref{eq:likelihoodFn} to find an optimal physical quantum state. Due to the problem formulation, the $t_i$ do not have unique solutions that minimize the negative log-likelihood. However, we can rest assured that the problem formulation will return a unique global optimal density matrix $\bm{\rho}$ with elements $A$, $B$, $\text{Re}C$, and $\text{Im}C$. This is because all local minima of the unconstrained parameterized problem $\mathcal{L}(\bm{\rho}(t_i)$) are equivalent in $\mathcal{L}(\bm{\rho})$. Thus, they are also the global minimizer for all convex $\mathcal{L}$ \cite{gonccalves2011local}. 

\subsubsection{Bayesian Framework}

Rather than proceeding with MLE, we add prior information and demonstrate how to model the problem according to a Bayesian framework. Bayesian analysis is incorporated by assigning probability distributions for each variable. A detailed coding example can be found within the appendix for clarification.

As with MLE, we assign the same likelihood function given in Eq. \ref{eq:likelihoodFn} used to model the detector counts $\bm{\mathcal{N}}$. In contrast to MLE, we also add prior information about the unknown variables and our measurement uncertainty. With little to no previous knowledge of the input state, the primary variables of interest $t_0$, $t_1$, $t_2$, and $t_3$ are each assigned a uniform distribution $\mathcal{U}(-1,1)$ with lower limit $-1$ and upper limit $1$.

Finally, we assign Gaussian distributions to the remaining variables with mean values using the desired settings in Eq. \ref{eq:SettingsLC} and experimentally estimate the standard deviations. To estimate the input photon flux for each set of measurements ($\mathcal{F}_{HV}$, $\mathcal{F}_{DA}$, and $\mathcal{F}_{LR}$), we effectively use the pseudoinverse of the crosstalk matrix to arrive at the input flux as
\begin{equation}
\mathcal{F}_{HV} = 
\begin{bmatrix}
1 & 1
\end{bmatrix} \cdot
\begin{bmatrix}
\mu T_H & \mu T_V \\
\nu R_H & \nu R_V
\end{bmatrix}^{-1} \cdot
\begin{bmatrix}
\mathcal{N}_{|H\rangle} \\
\mathcal{N}_{|V\rangle}
\end{bmatrix}.
\label{eq:InverseCrsstlak}
\end{equation}
Repeating the process in Eq. $\ref{eq:InverseCrsstlak}$ for $\mathcal{F}_{DA}$ and $\mathcal{F}_{LR}$, we obtain reasonable estimates for the input flux. We note that while we explicitly provide an estimate for the mean photon flux above, the PyMC model would arrive at the same result if we were to leave the mean input photon numbers as unknown random variables with distributions of their own. However, providing this estimate ourselves reduces the computation time by reducing the number of free variables.

In the model presented below, we neglect any covariance that could exist between the waveplate settings (as not all waveplates need adjustments between measurements). This assumption can easily be altered if desired by using multivariate-normal distributions. Using uniform distributions $\mathcal{U}(-1,1)$ with limits $-1$ and $1$, Gaussian distributions $\mathcal{G}(n,\sigma_n)$ with mean $n$, standard deviation $\sigma_n$ and limits (if applicable), previously defined variables are assigned probability distributions as follows:
\begin{align}
t_0 \rightarrow \mathcal{U}(-1,1)& \;\;\;
t_1 \rightarrow \mathcal{U}(-1,1) \nonumber \\
t_2 \rightarrow \mathcal{U}(-1,1)& \;\;\;
t_3 \rightarrow \mathcal{U}(-1,1) \nonumber \\
\bm{\eta_Q} \rightarrow \mathcal{G}(\bm{\eta_Q},\sigma)& \;\;\;
\bm{\eta_H} \rightarrow \mathcal{G}(\bm{\eta_H},\sigma) \nonumber \\
\bm{\theta_Q} \rightarrow \mathcal{G}(\bm{\theta_Q},\sigma_{\theta_Q})& \;\;\;
\bm{\theta_H} \rightarrow \mathcal{G}(\bm{\theta_H},\sigma_{\theta_H}) \nonumber \\
\mu\rightarrow \mathcal{G}(\mu,\sigma_{\mu})|_{\min = 0,\;\max = 1}&  \;\;\;
\nu\rightarrow \mathcal{G}(\nu,\sigma_\nu)|_{\min = 0,\;\max = 1} \nonumber \\
T_H\rightarrow \mathcal{G}(T_H,\sigma_{T_H})|_{\min = 0,\;\max = 1}& \;\;\;
T_V\rightarrow \mathcal{G}(T_V,\sigma_{T_V})|_{\min = 0,\;\max = 1}\nonumber \\
R_H\rightarrow \mathcal{G}(R_H,\sigma_{R_H})|_{\min = 0,\;\max = 1}& \;\;\;
R_V\rightarrow \mathcal{G}(R_V,\sigma_{R_H})|_{\min = 0,\;\max = 1} \nonumber \\
\bm{\mathcal{F}} &\rightarrow \mathcal{G}(\bm{\mathcal{F}},\sqrt{\bm{\mathcal{F}}}) \nonumber \\
\bm{\mathcal{N}}_{\text{likelihood}} &\rightarrow \mathcal{G}(\bm{\mathcal{N}},\sqrt{\bm{\mathcal{N}}})
\label{eq:OneQubitPyMC}
\end{align}

Note the lack of error associated with the $\mathbf{h}$ and $\mathbf{v}$ variables as there is no uncertainty associated with the polarization measurement from the PBS. 

Python's PyMC is used to calculate the posterior distribution and assign numerical trace distributions to each variable with sufficient sampling. From each variable's trace distribution, we calculate a mean and highest density interval (HDI). The HDI for a 95\% credible interval encompasses 95\% of the probability such that any point in the interval has a higher probability than any point outside the interval. While not always continuous, it is advantageous over the standard deviation in that it works well for asymmetric distributions (such as those that arise with pure states) and will not place an error estimate outside the Bloch sphere. Other credible intervals can be defined; e.g., 89\% is often regarded as being more stable or reliable if sampling is not sufficient. Rather than attempt to propagate the uncertainty from $\{t_0,t_1,t_2,t_3\}$ into $\{A,B,\text{Re}C,\text{Im}C\}$ ourselves, we have PyMC propagate the numerically calculated trace distributions directly. The same is true for the Stokes parameters and any value calculated from the resulting density matrix (such as entropy, expectation values, etc).

\subsection{Experimental single-qubit QST}

To demonstrate the feasibility of our mode, we experimentally prepare an ensemble of pure states and perform a Bayesian QST. Because we are using (to the best of our knowledge) the horizontal pure polarization state $|H\rangle$, our density matrix $\bm{\rho}$ should take the form
\begin{equation}
\bm{\rho} = 
\begin{bmatrix}
1 & 0  \\
0 & 0 \\
\end{bmatrix}
\label{eq:MaxMixedState}
\end{equation}
with a Stokes vector of $\bm{S} = [0,0,1]$. Thus, the state estimate should reside on the boundary of the Bloch sphere.

Using the settings provide by Eq. \ref{eq:SettingsLC} with variable liquid variable waveplates, a pulsed laser operating at 1550 nm attenuated down to $< 1$ photon per pulse, two InGaAs single photon detectors, and the transmission of a PBS to generate an ensemble of $|H\rangle$ photons, we experimentally perform a QST. 

\subsubsection{Measurement uncertainties}

To estimate our experimental uncertainties with each polarization projective measurement, we note that the dark-count rate for the detector in the PBS transmission arm (reflection arm) occurred at 0.046 (0.706) dark counts per detected photon. These dark counts are first subtracted from the raw counts. The detector sensitivity was also biased due to discriminator voltage settings. Not being able to calculate an overall single-photon detection efficiency, we obtain relative values and find that the detector in the PBS transmission arm detected photons 0.552 times as likely as the detector in the reflected PBS arm for the same flux inputs. Finally, the fiber-optic coupling efficiencies using multimode fiber amounted to the PBS transmission arm having a 0.76 coupling efficiency while the PBS reflection arm had a 0.75 coupling efficiency. Multiplying the relative detection efficiencies with the fiber coupling efficiencies and conservatively assuming we know the value to within $\pm 3\%$ based on power-meter readings, we obtain estimated distributions for $\mu$ and $\nu$ as 
\begin{align}
\mu &\rightarrow \mathcal{G}_{\mu}(0.42,\;0.03)|_{\min = 0,\;\max = 1} \nonumber \\
\nu &\rightarrow \mathcal{G}_{\nu}(0.75,\;0.03)|_{\min = 0,\;\max = 1}.
\end{align}

The PBS crosstalk error was obtained via power-meter measurements of transmitted and reflected $|H\rangle$ and $|V\rangle$ input states. Using power deviations, the crosstalk uncertainty is estimated at $\approx 1\%$. Thus, the model crosstalk parameters recorded are
\begin{align}
T_H &\rightarrow \mathcal{G}_{T_H}(0.973,\;0.01) \nonumber \\
R_H &\rightarrow \mathcal{G}_{R_H}(0.013,\;0.01) \nonumber \\
T_V &\rightarrow \mathcal{G}_{T_V}(0.987,\;0.01) \nonumber \\
R_V &\rightarrow \mathcal{G}_{R_V}(0.027,\;0.01).
\end{align}

Finally, the liquid-crystal arbitrary waveplates each have a voltage dependent phase retardance and an angle setting. Each arbitrary waveplate was measured independently by setting the other waveplate's phase retardance to 0 (using the maximum voltage). To estimate errors, PBS projections were used to measure the degree of polarization rotations for a known input state. The deviation from the desired state after the rotation was then attributed entirely to either an angular error or a phase-retardance error to arrive at an upper error-limit for each variable. In short, the angle settings were off by no more than 1 degree while the phase retardance settings were off by no more than 2 degrees. We arrive at the following distributions for the liquid-crystal waveplates as
\begin{align}
\bm{\eta_Q} &\rightarrow \mathcal{G}(\bm{\eta_Q},\;2\pi/180) \nonumber \\
\bm{\eta_H} &\rightarrow \mathcal{G}(\bm{\eta_H},\;2\pi/180) \nonumber \\
\bm{\theta_Q} &\rightarrow \mathcal{G}(\bm{\theta_Q},\;\pi/180) \nonumber \\
\bm{\theta_H} &\rightarrow \mathcal{G}(\bm{\theta_H},\;\pi/180).
\end{align}

\subsubsection{QST results}

\begin{figure}
  \includegraphics[width=.8\linewidth]{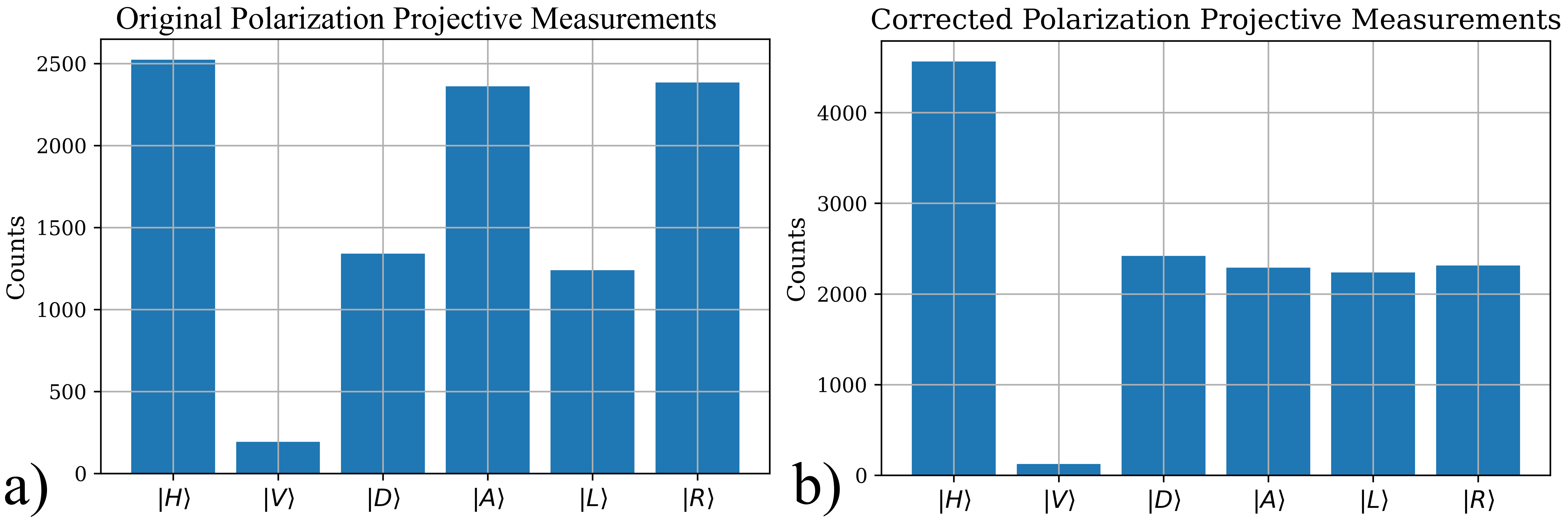}
  \caption{\textbf{(a)} Original (unaltered) photon counts per polarization projection for a known horizontal input state. \textbf{(b)} Processed photon counts when correcting for each detector's relative efficiency mismatch and fiber-coupling efficiencies. This corrected data follows intuition more closely (knowing the horizontal-state input).}
  \label{fig:Counts}
\end{figure}

After integrating for 100 seconds per measurement, single-photon counts as a function of the six polarization projections (having subtracting dark counts) can be found in Fig. \ref{fig:Counts} (a). Knowing the ``unknown'' state in advance, the ratio of counts are not what one would expect with a $|H\rangle$ polarization state. As a sanity check, if we correct for the detector bias and fiber coupling efficiencies, we arrive at Fig. \ref{fig:Counts} (b), which is more in line with our intuition.

\begin{figure}
  \includegraphics[width=.6\linewidth]{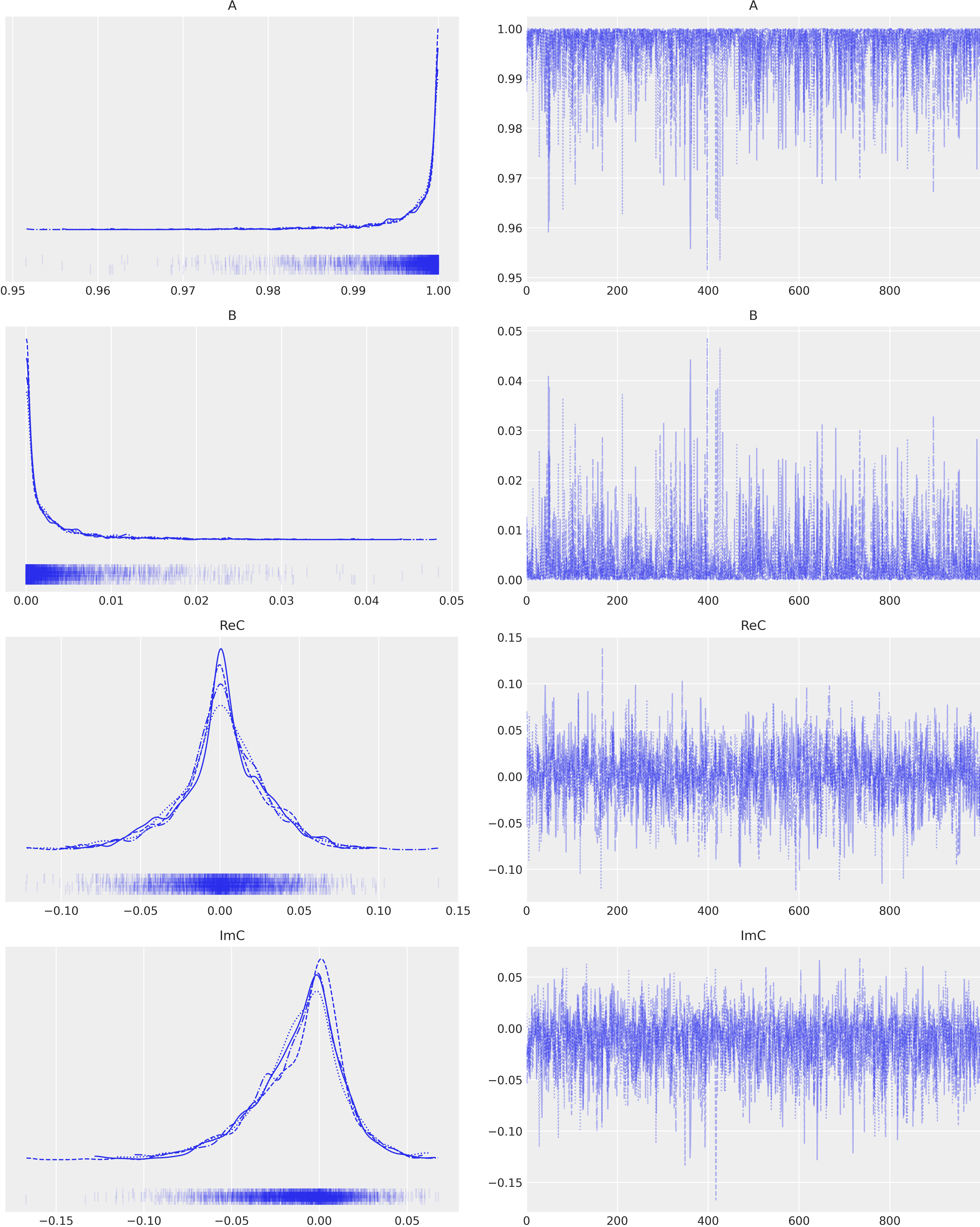}
  \caption{Raw traces output directly from the numerical sampling of the density matrix components. Histograms for each variable are presented on the left column while the raw samples are presented on the right column.}
  \label{fig:traceH}
\end{figure}

Using the code provided in the appendix, the dark-count corrected data together with the model distributions defined above are fed into the Bayesian model where the sampler is told draw 7200 samples distributed across 4 cores. Thus, after tuning and then discarding 800 samples per core, only 1000 draws per core for all variable are kept. Histograms of the raw traces for variables $A$, $B$, $\text{Re}C$, and $\text{Im}C$ are calculated from the $t_i$ traces and are presented in Fig. \ref{fig:traceH}. The BME and 95\% HDI are given as
\begin{equation}
\bm{\rho}_{\text{BME}} = 
\begin{bmatrix}
0.997           & 0.001 + 0.011i \\
0.001 - 0.011i & 0.003
\end{bmatrix},
\label{eq:BME}
\end{equation} 
while the 95\% HDIs for each element of $\bm{\rho}_{\text{BME}}$ are presented as
\begin{equation}
\bm{\rho}_{\text{HDI}_{95}} = 
\begin{bmatrix}
[0.986,\,1.0]                    & [-0.057,\,0.061] - [-0.069,\,0.034]i \\
[-0.057,\,0.061] + [-0.069,\,0.034]i & [0.000,\,0.001]
\end{bmatrix}.
\label{eq:rhoBMI}
\end{equation}

\begin{figure}
  \includegraphics[width=.5\linewidth]{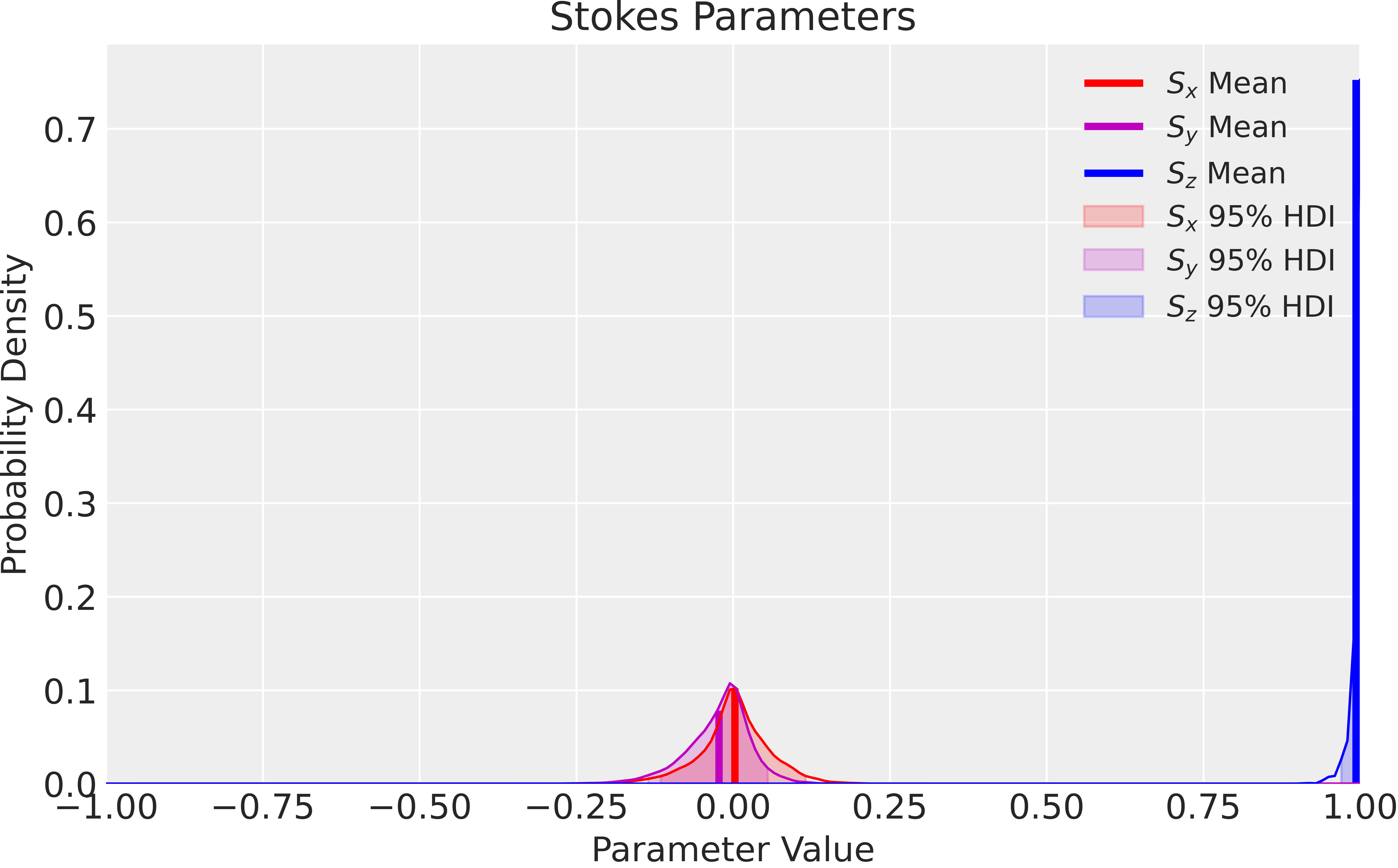}
  \caption{Distributions for each of the three Stokes parameters as output by PyMC. The width of each distribution is governed by uncertainties in the shot-noise, the waveplates (including angles and retardances), the fiber-coupling efficiencies, the PBS crosstalk, and the relative detector imbalances.}
  \label{fig:tomoH}
\end{figure}

Again, using the traces obtained for the $t_i$ model parameters, distributions for the Stokes parameters are presented in Fig. \ref{fig:tomoH}. We can immediately see the Stokes parameters BME is
\begin{equation}
\bm{S}_{\text{BME}}=
\begin{bmatrix}
0.003 & -0.023 & 0.994
\end{bmatrix}
\label{eq:StokesBMI}
\end{equation}
with the 95\% HDI being
\begin{equation}
\bm{S}_{\text{HDI}_{95}}=
\begin{bmatrix}
[-0.113,\,0.122] & [-0.138,\,0.073] & [0.972,\,1.000]
\end{bmatrix}.
\label{eq:StokesBMI}
\end{equation}

\begin{figure}
  \includegraphics[width=.8\linewidth]{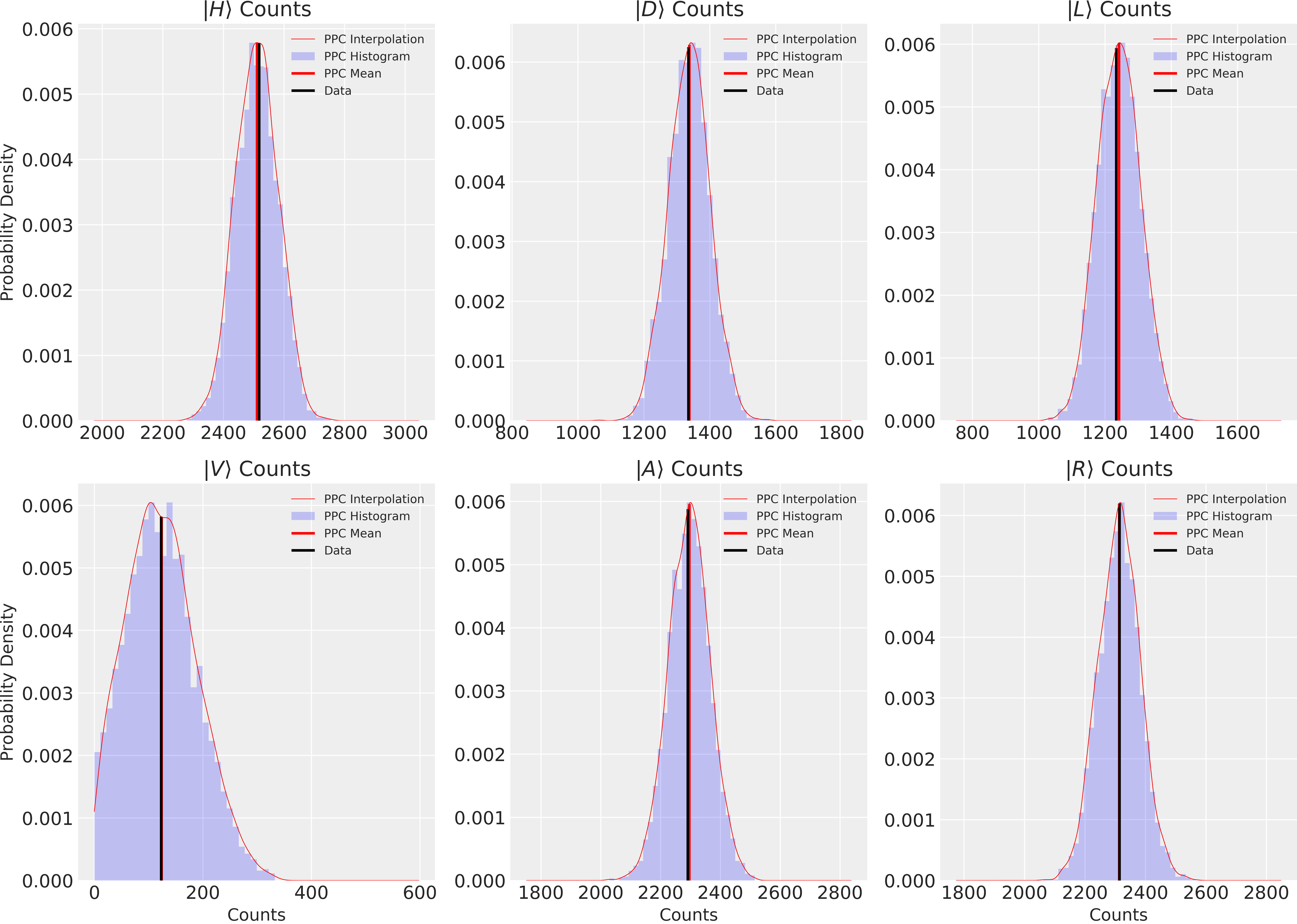}
  \caption{After obtaining our model parameters, the posterior distribution is sampled to synthesize data that should be comparable to the input data. Given the synthesized data is consistent with the original data, the posterior-predictive checks lends credence to the validity of the model and its parameters.}
  \label{fig:ppc}
\end{figure}

As a check to help verify that our model is reasonable, we use the obtained posterior distribution and our model to generate histograms of synthetic data and compare them with our real data. If the model is reliable, the mean of the synthetically generated data should reside near the actual data. Figure \ref{fig:ppc} plots the results of our posterior-predictive checks for each polarization projection. Indeed, the mean values for the synthetically generated data sets reside near the values recorded in our measurements. 

Note that these checks are similar to methods used to verify MLE methods. The difference here is that we are not using predictive checks to simulate data that is then fed through our algorithm to estimate errors based on the spread of solutions. Rather, we first propagate our uncertainties and then use the spread of solutions to check that our model is still consistent with the data.

\subsubsection{Low-flux experimental results}

\begin{figure}
  \includegraphics[width=.8\linewidth]{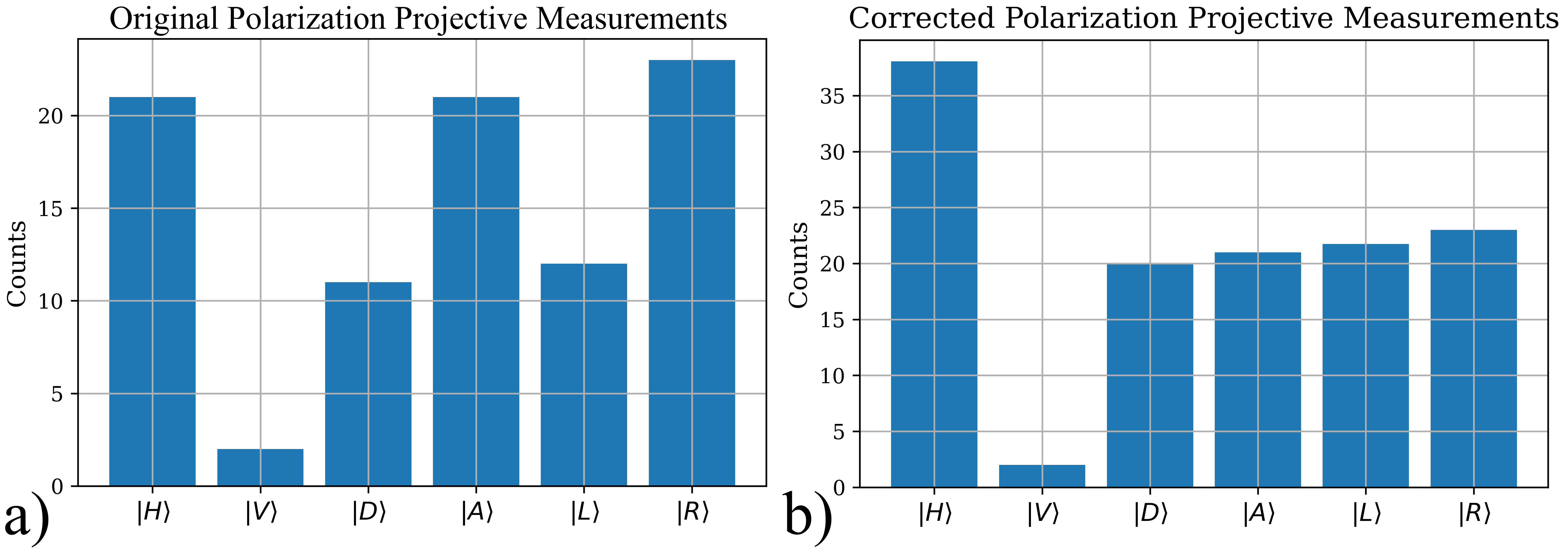}
  \caption{\textbf{(a)} Original (unaltered) photon counts per polarization projection for a known horizontal input state (with less photon flux). \textbf{(b)} Processed photon counts when correcting for each detector's relative efficiency mismatch and fiber-coupling efficiencies.}
  \label{fig:CountsLowFlux}
\end{figure}

To examine how additional noise will affect the final posterior distribution and the resulting BME, we consider the case where shot-noise is more prevalent. In the previous demonstration, approximately 1500 photon counts were observed by each detector. For a shot-noise limited signal (i.e., Poisson noise), the signal-to-noise (SNR) ratio is $\approx 39$. One would expect that the posterior distributions should grow wider with increasing shot-noise. As a test, we performed the experiment in which approximately 20 photons were observed by each detector per measurement such that the SNR is $\approx 4$ and observe how this affects the uncertainty in the BME. Raw photon counts from the experiment are shown in Fig. \ref{fig:CountsLowFlux} (a) while corrected counts when considering detector imbalance and fiber-coupling efficiencies are shown in Fig. \ref{fig:CountsLowFlux} (b).

\begin{figure}
  \includegraphics[width=.5\linewidth]{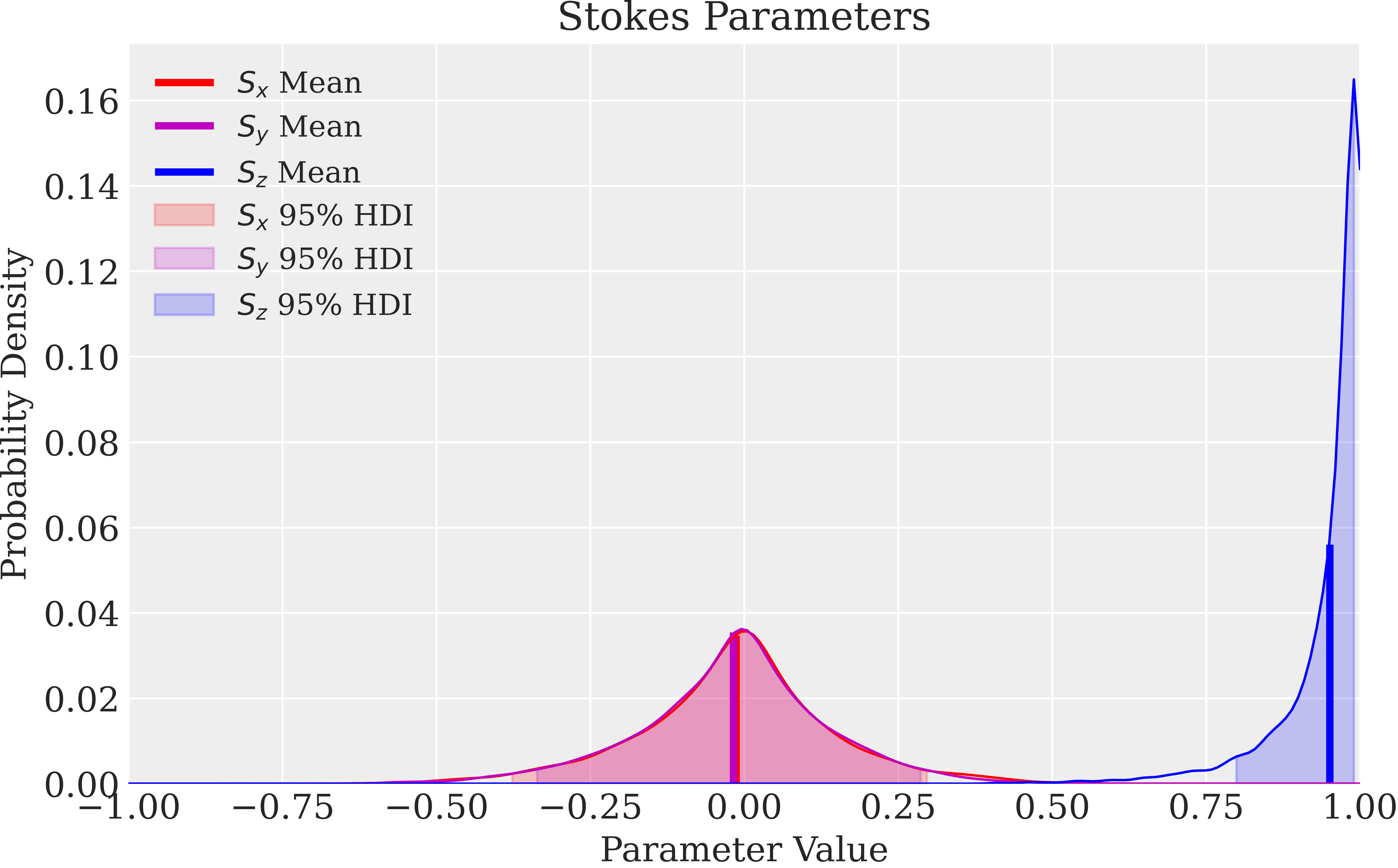}
  \caption{Distributions for each of the three Stokes parameters as output by PyMC. The width of each distribution is governed by uncertainties in the shot-noise and is less certain given the reduction in photon flux. All other experimental parameters are identical to the previous demonstration.}
  \label{fig:tomoHLowFlux}
\end{figure}

The reconstructed Stokes parameters after propagating the numerically obtained $t_i$ traces are shown in Fig. \ref{fig:tomoHLowFlux}. As expected, the distributions are wider (less certain) given the decrease in SNR. The BME for the reduced flux is
\begin{equation}
\bm{\rho}_{\text{BME}} = 
\begin{bmatrix}
.975           & -0.007 + 0.009i \\
-0.007 - 0.009i & 0.025
\end{bmatrix},
\label{eq:BME_LowFlux}
\end{equation} 
while the 95\% HDIs for each element of $\bm{\rho}_{\text{BME}}$ are presented as
\begin{equation}
\bm{\rho}_{\text{HDI}_{95}} = 
\begin{bmatrix}
[0.899,\,1.000]                    & [-0.187,\,0.152] - [-0.171,\,0.147]i \\
[-0.187,\,0.152] + [-0.171,\,0.147]i & [0.000,\,0.101]
\end{bmatrix}.
\label{eq:rhoBME_LowFlux}
\end{equation}
Calculating the Stokes parameters, we find the BME is
\begin{equation}
\bm{S}_{\text{BME}}=
\begin{bmatrix}
-0.014 & -0.018 & 0.951
\end{bmatrix}
\label{eq:StokesBMI}
\end{equation}
with the 95\% HDI being
\begin{equation}
\bm{S}_{\text{HDI}_{95}}=
\begin{bmatrix}
[-0.373,\,0.304] & [-0.344,\,0.305] & [0.798,\,1.000].
\end{bmatrix}
\label{eq:StokesBME_LowFlux}
\end{equation}

\begin{figure}
  \includegraphics[width=.8\linewidth]{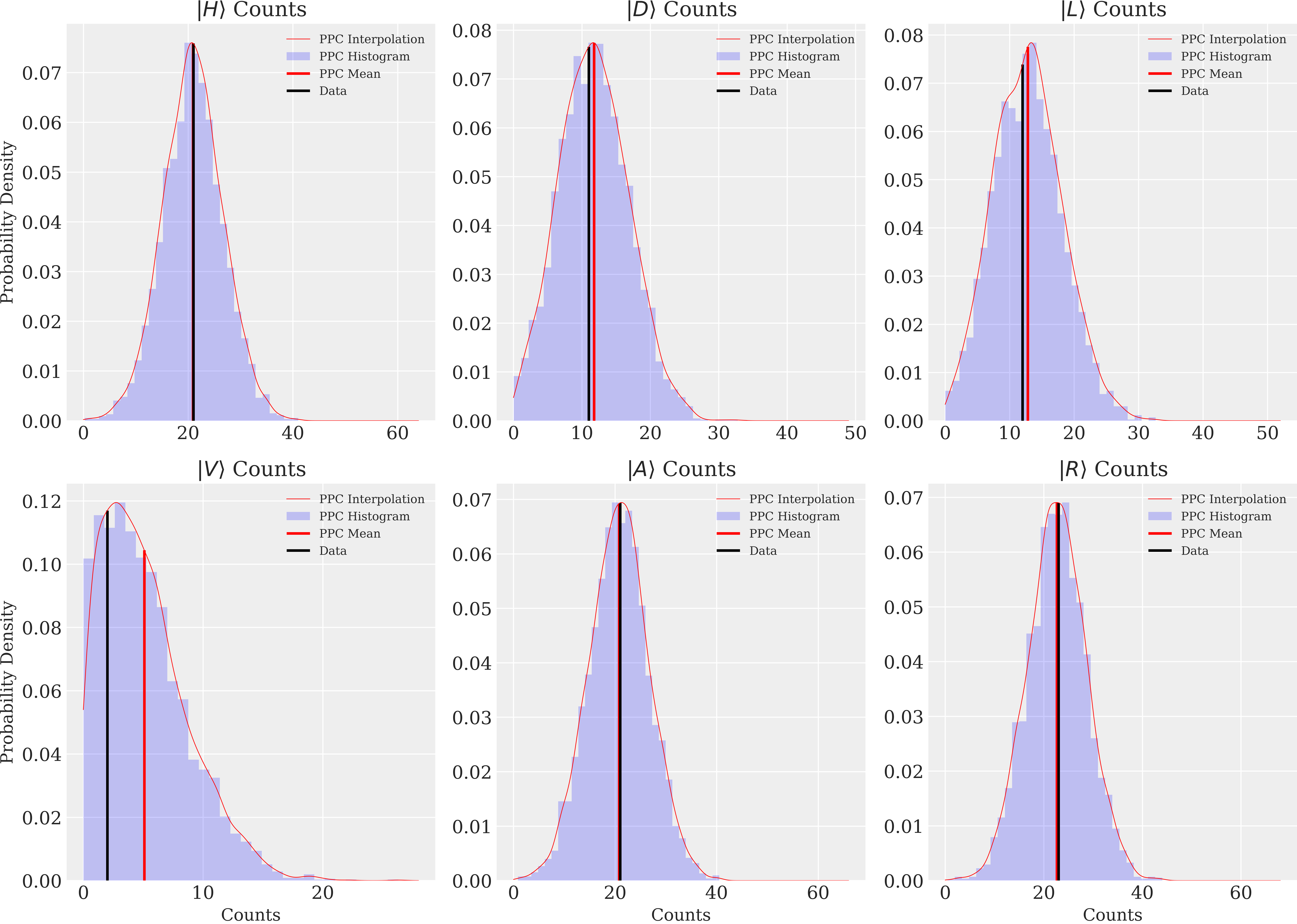}
  \caption{Posterior-predictive checks obtained by simulating data from the state estimate posterior distributions. While less accurate than the previous demonstration with more photon flux, the measured values still reside within the distributions of each simulated dataset and add supporting evidence to the model's validity.}
  \label{fig:ppcLowFlux}
\end{figure}

Thus far, the model behaves as expected. As one last check to help verify the model's validity, we again perform posterior-predictive checks and present the results in Fig. \ref{fig:ppcLowFlux}. Again, the distributions obtained by the simulated photon counts (using the $t_i$ parameter distributions in our model) are still consistent with the dark-count subtracted data. 

To conclude the single-qubit Bayesian QST section, we developed a measurement model and verified its use with experimentally obtained data. The model should prove useful to an experimentalist performing QST in a laboratory. By using the measurement settings and associated uncertainties and efficiencies, the code provided in the appendix demonstrates how to use PyMC to propagate the uncertainty to sample the posterior probability distribution and return trace distributions from which the Bayes mean estimate and highest density intervals may be calculated.

\section{Two-Qubit QST}

While we present a two-qubit formalism with orthogonal measurement-bases here, a more general framework extended to $n$-qubits is presented in \cite{ALTEPETER2005105}. The framework for generating a PyMC model with more qubits is similar to what is done above, except that the parameter space will result in a more time-consuming computation. 

As per the one-qubit case, we start by defining a two-qubit density matrix. A discrete bipartite system composed of two-dimensional qubits will have a density matrix following 
\begin{equation}
\bm{\rho} = \frac{1}{4}\sum\limits_{i,j=0}^3 S_{i,j}\,\bm{\sigma}_{i}\otimes\bm{\sigma}_{j},
\label{eq:twoqubitdensitymatrix}
\end{equation}
where the $\bm{\sigma}_i$ are once again the Pauli matrices, $S_{i,j}$ are the joint-space two-qubit Stokes parameters, and $\otimes$ is the Kronecker product. The objective is to find the ($4^2-1=15$) two-qubit Stokes parameters through projective measurements. Rather than record the measurement outcomes on each subsystem within the bipartite system individually, we must apply projective measurements on each subsystem while measuring in coincidence (only recording cases when both detectors appear to simultaneously fire).

The expectation values associated with the joint-space Pauli matrices yields the joint-space Stokes parameters as
\begin{equation}
\langle \bm{\sigma}_i\otimes \bm{\sigma}_j \rangle = \text{Tr}\left(\bm{\rho}\cdot\bm{\sigma}_i\otimes\bm{\sigma}_j\right) = S_i\otimes S_j \equiv S_{i,j},
\end{equation}
where the $S_i$ are the single-qubit Stokes parameters. With six measurements per photon in the pair, there exist 36 unique measurement settings needed to construct the bipartite Stokes parameters.

Again, the problem of simply measuring the Stokes parameters and not being guaranteed a positive-definite matrix with unity trace remains. Similar to how we constructed a physical model of the system via a likelihood function with priors to ensure the state estimate is physical, we apply the same technique to the two-qubit tomography in the next section.

\subsection{Two-Qubit Problem Formulation}

A two-qubit QST is much like the single-qubit case in that polarization projection measurements are performed on each qubit. However, there are now 15 unknown variables to solve for given the $4^n-1$ scaling (for $n$ qubits). Proceeding in a similar fashion, we define a density matrix $\bm{\rho} = \mathbf{T}^\dagger\mathbf{T}$ where
\begin{equation}
\mathbf{T} = \begin{bmatrix}
 t_0 & 0 & 0 & 0 \\
 t_1+it_2 & t_3 & 0 & 0 \\
 t_4+it_5 & t_6+it_7 & t_8 & 0 \\
 t_9+it_{10} & t_{11}+it_{12} & t_{13}+it_{14} & t_{15}
\end{bmatrix}.
\label{eq:2QubitTMatrix}
\end{equation}
The density matrix within our problem formulation has the structure
\begin{equation}
\bm{\rho} = \frac{1}{A+B+C+D}\begin{bmatrix}
  A & \text{Re}E - i\text{Im}E & \text{Re}F - i\text{Im}F & \text{Re}G - i\text{Im}G \\
  \text{Re}E + i\text{Im}E & B & \text{Re}H - i\text{Im}H & \text{Re}I - i\text{Im}I \\ 
  \text{Re}F + i\text{Im}F & \text{Re}H + i\text{Im}H & C & \text{Re}J - i\text{Im}J \\
  \text{Re}G + i\text{Im}G & \text{Re}I + i\text{Im}I & \text{Re}J + i\text{Im}J & D 
\end{bmatrix}
\label{eq:2QubitDensityMatrix}
\end{equation}
where 
\begin{align}
\mathcal{N} &= \sum\limits_{i=0}^{15}t_i^2 \nonumber \\
A &= \left( t_0^2+t_1^2+t_2^2+t_4^2+t_5^2+t_{9}^2+t_{10}^2 \right)/\mathcal{N}\nonumber \\
B &= \left(t_{3}^2+t_{6}^2+t_{7}^2+t_{11}^2+t_{12}^2 \right)/\mathcal{N}\nonumber \\
C &= \left( t_{8}^2+t_{13}^2+t_{14}^2 \right)/\mathcal{N} \nonumber \\
D &= t_{15}^2/\mathcal{N} \nonumber \\
\text{Re}E &= \left( t_{10}t_{12} + t_{1}t_{3}+t_{4}t_{6}+t_{5}t_{7}+t_{11}t_{9} \right)/\mathcal{N} \nonumber \\
\text{Im}E &= \left( t_{10}t_{12} + t_{1}t_{3}+t_{4}t_{6}-t_{5}t_{7}-t_{11}t_{9} \right)/\mathcal{N} \nonumber \\
\text{Re}F &= \left( t_{10}t_{14} + t_{4}t_{8}+t_{13}t_{9} \right)/\mathcal{N} \nonumber \\
\text{Im}F &= \left( t_{10}t_{13} + t_{5}t_{8}-t_{14}t_{9} \right)/\mathcal{N} \nonumber \\
\text{Re}G &= t_{15}t_{9}/\mathcal{N} \nonumber \\
\text{Im}G &= t_{15}t_{10}/\mathcal{N} \nonumber \\
\text{Re}H &= \left( t_{11}t_{13} + t_{12}t_{14} + t_{6}t_{8} \right)/\mathcal{N} \nonumber \\
\text{Im}H &= \left( t_{12}t_{13} - t_{11}t_{14} + t_{7}t_{8} \right)/\mathcal{N} \nonumber \\
\text{Re}I &= t_{11}t_{15}/\mathcal{N} \nonumber \\
\text{Im}I &= t_{12}t_{15}/\mathcal{N} \nonumber \\
\text{Re}J &= t_{13}t_{15}/\mathcal{N} \nonumber \\
\text{Im}J &= t_{14}t_{15}/\mathcal{N}.
\label{eq:TwoQubitTElements}
\end{align}
Notice that the above formalism suggests there are 16 unknowns when there should only be 15 unknowns. A unit trace density matrix would normally apply constraints to $t_{15}$ such that 
\begin{equation}
t_{15}^2 = 1 - \sum\limits_{i = 0}^{14}t_i^2,
\end{equation}
which would negate the need for calculating the normalization constant $\mathcal{N}$ in Eqn. \ref{eq:TwoQubitTElements}. However, PyMC appears to struggle with calculating the posterior distribution when $t_{15}$ is a dependent variable. Instead, allowing $t_{15}$ to be an independent variable results in the calculation completing within a small fraction of the original time.

With the above density matrix defined, we define a set of measurements to perform on each qubit separately. The two-qubit model deviates from the single-qubit measurement model in that we were unable to reduce the resulting equations affiliated with arbitrary waveplates (having both angle and phase retardance) into a form suitable for this article. Instead, we drop the phase-retardance degree of freedom, and we use the pure waveplate formalism having only angle dependences as in Eq. \ref{eq:WPOperations}. Labeling the subspace for each qubit as $A$ and $B$, we allow polarization rotations and projections to operate on each subspace independently. In keeping with the previous notation where the user needs to supply waveplate settings for each photon, we specify each subspace polarization-rotation operator as $\bm{U}_A(\theta_{Q\text{A}},\theta_{H\text{A}})$ and $\bm{U}_B(\theta_{Q\text{B}},\theta_{H\text{B}})$, where we must now keep track of four waveplate settings per measurement. When measuring in the polarization basis, this amounts to performing polarization-based projections on each photon and then measuring the results in coincidence. We allow for the same standard projective measurements ($|H\rangle$, $|V\rangle$, $|D\rangle$, $|A\rangle$, $|L\rangle$, and $|R\rangle$) used previously. Because the joint Hilbert-space in which our biphoton exists is $\in \mathbb{C}_{AB}^{4\times4}$, there now exist 36 possible combinations of mutually unbiased measurements. 

For a single projective coincidence measurement of a two-photon state, we rotate each photon as before to obtain desired state projections. Even though the state may actually be in a mixed state, the desired pure states for photon $A$ and photon $B$ take the form
\begin{align}
|\psi\rangle_\text{A} &= \bm{U}_\text{A}\left(\theta_{Q\text{A}},\theta_{H\text{A}}\right)\cdot [h_\text{A} \,\, v_\text{A}]^T \nonumber \\
|\phi\rangle_\text{B} &= \bm{U}_\text{B}\left(\theta_{Q\text{B}},\theta_{H\text{B}}\right)\cdot [h_\text{B} \,\, v_\text{B}]^T, 
\end{align}
(where $h$ and $v$ are horizontal and vertical polarization components). Letting the two-photon pure state be represented by $|\psi\rangle_\text{A}\otimes |\phi\rangle_\text{B} = |\psi \phi\rangle$, the probability of simultaneously measuring photon $A$ in pure state $|\psi\rangle_\text{A}$ while also measuring photon $B$ in pure state $|\phi\rangle_\text{B}$ is
\begin{equation}
\bm{P}_{\psi\phi} = \text{Tr}\left[ \bm{\rho} \cdot \bm{U}\left( \bm{\theta_{Q\text{A}}}, \bm{\theta_{H\text{A}}}, \bm{\theta_{Q\text{B}}}, \bm{\theta_{H\text{B}}} \right) \right],
\label{eq:TwoQubitProb}
\end{equation}
where
\begin{equation}
\bm{U}\left( \bm{\theta_{Q\text{A}}}, \bm{\theta_{H\text{A}}}, \bm{\theta_{Q\text{B}}}, \bm{\theta_{H\text{B}}} \right) = \bm{U}_\text{A}\left(\bm{\theta_{Q\text{A}}},\bm{\theta_{H\text{A}}}\right)\otimes \bm{U}_\text{B}\left(\bm{\theta_{Q\text{B}}},\bm{\theta_{H\text{B}}}\right)
\end{equation}
is the joint-space rotation operator.
The bold notation for settings and polarization components designates a vector $\in\mathbb{R}^{n}$ for $n$ measurements.
Due to the size of Eq. \ref{eq:TwoQubitProb}, the analytical expression is only presented in the appendix. 

When including a scalar biphoton flux $\mathcal{F}_{\psi\phi}$, scalar detection efficiencies for each beamsplitter in systems $A$ and $B$ as $\mu_{\text{A}}$ and $\mu_{\text{B}}$ (for transmitted components) with $\nu_{\text{A}}$ and $\nu_{\text{B}}$ (for reflected components), and PBS transmission and reflection coefficients for horizontal and vertical polarizations ($T_H$, $T_V$, $R_H$, $R_V$), we create a physical model of the anticipated photon counts $\mathcal{N}_{\psi\phi}$ (similar to Eq. \ref{eq:OneQubitModel}) by incorporating a bi-photon crosstalk matrix as
\begin{equation}
\bm{\mathcal{N}}_{\psi\phi} = 
\left\{
\left( \mathds{1}_{3\times 3} \otimes
\begin{bmatrix}
\mu_\text{A} T_{H\text{A}} & \mu_\text{A} T_{V\text{A}} \\
\nu_\text{A} R_{H\text{A}} & \nu_\text{A} R_{V\text{A}} 
\end{bmatrix}
\right)
\otimes
\left( \mathds{1}_{3\times 3} \otimes
\begin{bmatrix}
\mu_{\text{B}} T_{H\text{B}} & \mu {\text{B}} T_{V\text{B}} \\
\nu {\text{B}} R_{H\text{B}} & \nu {\text{B}} R_{V\text{B}} 
\end{bmatrix}
\right)
\right\}
\cdot
\left(
\mathcal{F}_{\psi\phi} \bm{P}_{\psi\phi}\right).
\label{eq:TwoQubitModel}
\end{equation}
Within Eq. \ref{eq:TwoQubitModel}, $\bm{\mathcal{N}}_{\psi\phi}$ and $\bm{P}_{\psi\phi}\in\mathbb{R}^{36}$ and $\mathds{1}_{3\times3}$ is the $3\times 3$ identity matrix.

Concerning the measurement settings for systems A and B, we present them as a Kronecker product to avoid printing two $36\times 8$ dimension matrices. The two matrices in Eq. \ref{eq:TwoQubitSettings} state how each subsystem should be configured and which polarized-beamsplitter output polarization will be measured (in coincidence) for each of the 36 measurements.

\begin{equation}
\text{Settings}_B = 
\kbordermatrix{                          & \bm{\theta_{Q\text{B}}} & \bm{\theta_{H\text{B}}} & \mathbf{h_B} & \mathbf{v_B} \cr
                \mathcal{N}_{H\text{B}}  &   0           &    0          &   1    &   0    \cr
                \mathcal{N}_{V\text{B}}  &   0           &    0          &   0    &   1    \cr
                \mathcal{N}_{D\text{B}}  &   \pi/4       &    \pi/8      &   1    &   0    \cr
                \mathcal{N}_{A\text{B}}  &   \pi/4       &    \pi/8      &   0    &   1    \cr
				\mathcal{N}_{L\text{B}}  &   \pi/4       &    0          &   1    &   0    \cr
				\mathcal{N}_{R\text{B}}  &   \pi/4       &    0          &   0    &   1    }
\otimes
\begin{bmatrix}
1 \\
1 \\
1 \\
1 \\
1 \\
1 \\
\end{bmatrix}, \,\,\, \& \,\,\,
\text{Settings}_A = 
\begin{bmatrix}
1 \\
1 \\
1 \\
1 \\
1 \\
1 \\
\end{bmatrix} \otimes
\kbordermatrix{                          & \bm{\theta_{Q\text{A}}} & \bm{\theta_{H\text{A}}} & \mathbf{h_A} & \mathbf{v_A} \cr
                \mathcal{N}_{H\text{A}}  &   0           &    0          &   1    &   0    \cr
                \mathcal{N}_{V\text{A}}  &   0           &    0          &   0    &   1    \cr
                \mathcal{N}_{D\text{A}}  &   \pi/4       &    \pi/8      &   1    &   0    \cr
                \mathcal{N}_{A\text{A}}  &   \pi/4       &    \pi/8      &   0    &   1    \cr
				\mathcal{N}_{L\text{A}}  &   \pi/4       &    0          &   1    &   0    \cr
				\mathcal{N}_{R\text{A}}  &   \pi/4       &    0          &   0    &   1    }
\label{eq:TwoQubitSettings}
\end{equation}

In situations where the biphoton flux is nonscalar, unknown, and must be estimated from the data, we again use the dark-count adjusted data and the pseudoinverse of the two-photon crosstalk matrix to obtain estimates of the two-photon input flux.
Knowing that the probability of measuring orthogonal polarizations on a single subsystem ($A$ or $B$) is unitary; i.e., $P_{\psi} + P_{\psi^\top} = 1$, where again $\langle \psi |\psi^\top \rangle = 0$, we calculate the joint-space probability contribution as
\begin{align}
\left(P_{\psi} + P_{\psi^\top}\right)_A\left(P_{\phi} + P_{\phi^\top}\right)_B &= 1 \nonumber\\
P_{\psi\phi} + P_{\psi\phi^\top} + P_{\psi^\top\phi} + P_{\psi^\top\phi^\top} &= 1.
\end{align}
Thus, we can find estimates for the photon flux of two polarizations $\psi$ and $\phi$ for all combinations of $\psi \in \{H,V,D,A,L,R\}$ and $\phi \in \{H,V,D,A,L,R\}$ such that
\begin{equation}
\mathcal{F}_{\psi\phi} = 
\begin{bmatrix}
1 & 1 & 1 & 1
\end{bmatrix} \cdot
\left\{
\left(
\begin{bmatrix}
\mu_\text{A} T_{H\text{A}} & \mu_\text{A} T_{V\text{A}} \\
\nu_\text{A} R_{H\text{A}} & \nu_\text{A} R_{V\text{A}} 
\end{bmatrix}
\right)
\otimes
\left( 
\begin{bmatrix}
\mu_{\text{B}} T_{H\text{B}} & \mu_{\text{B}} T_{V\text{B}} \\
\nu_{\text{B}} R_{H\text{B}} & \nu_{\text{B}} R_{V\text{B}} 
\end{bmatrix}
\right)
\right\}^{-1} \cdot
\begin{bmatrix}
\mathcal{N}_{\psi\phi} \\
\mathcal{N}_{\psi\phi^{\top}} \\
\mathcal{N}_{\psi^{\top}\phi} \\
\mathcal{N}_{\psi^{\top}\phi^{\top}}
\end{bmatrix},
\label{eq:2photonInvCrsstlk}
\end{equation}
where $\mathcal{F}_{\psi\phi}$ is scalar and $\mathcal{N}_{\psi\phi}$ are experimentally measured photon counts for a biphoton with polarization projections of $\psi$ for system A and $\phi$ for system B. 
Using Eq. \ref{eq:2photonInvCrsstlk}, we obtain mean two-photon input flux estimates for the variables $\mathcal{F}_{\text{HH}}$, $\mathcal{F}_{\text{HD}}$, $\mathcal{F}_{\text{HL}}$, $\mathcal{F}_{\text{DH}}$, $\mathcal{F}_{\text{DD}}$, $\mathcal{F}_{\text{DL}}$, $\mathcal{F}_{\text{LH}}$, $\mathcal{F}_{\text{LD}}$, and $\mathcal{F}_{\text{LL}}$. For clarity when using Eq. \ref{eq:2photonInvCrsstlk}, the subscripts $\psi\phi = HH$ correspond to $\psi\phi^{\top} = HV$, $\psi^\top\phi = VH$, and $\psi^\top\phi^\top = VV$. Thus, using the 36 measurements, we obtain 9 input biphoton flux estimates.

With the physics model and measurement settings complete, we structure the problem for PyMC in a Bayesian framework by building a likelihood function and prior distributions that capture our uncertainty in the measurements and settings. The formulation is analogous to Eq. \ref{eq:OneQubitPyMC} used by PyMC (where a distribution is defined for each variable with Gaussian $\mathcal{G}(\text{mean},\sigma)$ and Uniform $\mathcal{U}(\text{-1,1})$ distributions).
\begin{align}
t_{0} \rightarrow \mathcal{U}(-1,1)& \nonumber \,\,\,\,\,
t_{1} \rightarrow \mathcal{U}(-1,1) \nonumber \\
t_{2} \rightarrow \mathcal{U}(-1,1)& \nonumber \,\,\,\,\,
t_{3} \rightarrow \mathcal{U}(-1,1) \nonumber \\
t_{4} \rightarrow \mathcal{U}(-1,1)& \nonumber \,\,\,\,\,
t_{5} \rightarrow \mathcal{U}(-1,1) \nonumber \\
t_{6} \rightarrow \mathcal{U}(-1,1)& \nonumber \,\,\,\,\,
t_{7} \rightarrow \mathcal{U}(-1,1) \nonumber \\
t_{8} \rightarrow \mathcal{U}(-1,1)& \nonumber \,\,\,\,\,
t_{9} \rightarrow \mathcal{U}(-1,1) \nonumber \\
t_{10} \rightarrow \mathcal{U}(-1,1)& \nonumber \,\,\,\,\,
t_{11} \rightarrow \mathcal{U}(-1,1) \nonumber \\
t_{12} \rightarrow \mathcal{U}(-1,1)& \nonumber \,\,\,\,\,
t_{13} \rightarrow \mathcal{U}(-1,1) \nonumber \\
t_{14} \rightarrow \mathcal{U}(-1,1)& \nonumber \,\,\,\,\,
t_{15} \rightarrow \mathcal{U}(-1,1) \nonumber \\
\bm{\theta}_{Q\text{A}} \rightarrow \mathcal{G}(\bm{\theta}_{Q\text{A}},\sigma_{\theta_{Q\text{A}}})& \nonumber \,\,\,\,\,
\bm{\theta}_{H\text{A}} \rightarrow \mathcal{G}(\bm{\theta}_{H\text{A}},\sigma_{\theta_{Q\text{A}}}) \nonumber \\
\bm{\theta}_{Q\text{B}} \rightarrow \mathcal{G}(\bm{\theta}_{Q\text{B}},\sigma_{\theta_{Q\text{B}}})& \nonumber \,\,\,\,\,
\bm{\theta}_{H\text{B}} \rightarrow \mathcal{G}(\bm{\theta}_{H\text{B}},\sigma_{\theta_{Q\text{B}}}) \nonumber \\
\mu_{A} \rightarrow \mathcal{G}(\mu_{\text{A}},\sigma_{\mu_{\text{A}}})|_{\min = 0,\;\max = 1}& \nonumber \,\,\,\,\,
\nu_{A} \rightarrow \mathcal{G}(\nu_{\text{A}},\sigma_{\nu_{\text{A}}})|_{\min = 0,\;\max = 1} \nonumber \\
\mu_{B} \rightarrow \mathcal{G}(\mu_{\text{B}},\sigma_{\mu_{\text{B}}})|_{\min = 0,\;\max = 1}& \nonumber \,\,\,\,\,
\nu_{B} \rightarrow \mathcal{G}(\nu_{\text{B}},\sigma_{\nu_{\text{B}}})|_{\min = 0,\;\max = 1} \nonumber \\
T_{H\text{A}}\rightarrow \mathcal{G}(T_{H\text{A}},\sigma_{T_{H\text{A}}})|_{\min = 0,\;\max = 1}&\nonumber \,\,\,\,\,
T_{V\text{A}}\rightarrow \mathcal{G}(T_{V\text{A}},\sigma_{T_{V\text{A}}})|_{\min = 0,\;\max = 1}\nonumber \\
R_{H\text{A}}\rightarrow \mathcal{G}(R_{H\text{A}},\sigma_{R_{H\text{A}}})|_{\min = 0,\;\max = 1}&\nonumber \,\,\,\,\,
R_{V\text{A}}\rightarrow \mathcal{G}(R_{V\text{A}},\sigma_{R_{V\text{A}}})|_{\min = 0,\;\max = 1}\nonumber \\
T_{H\text{B}}\rightarrow \mathcal{G}(T_{H\text{B}},\sigma_{T_{H\text{B}}})|_{\min = 0,\;\max = 1}&\nonumber \,\,\,\,\,
T_{V\text{B}}\rightarrow \mathcal{G}(T_{V\text{B}},\sigma_{T_{V\text{B}}})|_{\min = 0,\;\max = 1}\nonumber \\
R_{H\text{B}}\rightarrow \mathcal{G}(R_{H\text{B}},\sigma_{R_{H\text{B}}})|_{\min = 0,\;\max = 1}&\nonumber \,\,\,\,\,
R_{V\text{B}}\rightarrow \mathcal{G}(R_{V\text{B}},\sigma_{R_{V\text{B}}})|_{\min = 0,\;\max = 1}\nonumber \\
\bm{\mathcal{F}} &\rightarrow \mathcal{G}(\bm{\mathcal{F}},\sqrt{\bm{\mathcal{F}}}) \nonumber \\
\bm{\mathcal{N}}_{\text{likelihood}} &\rightarrow \mathcal{G}(\bm{\mathcal{N}},\sqrt{\bm{\mathcal{N}}})
\label{eq:TwoQubitPyMC}
\end{align}

\subsection{Two-qubit simulations}

From here, we perform simulations in coincidence counts (as opposed to the experimentally obtained data in the single-qubit examples) to show how uncertainty should propagate through to the final density matrix.

Our state of choice for the QST simulation is the maximally entangled Bell-singlet state $|\Psi^-\rangle$ in the polarization basis where
\begin{equation}
 |\Psi^-\rangle = \frac{1}{\sqrt{2}}\left(|HV\rangle-|VH\rangle\right).
\end{equation}
Thus, the density matrix $\bm{\rho}$ we wish to test is
\begin{equation}
\bm{\rho} = |\Psi^-\rangle\langle \Psi^-| = 
\frac{1}{2}\begin{bmatrix}
0 & 0 & 0 & 0 \\
0 & 1 & -1 & 0 \\
0 & -1 & 1 & 0 \\
0 & 0 & 0 & 0 \\
\end{bmatrix}.
\label{eq:BellSinglet}
\end{equation}

Systematic errors were used in the simulation for all uncertainties. Thus, all errors in the simulated measurements (apart from the Poisson noise) were randomly drawn from normal distributions and shown in Eq. \ref{eq:SimUncerts}.
\begin{align}
\bm{\theta_{Q\text{A}}} \rightarrow \mathcal{G}(\bm{\theta_{Q\text{A}}},2\pi/180)& \,\,\,\,\,
\bm{\theta_{H\text{A}}} \rightarrow \mathcal{G}(\bm{\theta_{H\text{A}}},2\pi/180) \nonumber \\
\bm{\theta_{Q\text{B}}} \rightarrow \mathcal{G}(\bm{\theta_{Q\text{B}}},2\pi/180)&  \,\,\,\,\,
\bm{\theta_{H\text{B}}} \rightarrow \mathcal{G}(\bm{\theta_{H\text{B}}},2\pi/180) \nonumber \\
\mu_{A} \rightarrow \mathcal{G}(0.6,\,0.02)|_{\min = 0,\max = 1}&  \,\,\,\,\,
\nu_{A} \rightarrow \mathcal{G}(0.7,\,0.02)|_{\min = 0,\max = 1} \nonumber \\
\mu_{B} \rightarrow \mathcal{G}(0.7,\,0.02)|_{\min = 0,\max = 1}&  \,\,\,\,\,
\nu_{B} \rightarrow \mathcal{G}(0.8,\,0.02)|_{\min = 0,\max = 1} \nonumber \\
T_{H\text{A}}\rightarrow \mathcal{G}(0.98,\;0.02)|_{\min = 0,\;\max = 1}& \,\,\,\,\,
T_{V\text{A}}\rightarrow \mathcal{G}(0.01,\;0.02)|_{\min = 0,\;\max = 1}\nonumber \\
R_{H\text{A}}\rightarrow \mathcal{G}(0.01,\;0.02)|_{\min = 0,\;\max = 1}& \,\,\,\,\,
R_{V\text{A}}\rightarrow \mathcal{G}(0.97,\;0.02)|_{\min = 0,\;\max = 1}\nonumber \\
T_{H\text{B}}\rightarrow \mathcal{G}(0.96,\;0.02)|_{\min = 0,\;\max = 1}& \,\,\,\,\,
T_{V\text{B}}\rightarrow \mathcal{G}(0.01,\;0.02)|_{\min = 0,\;\max = 1}\nonumber \\
R_{H\text{B}}\rightarrow \mathcal{G}(0.01,\;0.02)|_{\min = 0,\;\max = 1}& \,\,\,\,\,
R_{V\text{B}}\rightarrow \mathcal{G}(0.97,\;0.02)|_{\min = 0,\;\max = 1}\nonumber \\
\label{eq:SimUncerts}
\end{align}

\begin{figure}
  \includegraphics[width=.8\linewidth]{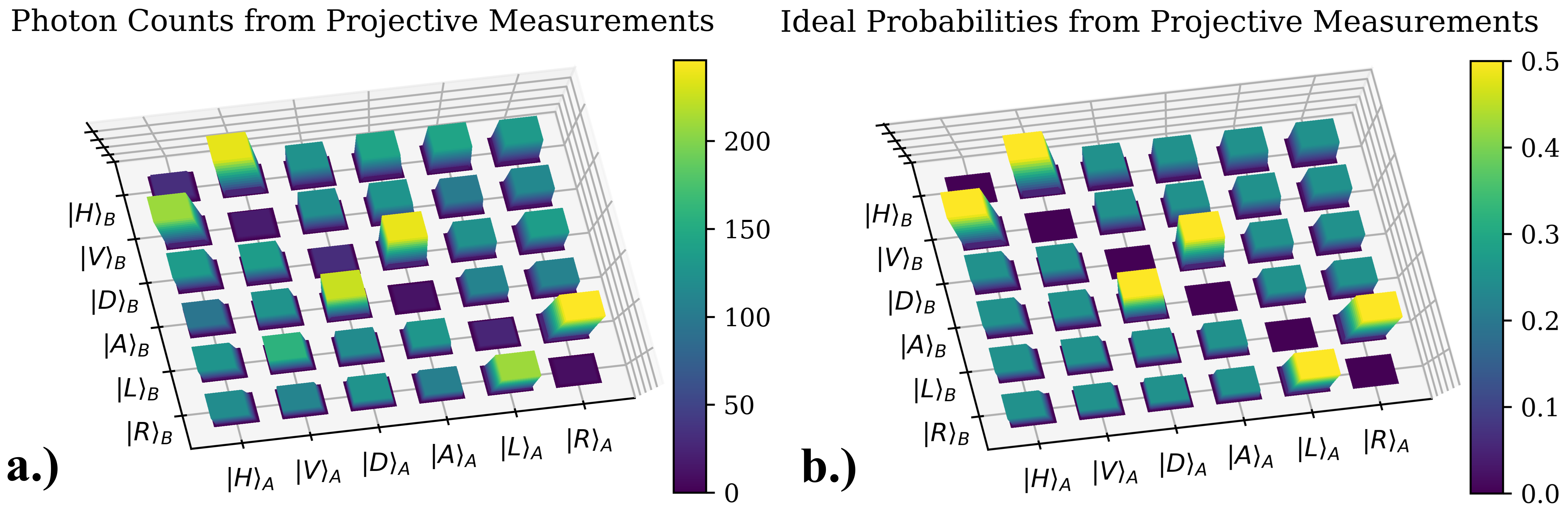}
  \caption{\textbf{a.)} Simulated coincidence measurements (where the biphoton flux is $\mathcal{N} = 1000$ biphotons/sec). We project each particle within the bipartite state onto the states $|H\rangle$, $|V\rangle$, $|D\rangle$, $|A\rangle$, $|L\rangle$, and $|R\rangle$. Given the six measurement settings per particle (in subspaces $A$ and $B$), there are 36 measurement settings in total. \textbf{b.)} Anticipated probabilities for an ideal measurement. With a biphoton flux of $\mathcal{N} = 1000$, there should be a maximum of 500 coincidences. Due to PBS crosstalk errors, waveplate angle errors, Poisson noise, and losses, the simulated data is clearly altered and makes a recovery of the true state more difficult.}
  \label{fig:LowFlux2QubitCounts}
\end{figure}

Beginning with a biphoton flux of $\mathcal{N}=1000$ biphotons/measurement, simulated photon counts with Poisson-distributed counting uncertainties are plotted in Fig. \ref{fig:LowFlux2QubitCounts}.

Running the Bayesian analysis PyMC code found in the appendix using four computing cores, 1000 tuning samples to be discarded per core, and 1500 draws per core (for a total of 4000 tuning samples and 6000 recorded samples), we obtain the density matrix below:
\begin{equation}
\hat{\bm{\rho}} = \begin{bmatrix}
 0.056               & 0.002-i0.016      & 0.022-i0.001     & 0.001+i0.001 \\
 0.002+i0.016        & 0.483              & -0.430-i0.015    & -0.007-i0.013 \\
 0.022+i0.001        & -0.430+i0.015      & 0.458            & 0.007+i0.012 \\
 0.001-i0.001        & -0.007+i0.013      & 0.007-i0.013     & 0.004 .
\end{bmatrix}.
\label{eq:biphotonDnstyMtrx}
\end{equation}
The 95\% HDI for a density matrix having the form
\begin{equation}
\begin{bmatrix}
A                      & \text{Re}E-i\text{Im}E & \text{Re}F-i\text{Im}F & \text{Re}G-i\text{Im}G \\
\text{Re}E+i\text{Im}E & B                      & \text{Re}H-i\text{Im}H & \text{Re}I-i\text{Im}I \\
\text{Re}F+i\text{Im}F & \text{Re}H+i\text{Im}H & C                      & \text{Re}J-i\text{Im}J  \\
\text{Re}G+i\text{Im}G & \text{Re}I+i\text{Im}I & \text{Re}J+i\text{Im}J & D 
\end{bmatrix} \nonumber
\end{equation}
has the following HDIs
\begin{align}
A_{\text{HDI}_{95}} \in [0.016,\;0.096]& \;\;\;
B_{\text{HDI}_{95}} \in [0.443,\;0.524] \nonumber\\
C_{\text{HDI}_{95}} \in [0.415,\;0.501]& \;\;\;
D_{\text{HDI}_{95}} \in [0.000,\;0.001] \nonumber\\
\text{Re}E_{\text{HDI}_{95}} \in [-0.048,\;0.051]& \;\;\;
\text{Im}E_{\text{HDI}_{95}} \in [-0.034,\;0.064] \nonumber\\
\text{Re}F_{\text{HDI}_{95}} \in [-0.030,\;0.072]& \;\;\;
\text{Im}F_{\text{HDI}_{95}} \in [-0.046,\;0.047] \nonumber\\
\text{Re}G_{\text{HDI}_{95}} \in [-0.007,\;0.012]& \;\;\;
\text{Im}G_{\text{HDI}_{95}} \in [-0.010,\;0.008] \nonumber\\
\text{Re}H_{\text{HDI}_{95}} \in [-0.475,\;-0.382]& \;\;\;
\text{Im}H_{\text{HDI}_{95}} \in [-0.028,\;0.160] \nonumber\\
\text{Re}I_{\text{HDI}_{95}} \in [-0.048,\;0.032]& \;\;\;
\text{Im}I_{\text{HDI}_{95}} \in [-0.023,\;0.051] \nonumber\\
\text{Re}J_{\text{HDI}_{95}} \in [-0.036,\;0.050]& \;\;\;
\text{Im}J_{\text{HDI}_{95}} \in [-0.053,\;0.027].
\label{eq:biphotonDnstyMtrxHDI}
\end{align}.

\begin{figure}
  \includegraphics[width=.5\linewidth]{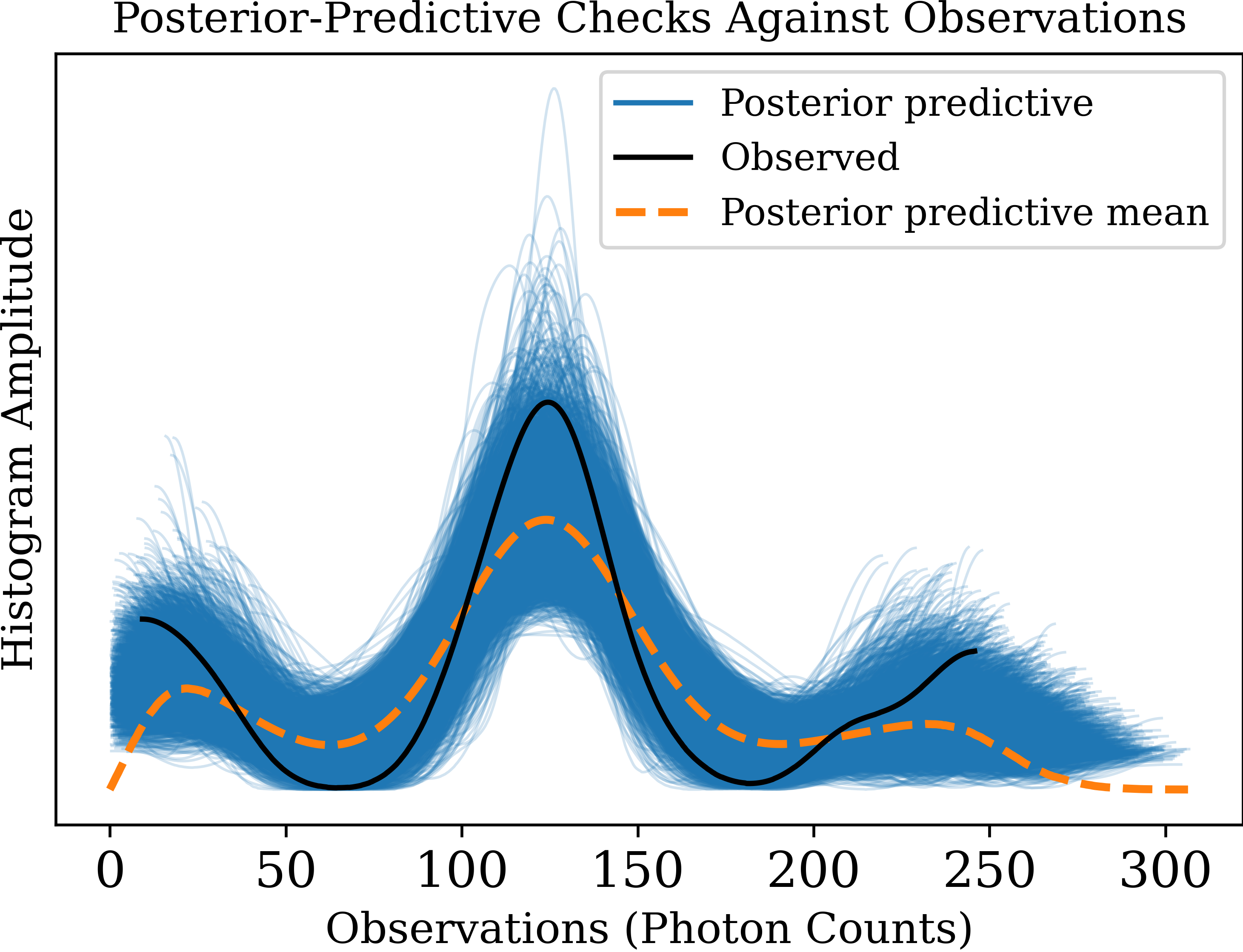}
  \caption{Posterior-predictive checks displaying smoothed histograms of generated data together with the input data. The variance in the generated data is enough to capture all possible values of the input data (also a smoothed histogram shown in black).}
  \label{fig:PPCTwoPhoton}
\end{figure}

While the true state is not within the 95\% HDI for all matrix elements, it is still a reasonable estimate given its ability to model the data input with the model uncertainty as confirmed with the posterior-predictive checks shown in Fig. \ref{fig:PPCTwoPhoton}. Figure \ref{fig:PPCTwoPhoton} shows a histogram of the 36 input photon counts together with histograms of simulated data. Clearly the Poisson noise, PBS crosstalk uncertainty, and waveplate errors induce a broadening in the generated histograms. Yet, the general trend is consistent with the observed input coincidence counts. Finally, the matrix component $H$ and the affiliated HDIs in Eq. \ref{eq:biphotonDnstyMtrxHDI} highlights the state's inability to yield perfect two-photon interference in coincidences (even when considering the spread in values) as $\text{Re}H=-1/2$ is not within the 95\% HDI. This is consistent with the fact that the losses in our PBS crosstalk matrix should prevent one from witnessing perfect interference.

\section{Conclusion}

In summary, we present a mathematical model (complete with Python-3 code listed in the appendix) that we believe will be helpful to any laboratory performing polarization-based QST on either single or two qubit systems. The models merely require that the user input the detected photon counts, the waveplate settings (with angles defined with respect to the horizontal axis), and the uncertainty and detection efficiency estimates. The PyMC library package will then numerically calculate the posterior distributions and return Bayes mean estimates and HDIs after brute-force propagating the distributions. Distributions for additional metrics, such as the entropy or Stokes parameters, can be easily calculated where PyMC will propagate the error with an appropriate number of samples which can help the user to avoid mishandling error propagation.

While we focus on polarization-based tomography in this article, it should be clear that any tomography is possible as long as the model can be formulated in a similar manner presented here. We hope that this work will prove useful to those labs performing QST and serve as an additional reference to those seeking to add Bayesian methods to their list of laboratory tools.


\bibliography{BayesQST_v2.bib}

\newpage
\section{Appendix}

\subsection{Building models in PyMC}

At the heart of PyMC is the establishment of a Bayesian model. As a quick example, consider a simple Poisson regression problem in which the log of the expectation value $\log[\mathrm{E}(n|X)]$ for $n$ observations $y_i \in Y$ and sample points $x_i\in X$ is a linear function such that
\begin{equation}
\log[\mathrm{E}(Y|X)] = \alpha + \beta X.
\end{equation}
This implies $\mathrm{E}(Y|X) = \text{exp}\left(\alpha + \beta X\right)$ and ensures that the average values used in our Poisson model are strictly positive. Given that a Poisson distribution $\mathcal{P}$ with average value $\lambda$ follows
\begin{equation}
\mathcal{P}(y|\lambda) = \frac{\lambda^y e^{-\lambda}}{y!},
\end{equation}
the outcome counts from our measurements $Y$ follow the Poisson regression likelihood function
\begin{equation}
\mathcal{P}(Y|\alpha,\beta,X ) = \prod_{i=1}^n \frac{e^{y_i\left(\alpha +\beta x_i\right)}e^{-e^{\alpha + \beta x_i}}}{y_i!}.
\end{equation}
The negative log-likelihood function maximized in MLE is
\begin{align}
-\log\left[\mathcal{P}(Y|\alpha,\beta,X)\right] & = -\sum\limits_{i=1}^n \log\left( \frac{e^{y_i\left(\alpha +\beta x_i\right)}e^{-e^{\alpha + \beta x_i}}}{y_i!}\right) \nonumber \\
&= -\sum\limits_{i=1}^n \left[ y_i\left(\alpha +\beta x_i\right) - e^{\alpha + \beta x_i} - \log\left(y_i!\right) \right]
\end{align}
and partial derivatives with respect to $\alpha$ and $\beta$ can be easily maximized numerically and used as a starting point for MCMC sampling algorithms. 

Since we are interested in a Bayesian model, we slightly alter the likelihood model to include prior information about our unknown variables $\alpha$ and $\beta$ such that our prior distribution $P(Y|\alpha,\beta,X)$ reads
\begin{align}
P(Y|\alpha,\beta,X) &\propto \mathcal{P}(X|Y)P(\alpha,\beta) \nonumber \\
&\propto \prod\limits_{i=1}^n \frac{ e^{y_i\left(\alpha +\beta x_i\right)}e^{-e^{\alpha + \beta x_i}} }{ y_i! } e^{-\frac{\left(\alpha - \alpha_0\right)^2}{2\sigma_{\alpha}^2}-\frac{\left(\beta - \beta_0\right)^2}{2\sigma_{\beta}^2}}
\end{align}
where $P(\alpha,\beta)$ is a normally distributed prior probability distributions for $\alpha$\,($\beta$) having mean $\alpha_0$\,($\beta_0$) and standard deviation $\sigma_{\alpha}$\,($\sigma_{\beta}$). We should provide relatively uninformative priors (large standard deviations) for our variables if we are not fairly certain of their values or ranges. 

PyMC is accessed by our program by first importing \texttt{pymc}. We also import the \texttt{arviz} library to help displaying the traces returned from PyMC and grants us access to the HDI function \texttt{arviz.hdi()}. The \texttt{numpy} library is used for matrix-vector manipulation and provides access to the \texttt{numpy.mean()} function. The \texttt{matplotlib} library allows us to make plots. The program below details how to define a probabilistic model of our measurements using \texttt{pymc.model()} that we name \texttt{PoissonRegression}.

\begin{lstlisting}[style = mystyle]
import pymc as pm
import pymc.math as pmm
import arviz as az
import numpy as np
import matplotlib.pyplot as plt

# x must be previously defined as an array of sample points 
# y must be previously defined as an array of count data
# y = Poisson(alpha + beta*x)

with pm.Model() as PoissonRegression:

	# Define Priors
	sig   = pm.Exponential("sig",1.)
	alpha = pm.Normal("alpha", 0, sigma=sig)
	beta  = pm.Normal("beta" , 0, sigma=sig)

	# Define the Poisson model and supply measurement values 
	likelihood = pm.Poisson("likelihood", mu=pmm.exp(alpha + beta*x), observed = y)
	trace = pm.sample(draws=3000, tune=1000, cores=4, target_accept = 0.98)
	
	# generate posterior-predictive checks
	ppc = pm.sample_posterior_predictive(trace, model=PoissonRegression, var_names = ["likelihood"])
	
	# Plot trace
	print(az.summary(trace,round_to=2))
	az.plot_trace(trace, compact=True); plt.show()
	
	# Plot posterior-predictive checks
	az.plot_ppc(data = ppc);plt.show()

# convert traces from all cores to numpy arrays
alphaTrace = np.asarray(trace.posterior["alpha"]).flatten()
betaTrace = np.asarray(trace.posterior["beta"]).flatten()

# get averages
alphaAvg = alphaTrace.mean()
betaAvg  = beatTrace.mean()

# get 95-HDI
print("95 hdi alpha: ", az.hdi(alphaTrace, hdi_prob=0.95,  multimodal=True))
print("95 hdi beta:  ", az.hdi(betaTrace , hdi_prob=0.95,  multimodal=True))
\end{lstlisting}

PyMC provides a library of predefined probability distributions with the capability of defining our own. For simplicity, we use Gaussian (Normal) distributions for $\alpha$ and $\beta$ when defining priors as shown in the code below.
\begin{lstlisting}[style = mystyle]
with pm.Model() as PoissonRegression:
	# Define Priors
	sig   = pm.Exponential("sig",1.)
	alpha = pm.Normal("alpha", 0, sigma=sig)
	beta  = pm.Normal("beta" , 0, sigma=sig)
\end{lstlisting}
Note that each Gaussian's standard deviation \texttt{sig} was not explicitly defined. Instead, we allow the standard deviations to follow an exponential probability distribution having an average of unity. In this manner, we request the standard deviation of each variable be as small as necessary while remaining positive and not fixing the model at strict values. The variable \texttt{likelihood} represents the likelihood of observing the data given parameters $\alpha$ and $\beta$ and will be compared against our observations $y$. The likelihood argument \texttt{observed = y} within the line
\begin{lstlisting}[style = mystyle]
likelihood = pm.Poisson("likelihood", mu=pmm.exp(alpha + beta*x), observed = y)
\end{lstlisting}
designates that this variable cannot be varied; it is a fixed point with respect to the data. 

Finally, the line
\begin{lstlisting}[style = mystyle]
trace = pm.sample(draws=3000, tune=1000, cores=4, target_accept = 0.98)
\end{lstlisting}
tells PyMC to sample the posterior distribution by tuning with 1000 samples/core and to keep 3000 samples/core after tuning. The 1000 ``burn-in'' samples help to ensure convergence with the posterior probability distribution and are automatically discarded before analyzing the results. With more samples, the better we approximate the posterior distribution. The \texttt{target\_accept} parameter is kept close to 1 to tune the step size for problematic priors.

The call to 
\begin{lstlisting}[style = mystyle]
# generate posterior-predictive checks
ppc = pm.sample_posterior_predictive(trace, model=PoissonRegression, var_names = ["likelihood"])
\end{lstlisting}
will use the posterior distribution of our state estimate and model parameters to help validate the model by simulating distributions of possible input data. A well behaved model will follow the general trend of the original input data \texttt{y}.

The lines
\begin{lstlisting}[style = mystyle]
# Plot trace
print(az.summary(trace,round_to=2))
az.plot_trace(trace, compact=True); plt.show()
# Plot posterior-predictive checks
az.plot_ppc(data = ppc);plt.show()
\end{lstlisting}
display summary information and plot histograms of the samples (one plot for the trace variables and another for the posterior-predictive checks).

Finally, we extract averages and HDIs from our variables with the lines immediately below.
\begin{lstlisting}[style = mystyle]
# convert traces from all cores to numpy arrays
alphaTrace = np.asarray(trace.posterior["alpha"]).flatten()
betaTrace = np.asarray(trace.posterior["beta"]).flatten()
# get averages
alphaAvg = alphaTrace.mean()
betaAvg  = beatTrace.mean()
# get 95-HDI
print("95 hdi alpha: ", az.hdi(alphaTrace, hdi_prob=0.95,  multimodal=True))
print("95 hdi beta:  ", az.hdi(betaTrace , hdi_prob=0.95,  multimodal=True))
\end{lstlisting}

The above code serves as the basis for our QST reconstruction algorithms. In both the single and double qubit sections, variables are assigned uniform and Gaussian probability distributions. These variables are then used to calculate average photon expectation values that are used in a Gaussian likelihood model for comparison against data. In contrast to the code above, we do not need the log of our expectation values to be a linear function of our unknowns as we ensure our average values are strictly positive and real from the beginning.

\subsection{Single-Qubit QST (Detailed)}

Any single-qubit density matrix can be written as
\begin{equation}
\bm{\rho} = 
\begin{bmatrix} 
A & B e^{i\phi}\\ 
B e^{-i\phi} & 1-A
\end{bmatrix}, \nonumber
\end{equation}
where $A,B,\phi\in\mathbb{R}$ and $B\leq\sqrt{A(1-A)}$ ensures all eigenvalues are real and $\geq 0$. This shows that any single-qubit state is completely defined by three variables. Hence, a minimum of three properly chosen measurements are needed to form a complete set of equations having a unique solution. In general, any density matrix of $n$ qubits will have $4^n-1$ independent variables.

A useful way to parameterize a single-qubit density matrix using the basis vectors in Eq. \ref{eq:polstates} is via the Pauli matrices:
\begin{align}
\bm{\sigma}_0 &= |H\rangle\langle H| + |V\rangle\langle V| = 
\begin{bmatrix}
1 & 0 \\
0 & 1
\end{bmatrix} \nonumber\\
\bm{\sigma}_1 &= |D\rangle\langle D| - |A\rangle\langle A| = 
\begin{bmatrix}
0 & 1 \\
1 & 0
\end{bmatrix} \nonumber\\
\bm{\sigma}_2 &= | L \rangle\langle L| - |R\rangle\langle R| = 
\begin{bmatrix}
0 & -i \\
i & 0
\end{bmatrix} \nonumber\\
\bm{\sigma}_3 &= |H\rangle\langle H| - |V\rangle\langle V| = 
\begin{bmatrix}
1 & 0 \\
0 & -1
\end{bmatrix}.
\end{align} 
Additionally, the Pauli matrices can be used to represent any single-qubit density matrix as a linear sum with each Pauli matrix weighted by a Stokes parameter $S_i$:
\begin{equation}
\bm{\rho} = \frac{1}{2}\sum\limits_{i=0}^3 S_i\bm{\sigma_i},
\label{eq:stokesdensitymatrix}
\end{equation}
where the Stokes parameters are used to visualize the qubit on the Bloch sphere (or Poincar\'e sphere for polarization states).
The expectation value of an observable $\bm{X}$ can be determined using the density matrix as 
\begin{equation}
\langle \bm{X} \rangle = \sum\limits_i p_i \langle \psi_i | \bm{X} |\psi_i \rangle = \sum\limits_i p_i \text{Tr}\left( |\psi_i \rangle\langle \psi_i | \bm{X} \right) = \sum\limits_i \text{Tr}\left( p_i  |\psi_i \rangle\langle \psi_i | \bm{X} \right) = \text{Tr}\left(\bm{\rho}\cdot\bm{X}\right).
\end{equation}
Noticing that
\begin{equation}
\text{Tr}\left[\bm{\sigma}_i\cdot\bm{\sigma}_j\right] = 2\delta_{ij},
\end{equation}
where $\delta_{ij}$ is the Kronecker delta function, the expectation value of a Pauli operator $\bm{\sigma}_i$ is
\begin{equation}
\langle \bm{\sigma}_i\rangle = \text{Tr}\left[\bm{\sigma}_i\bm{\rho}\right] = \text{Tr}\left[\bm{\sigma}_i \frac{1}{2}\sum\limits_{j=0}^3 S_j \bm{\sigma}_j \right] = \frac{1}{2}\sum\limits_{j=0}^3 S_j \text{Tr}\left[ \bm{\sigma}_i\cdot\bm{\sigma}_j \right] = \frac{1}{2}\sum\limits_{j=0}^3 S_j 2\delta_{ij} = S_i.
\label{eq:StokesExpectationRelation}
\end{equation}
We can immediately see that the Stokes parameters can be found by measuring the projection probabilities
\begin{align}
P_H &= \text{Tr}\left(|H\rangle\langle H |\cdot\bm{\rho}\right) = \langle H |\bm{\rho}|H \rangle \nonumber\\
P_V &= \text{Tr}\left(|V\rangle\langle V |\cdot\bm{\rho}\right) = \langle V |\bm{\rho}|V \rangle \nonumber\\
P_D &= \text{Tr}\left(|D\rangle\langle D |\cdot\bm{\rho}\right) = \langle D |\bm{\rho}|D \rangle \nonumber\\
P_A &= \text{Tr}\left(|A\rangle\langle A |\cdot\bm{\rho}\right) = \langle A |\bm{\rho}|A \rangle \nonumber\\
P_L &= \text{Tr}\left(|L\rangle\langle L |\cdot\bm{\rho}\right) = \langle L |\bm{\rho}|L \rangle \nonumber\\
P_R &= \text{Tr}\left(|R\rangle\langle R |\cdot\bm{\rho}\right) = \langle R |\bm{\rho}|R \rangle
\end{align}
and then constructing the appropriate Stokes parameters as
\begin{align}
S_0 &= P_H + P_V \nonumber\\
S_1 &= P_D - P_A \nonumber\\
S_2 &= P_L - P_R \nonumber\\
S_3 &= P_H - P_V. \label{eq:stokesParemeters}
\end{align}

In an optical polarization based QST, we require a quarter-waveplate operator $\bm{Q}$ and half-waveplate operator $\bm{H}$ that will allow the horizontal component of an electric field to advance a maximum of either a quarter wave or half wave (respectively) ahead of the vertical component. These operators take the form
\begin{align}
\bm{Q} &= e^{-i\frac{\pi}{4}}
\begin{bmatrix}
1 & 0 \nonumber\\
0 & i 
\end{bmatrix} \\
\bm{H} &= e^{-i\frac{\pi}{2}}
\begin{bmatrix}
1 & 0 \\
0 & -1 
\end{bmatrix}.
\end{align}
Rotating these operators with a rotation matrix $\bm{R}(\theta)$, where 
\begin{equation}
\bm{R}(\theta) = 
\begin{bmatrix}
\cos(\theta) & -\sin(\theta) \\
\sin(\theta) & \cos(\theta)
\end{bmatrix},
\end{equation}
results in the general polarization rotation operations
\begin{align}
\bm{Q}(\theta_Q) &= \bm{R}(\theta_Q)\cdot\bm{Q}\cdot\bm{R}(\theta_Q)^T = e^{-i\frac{\pi}{4}}
\begin{bmatrix}
\cos^2(\theta_Q)+ i\sin^2(\theta_Q) & \left(1-i\right)\cos(\theta_Q)\sin(\theta_Q) \\
\left(1-i\right)\cos(\theta_Q)\sin(\theta_Q)     & \sin^2(\theta_Q)+i\cos^2(\theta_Q)
\end{bmatrix} \\
\bm{H}(\theta_H) &= \bm{R}(\theta_H)\cdot\bm{Q}\cdot\bm{Q}\cdot\bm{R}(\theta_H)^T = -i
\begin{bmatrix}
\cos(2\theta_H) & 2\cos(\theta_H)\sin(\theta_H) \\
2\cos(\theta_H)\sin(\theta_H) & -\cos(2\theta_H) 
\end{bmatrix} \\
\bm{U}(\theta_Q,\theta_H) &= \bm{H}(\theta_H)\cdot \bm{Q}(\theta_Q) = \frac{1}{\sqrt{2}}
\begin{bmatrix}
-i\cos(2\theta_H)-\cos\left(2\left(\theta_H-\theta_Q\right)\right) & -i\sin(2\theta_H)+\sin\left(2\left(\theta_H-\theta_Q\right)\right) \\
-i\sin(2\theta_H)-\sin\left(2\left(\theta_H-\theta_Q\right)\right) &  i\cos(2\theta_H)-\cos\left(2\left(\theta_H-\theta_Q\right)\right)
\end{bmatrix}.
\end{align}

Alternatively, we can use variable waveplates as previously discussed within the main article. Having a phase retardance $\eta$ and a fast axis set to $\theta$ radians with respect to the horizontal axis, these arbitrary waveplates take the form.
\begin{equation}
\bm{AWP}(\eta,\theta) = e^{-i\frac{\eta}{2}}
\begin{bmatrix}
\cos^2(\theta)+e^{i\eta}\sin^2(\theta) & 1-e^{i\eta}\cos(\theta)\sin(\theta) \\
1-e^{i\eta}\cos(\theta)\sin(\theta)  & \sin^2(\theta)+e^{i\eta}\cos^2(\theta)
\end{bmatrix},
\end{equation}
where
\begin{align}
\bm{AWP}(\eta=\pi,\theta) &= \bm{H}(\theta) \\
\bm{AWP}(\eta=\pi/2,\theta) &= \bm{Q}(\theta).
\end{align} 

Using the settings defined in the main article at Eqs. \ref{eq:SettingsLC} and \ref{eq:SettingsQH} (depending on whether waveplates or variable-phase retarders are used), we can obtain the six measurements needed to identify our unknown quantum state.

\subsection{Single-Qubit Python Code}

In the following code snippets, we use the same Python-3 libraries listed below.
\begin{lstlisting}[language=Python]
import arviz as az
import numpy as np
import pymc as pm
import pymc.math as pmm
import matplotlib.pyplot as plt
import aesara.tensor as at
\end{lstlisting}

The function \texttt{main()}
prepares the problem formulation by defining the measurement settings, detector counts, efficiencies, and uncertainties. As we used a variable waveplate, the measurement-setting angles (\texttt{Mments}) are fixed while the phase retardance is adjusted with the waveplate controller's voltage to produce half- and quarter-wave polarization rotations. The PBS transmission and reflection efficiencies for horizontal and vertical polarization (\texttt{Th}, \texttt{Tv}, \texttt{Rh}, \texttt{Rv}) are needed to calculated the crosstalk component to our observed counts. The function call
\begin{lstlisting}[language=Python]
[trace,ppc,poisson_model] = ReconstructState(counts,params)
\end{lstlisting}
takes the dark-rate adjusted counts and the parameter dictionary (\texttt{params}) and returns the trace object (containing the trace of all model parameters), the posterior-predictive checks (\texttt{ppc}), and the actual model (\texttt{poisson\_model}) used by PyMC.

\begin{lstlisting}[language=Python]
def main():
	# seed random generator to duplicate results easily
	RANDOM_SEED = 12345
	np.random.seed(RANDOM_SEED)
	# measurements follow [HWP_eta, QWP_eta,    HWP_theta,  QWP_theta, h_meas, v_meas]
	Mments = np.array([ [0    ,    0      ,     np.pi/8,   np.pi/4   ,  1,  0], \
						[0    ,    0      ,     np.pi/8,   np.pi/4   ,  0,  1], \
						[np.pi,    0      ,     np.pi/8,   np.pi/4   ,  1,  0], \
						[np.pi,    0      ,     np.pi/8,   np.pi/4   ,  0,  1], \
						[0    ,    np.pi/2,     np.pi/8,   np.pi/4   ,  1,  0], \
						[0    ,    np.pi/2,     np.pi/8,   np.pi/4   ,  0,  1] ])  

	# density matrix has form:
	# [   A      ReC-i*ImC]
	# [ReC+i*ImC    B     ]

	# Detection efficiencies

	# fiber coupling and transmission to detector efficiency
	Mu = .76
	Nu = .75
	# detection efficiency std
	# assume we know this to within +/- 3%
	Mu_std = .03
	Nu_std = .03

	# detector bias / efficiencies
	# the horizontal detector measured roughly 1/2 the number of vertical counts
	# when both detectors had equal power entering (from same fiber)
	DetBias = np.zeros(2)
	DetBias[0] = 0.552
	DetBias[1] = 1.0

	# adjust arm efficiency by detector efficiency
	Mu = Mu*DetBias[0]
	Nu = Nu*DetBias[1]

	# assume QWP setting uncertainty is +/- 1 degree based on measurements
	QWP_std = 1.*np.pi/180.
	HWP_std = 1*np.pi/180.
	
	# assume phase retardance is known with within +/- 2 degrees based on measurements
	HWP_eta_std = 2.*np.pi/180.
	QWP_eta_std = 2.*np.pi/180.

	# Include PBS crosstalk efficiency (measured experimentally)
	Th = .973
	Rh = .013
	Rv = .987
	Tv = .027

	# assume values known to within +/- 1 percent based on power fluctuations
	Th_std = .01
	Rh_std = .01
	Rv_std = .01
	Tv_std = .01
	
	# experimentally measured counts (after subtracting background + dark counts)
	counts = np.array([2518, 123, 1335, 2291, 1234, 2314]) # high-flux data
	#counts = np.array([21, 2, 11, 21, 12, 23]) # low-flux data
	
	# define dictionary of model parameters and uncertainties
	Uncerts = {"Mu": Mu,
				"Nu": Nu,
				"Th": Th,
				"Tv": Tv,
				"Rh": Rh,
				"Rv": Rv,
				"Mu_std": Mu_std,
				"Nu_std": Nu_std,
				"HWP_std": HWP_std,
				"QWP_std": QWP_std,
				"HWP_eta_std": HWP_eta_std,
				"QWP_eta_std": QWP_eta_std,
				"Th_std": Th_std,
				"Tv_std": Tv_std,
				"Rh_std": Rh_std,
				"Rv_std": Rv_std}
				
	# Group Mments and Uncerts into dictionary of params for cleaner arguments
	params = {"Mments": Mments,
				"Uncerts": Uncerts}
	
	[trace,ppc,poisson_model] = ReconstructState(counts,params)

if __name__ == "__main__":
	main()
\end{lstlisting}
Within the function \texttt{ReconstructState()}, we use the function \texttt{getInputFlux()} to estimate the input flux by building a PBS crosstalk matrix, inverting it, and then apply the inverted form to the detector counts.

\begin{lstlisting}[language=Python]
def getInputFlux(counts, params):
	# get PBS crosstalk efficiencies and fiber coupling efficiencies
	Uncerts = params["Uncerts"]
	Mu = Uncerts["Mu"]
	Nu = Uncerts["Nu"]
	Th = Uncerts["Th"]
	Tv = Uncerts["Tv"]
	Rh = Uncerts["Rh"]
	Rv = Uncerts["Rv"]
		
	# generate crosstalk and coupling efficiency matrix to generate estimate of input flux
	Crsstlk = np.zeros([2,2])
	Crsstlk[0,0] = Mu*Th
	Crsstlk[0,1] = Mu*Tv
	Crsstlk[1,0] = Nu*Rh
	Crsstlk[1,1] = Nu*Rv
	CrsstlkMat = np.kron(np.eye(3),Crsstlk)
	
	# generate matrix inverse of crosstalk matrix
	InvCrsstlkMat = np.linalg.pinv(CrsstlkMat)
	
	# apply inverse crosstalk matrix to data counts
	InputCounts = np.dot(InvCrsstlkMat,counts)
	
	# sum flux estimates from orthogonal measurements to yield 3 flux estimates
	# one each for (H,V), (D,A), and (L,R) measurements
	InputFlux = InputCounts[0:len(InputCounts):2] + InputCounts[1:len(InputCounts):2]
	
	return InputFlux
\end{lstlisting}

After estimating the input flux, we build our PyMC model as outlined in the main article. For computational reasons, we split the unknown mean and standard deviation components from the normal distribution priors by using the fact that $\mathcal{N}(\bar{x},\sigma_{x}) = \bar{x}+\sigma_x \mathcal{N}(0,1)$ where $\bar{x}$ is the mean value of the normal distribution while $\sigma_x$ is the standard deviation. In situations where both $\bar{x}$ and $\sigma_x$ are unknown, the method used here is computationally more efficient. Because we prove the mean values a-priori, splitting the mean and standard deviation is likely no more efficient than simply calculating $\mathcal{N}(\bar{x},\sigma_{x})$. However, we use the method in the event a user decides not to supply mean values.

To reduce the number of traces that need to be taken, we combine the standard deviation variables \texttt{Mu\_std}, \texttt{Nu\_std}, \texttt{Th\_std}, \texttt{Rh\_std}, \texttt{Th\_std}, and \texttt{Tv\_std} by adding them in quadrature to form a PBS standard deviation term \texttt{PBS\_std}.

The variable \texttt{n\_obs} serves as our likelihood function and the line
\begin{lstlisting}[language=Python]
trace = pm.sample(draws = 1000, tune = 800, cores=4, target_accept = 0.98)
\end{lstlisting}
tells the sampler to use 4 CPU cores to draw 1000 samples per core after tuning with 800 samples per core. The variable \texttt{target\_accept} is set close to 1 to tune the steps size of the sampler with a high acceptance rate and has shown reliable performance for these particular posterior distributions.   

\begin{lstlisting}[language=Python]
def ReconstructState(counts,params):
	# extract measurement settings and uncertainty dictionary from params
	Mments = params["Mments"]
	Uncerts = params["Uncerts"]
	
	Mu = Uncerts["Mu"]
	Nu = Uncerts["Nu"]
	Th = Uncerts["Th"]
	Tv = Uncerts["Tv"]
	Rh = Uncerts["Rh"]
	Rv = Uncerts["Rv"]
	Mu_std = Uncerts["Mu_std"]
	Nu_std = Uncerts["Nu_std"]
	HWP_std = Uncerts["HWP_std"]
	QWP_std = Uncerts["QWP_std"]
	HWP_eta_std = Uncerts["HWP_eta_std"]
	QWP_eta_std = Uncerts["QWP_eta_std"]
	Th_std = Uncerts["Th_std"]
	Tv_std = Uncerts["Tv_std"]
	Rh_std = Uncerts["Rh_std"]
	Rv_std = Uncerts["Rv_std"]
	
	# Parse Measurement Settings into vectors
	EtaH = Mments[:,0]
	EtaQ = Mments[:,1]
	ThetaH = Mments[:,2]
	ThetaQ = Mments[:,3]
	h = Mments[:,4]
	v = Mments[:,5]

	InputFlux = getInputFlux(counts, params)

	poisson_model = pm.Model()
	with poisson_model:
		# flux values are correlated (H,V)(D,A)(L,R)
		# only 3 flux values to calculate 
		stdFlux = pm.Normal("stdFlux",np.zeros(3),sigma=1,size = 3)
		fluxDist = pmm.abs_(InputFlux.flatten() + stdFlux*np.sqrt(np.max(InputFlux)))
		flux = fluxDist[[0,0,1,1,2,2]]
		
		t0 = pm.Uniform("t0", lower = -1 , upper = 1)
		t1 = pm.Uniform("t1", lower = -1, upper = 1)
		t2 = pm.Uniform("t2", lower = -1, upper = 1)
		t3 = pm.Uniform("t3", lower = -1 , upper = 1)
		
		zEta = pm.Normal("zEta",np.zeros(2),sigma = 1,size = (2))
		EtaH =  Mments[:,0] + HWP_eta_std*zEta[0]
		EtaQ =  Mments[:,1] + QWP_eta_std*zEta[1]
		
		zWP = pm.Normal("zWP",np.zeros(2),sigma = 1,size = (2))
		ThetaQ =  Mments[:,3] + QWP_std*zWP[0]
		ThetaH =  Mments[:,2] + HWP_std*zWP[1]
		
		MuNu_std = np.sqrt(Mu_std**2 + Nu_std**2)
		PBS_std = np.sqrt(Th_std**2 + Tv_std**2 + Rh_std**2 + Rv_std**2)
		combPBS_std = np.sqrt((PBS_std**2+MuNu_std**2))
		zPBS = pm.Normal("zPBS",np.zeros(4),sigma = 1,size = (4))
		ThDist = pmm.abs_(Th*Mu + combPBS_std*zPBS[0])  
		TvDist = pmm.abs_(Tv*Mu + combPBS_std*zPBS[1]) 
		RhDist = pmm.abs_(Rh*Nu + combPBS_std*zPBS[2])  
		RvDist = pmm.abs_(Rv*Nu + combPBS_std*zPBS[3])
		
		PBScrsstlk = at.stack([ThDist,TvDist,RhDist,RvDist]).reshape((2,2))
		CrsstlkMatDist = at.slinalg.kron(np.eye(3),PBScrsstlk)
		
		# Deterministic variables have a recorded trace but do not add randomness to model
		tr = pm.Deterministic("tr", pm.math.sqr(t0)+pm.math.sqr(t1)+ pm.math.sqr(t2)+ pm.math.sqr(t3) )
		A = pm.Deterministic("A", ( pm.math.sqr(t0)+pm.math.sqr(t1)+pm.math.sqr(t2) )/tr )
		ReC = pm.Deterministic("ReC", ( t1*t3 )/tr )
		ImC = pm.Deterministic("ImC", ( t2*t3 )/tr )
		B = pm.Deterministic("B",  pm.math.sqr(t3)/tr)
		X = pm.Deterministic("X", (2*ReC) )
		Y = pm.Deterministic("Y", (2*ImC) )
		Z = pm.Deterministic("Z",  (A-B)  )
		
		# probability of a measurement outcome
		prob = 1/32*(4*(5*A*h+3*B*h+3*A*v+5*B*v)+2*(A-B)*(h-v)*pm.math.cos(EtaH-EtaQ)+(h-v)
			*(2*(A-B)*pm.math.cos(EtaH+EtaQ)+4*pm.math.cos(EtaQ)*(A-B+8*ImC*pm.math.sin(EtaH) \
			*pm.math.sin(2*ThetaH))+16*pm.math.cos(EtaQ/2)**2*pm.math.sin(EtaH/2)**2*((A-B) \
			*pm.math.cos(4*ThetaH)+2*ReC*pm.math.sin(4*ThetaH))+32*pm.math.cos(2*ThetaQ) \
			*pm.math.sin(EtaQ)*(ReC*pm.math.sin(EtaH)*pm.math.sin(2*ThetaH)-ImC*pm.math.sin(EtaH/2)**2 \
			*pm.math.sin(4*ThetaH))+8*pm.math.cos(4*ThetaQ)*pm.math.sin(EtaQ/2)**2*(A-B \
			+2*pm.math.sin(EtaH/2)**2*((A-B)*pm.math.cos(4*ThetaH)-2*ReC*pm.math.sin(4*ThetaH))) \
			-8*pm.math.sin(EtaQ)*(-4*ImC*pm.math.cos(2*ThetaH)**2+2*ImC*pm.math.cos(EtaH) \
			*pm.math.cos(4*ThetaH)+2*(A-B)*pm.math.sin(EtaH)*pm.math.sin(2*ThetaH))*pm.math.sin(2*ThetaQ)\
			-4*pm.math.sin(EtaQ/2)**2*(-4*ReC-4*pm.math.sin(EtaH/2)**2*(2*ReC*pm.math.cos(4*ThetaH) \
			+(A-B)*pm.math.sin(4*ThetaH)))*pm.math.sin(4*ThetaQ))+4*(h-v)*pm.math.cos(EtaH) \
			*(A-B+4*ImC*pm.math.sin(EtaQ)*pm.math.sin(2*ThetaQ)+2*pm.math.sin(EtaQ/2)**2 \
			*((A-B)*pm.math.cos(4*ThetaQ)+2*ReC*pm.math.sin(4*ThetaQ))))
		
		# original expected counts in the event of no  crosstalk matrix
		OrigCounts = prob*flux
		
		# counts altered by crosstalk matrix
		Ncounts = pm.math.dot(CrsstlkMatDist,OrigCounts.reshape(6,1))

		# likelihood function for number of observed counts
		n_obs = pm.TruncatedNormal("n_obs" , mu=Ncounts.flatten(), sigma=np.sqrt(np.max(counts)), \
			lower = 0,observed = counts.flatten())

		# take trace with 1000 samples per core
		trace = pm.sample(draws = 1000, tune = 800, cores=4, target_accept = 0.98)

		# posterior-predictive checks
		ppc = pm.sample_posterior_predictive(trace, model=poisson_model, var_names = ["n_obs"])
		
	az.plot_trace(trace,var_names = ["A","B","ReC","ImC","X","Y","Z"])

	return [trace,ppc,poisson_model]
\end{lstlisting}

\subsection{Two-Qubit QST (Detailed)}

While we present a two-qubit formalism with orthogonal measurement-bases here, a more general framework extending to $n$-qubits is presented in \cite{ALTEPETER2005105}. The framework for generating a PyMC model with more qubits is similar to what is done here, except the parameter space will result in a more time-consuming computation. 

As per the one-qubit case, we start by defining a two-qubit density matrix. A discrete bipartite system composed of two-dimensional qubits will have a density matrix following 
\begin{equation}
\bm{\rho} = \frac{1}{4}\sum\limits_{i,j=0}^3 S_{i,j}\,\bm{\sigma}_{i}\otimes\bm{\sigma}_{j},
\label{eq:twoqubitdensitymatrix}
\end{equation}
where the $\bm{\sigma}_i$ are once again the Pauli matrices and $S_{i,j}$ are the joint-space two-qubit Stokes parameters. The objective is to find the ($4^2-1=15$) two-qubit Stokes parameters through projective measurements. Rather than record the measurement outcomes on each subsystem within the bipartite system individually, we must apply projective measurements on each subsystem while measuring in coincidence (only recording cases when both detectors appear to simultaneously fire).

In a similar manner to the relation in Eq. \ref{eq:StokesExpectationRelation}, the expectation values associated with the joint-space Pauli matrices yields the joint-space Stokes parameters as
\begin{equation}
\langle \bm{\sigma}_i\otimes \bm{\sigma}_j \rangle = \text{Tr}\left(\bm{\rho}\cdot\bm{\sigma}_i\otimes\bm{\sigma}_j\right) = S_i\otimes S_j \equiv S_{i,j},
\end{equation}
where the $S_i$ are the single-qubit Stokes parameters presented in Eq. \ref{eq:stokesParemeters}. Thus, all the necessary projections can be combined as
\begin{align}
S_{0,0} &= \left(P_H+P_V\right)\otimes\left(P_H+P_V\right) = P_{HH}+P_{HV}+P_{VH}+P_{VV} \nonumber\\
S_{0,1} &= \left(P_H+P_V\right)\otimes\left(P_D-P_A\right) = P_{HD}-P_{HA}+P_{VD}-P_{VA} \nonumber\\
S_{0,2} &= \left(P_H+P_V\right)\otimes\left(P_L-P_R\right) = P_{HL}-P_{HR}+P_{VL}-P_{VR} \nonumber\\
S_{0,3} &= \left(P_H+P_V\right)\otimes\left(P_H-P_V\right) = P_{HH}-P_{HV}+P_{VH}-P_{VV} \nonumber\\
S_{1,0} &= \left(P_D-P_A\right)\otimes\left(P_H+P_V\right) = P_{DH}+P_{DV}-P_{AH}-P_{AV} \nonumber\\
S_{1,1} &= \left(P_D-P_A\right)\otimes\left(P_D-P_A\right) = P_{DD}-P_{DA}-P_{AD}+P_{AA} \nonumber\\
S_{1,2} &= \left(P_D-P_A\right)\otimes\left(P_L-P_R\right) = P_{DL}-P_{DR}-P_{AL}+P_{AR} \nonumber\\
S_{1,3} &= \left(P_D-P_A\right)\otimes\left(P_H-P_V\right) = P_{DH}-P_{DV}-P_{AH}+P_{AV} \nonumber\\
S_{2,0} &= \left(P_L-P_R\right)\otimes\left(P_H+P_V\right) = P_{LH}+P_{LV}-P_{RH}-P_{RV} \nonumber\\
S_{2,1} &= \left(P_L-P_R\right)\otimes\left(P_D-P_A\right) = P_{LD}-P_{LA}-P_{RD}+P_{RA} \nonumber\\
S_{2,2} &= \left(P_L-P_R\right)\otimes\left(P_L-P_R\right) = P_{LL}-P_{LR}-P_{RL}+P_{RR} \nonumber\\
S_{2,3} &= \left(P_L-P_R\right)\otimes\left(P_H-P_V\right) = P_{LH}-P_{LV}-P_{RH}+P_{RV} \nonumber\\
S_{3,0} &= \left(P_H-P_V\right)\otimes\left(P_H+P_V\right) = P_{HH}+P_{HV}-P_{VH}-P_{VV} \nonumber\\
S_{3,1} &= \left(P_H-P_V\right)\otimes\left(P_D-P_A\right) = P_{HD}-P_{HA}-P_{VD}+P_{VA} \nonumber\\
S_{3,2} &= \left(P_H-P_V\right)\otimes\left(P_L-P_R\right) = P_{HL}-P_{HR}-P_{VL}+P_{VR} \nonumber\\
S_{3,3} &= \left(P_H-P_V\right)\otimes\left(P_H-P_V\right) = P_{HH}-P_{HV}-P_{VH}+P_{VV}. 
\end{align}
Note that there are only 36 unique measurement settings needed to construct the bipartite Stokes parameters.

We list the full analytical expression for $\bm{P}_{\psi,\phi}$ that was first presented in Eq. \ref{eq:TwoQubitProb} here. As a refresher, we calculate the probability of measuring polarization projections for quarter- and half-waveplate settings where the operators $\bm{U}_\text{A}\left(\bm{\theta_{Q\text{A}}},\bm{\theta_{H\text{A}}}\right)$ and $\bm{U}_\text{B}\left(\bm{\theta_{Q\text{B}}},\bm{\theta_{H\text{B}}}\right)$ rotate each photon (labeled as $A$ and $B$) within the biphoton independently before taking polarization projections with a PBS. Thus, we calculate
\begin{align}
\bm{U}\left( \bm{\theta_{Q\text{A}}}, \bm{\theta_{H\text{A}}}, \bm{\theta_{Q\text{B}}}, \bm{\theta_{H\text{B}}} \right) &= \bm{U}_\text{A}\left(\bm{\theta_{Q\text{A}}},\bm{\theta_{H\text{A}}}\right)\otimes \bm{U}_\text{B}\left(\bm{\theta_{Q\text{B}}},\bm{\theta_{H\text{B}}}\right) \nonumber \\
\bm{P}_{\psi,\phi} &= \text{Tr}\left[ \bm{\rho} \cdot \bm{U}\left( \bm{\theta_{Q\text{A}}}, \bm{\theta_{H\text{A}}}, \bm{\theta_{Q\text{B}}}, \bm{\theta_{H\text{B}}} \right) \right],
\end{align}
where
\begin{align}
\bm{P}_{\psi,\phi} &= \frac{1}{32} \bigg(8 A \mathbf{h_A} \mathbf{h_B}+8 B \mathbf{h_A} \mathbf{h_B}+8 C \mathbf{h_A} \mathbf{h_B} \nonumber +8 D \mathbf{h_A} \mathbf{h_B}+8 A \mathbf{h_B} \mathbf{v_A}+8 B \mathbf{h_B} \mathbf{v_A} +8 C \mathbf{h_B} \mathbf{v_A} \nonumber \\
&+8 D \mathbf{h_B} \mathbf{v_A}+8 A \mathbf{h_A} \mathbf{v_B}+8 B \mathbf{h_A} \mathbf{v_B}+8 C \mathbf{h_A} \mathbf{v_B}+8 D \mathbf{h_A} \mathbf{v_B}+8 A \mathbf{v_A} \mathbf{v_B}+8 B \mathbf{v_A} \mathbf{v_B} \nonumber \\
&+8 C \mathbf{v_A} \mathbf{v_B}+8 D \mathbf{v_A} \mathbf{v_B} +4 (A+B-C-D) (\mathbf{h_A}-\mathbf{v_A}) (\mathbf{h_B}+\mathbf{v_B}) \cos(4 \bm{\theta_{H\text{A}}}) \nonumber \\
&+(A-B-C+D+2 \text{Re}G+2 \text{Re}H) (\mathbf{h_A}-\mathbf{v_A}) (\mathbf{h_B}-\mathbf{v_B}) \cos(4 \bm{\theta_{H\text{A}}}-4 \bm{\theta_{H\text{B}}}) \nonumber \\
&+4 (A-B+C-D) (\mathbf{h_A}+\mathbf{v_A}) (\mathbf{h_B}-\mathbf{v_B}) \cos(4 \bm{\theta_{H\text{B}}}) \nonumber \\
&+(A-B-C+D-2 \text{Re}G-2 \text{Re}H) (\mathbf{h_A}-\mathbf{v_A}) (\mathbf{h_B}-\mathbf{v_B}) \cos(4 (\bm{\theta_{H\text{A}}}+\bm{\theta_{H\text{B}}})) \nonumber \\
&+4 (A+B-C-D) (\mathbf{h_A}-\mathbf{v_A}) (\mathbf{h_B}+\mathbf{v_B}) \cos(4 \bm{\theta_{H\text{A}}}-4 \bm{\theta_{Q\text{A}}}) \nonumber \\
&+(A-B-C+D-2 \text{Re}G-2 \text{Re}H) (\mathbf{h_A}-\mathbf{v_A}) (\mathbf{h_B}-\mathbf{v_B}) \cos(4 \bm{\theta_{H\text{A}}}-4 \bm{\theta_{H\text{B}}}-4 \bm{\theta_{Q\text{A}}}) \nonumber \\
&-4 (\text{Im}G+\text{Im}H) (\mathbf{h_A}-\mathbf{v_A}) (\mathbf{h_B}-\mathbf{v_B}) \cos(4 \bm{\theta_{H\text{A}}}-4 \bm{\theta_{H\text{B}}}-2 \bm{\theta_{Q\text{A}}}) \nonumber \\
&+4 (\text{Im}G+\text{Im}H) (\mathbf{h_A}-\mathbf{v_A}) (\mathbf{h_B}-\mathbf{v_B}) \cos(4 \bm{\theta_{H\text{A}}}+4 \bm{\theta_{H\text{B}}}-2 \bm{\theta_{Q\text{A}}}) \nonumber \\
&+(A-B-C+D+2 \text{Re}G+2 \text{Re}H) (\mathbf{h_A}-\mathbf{v_A}) (\mathbf{h_B}-\mathbf{v_B}) \cos(4 (\bm{\theta_{H\text{A}}}+\bm{\theta_{H\text{B}}}-\bm{\theta_{Q\text{A}}})) \nonumber \\
&+4 (A-B+C-D) (\mathbf{h_A}+\mathbf{v_A}) (\mathbf{h_B}-\mathbf{v_B}) \cos(4 \bm{\theta_{H\text{B}}}-4 \bm{\theta_{Q\text{B}}}) \nonumber \\
&-4 (\text{Im}G+\text{Im}H) (\mathbf{h_A}-\mathbf{v_A}) (\mathbf{h_B}-\mathbf{v_B}) \cos(4 \bm{\theta_{H\text{A}}}+4 \bm{\theta_{H\text{B}}}-2 \bm{\theta_{Q\text{A}}}-4 \bm{\theta_{Q\text{B}}}) \nonumber \\
&+4 (\text{Im}G-\text{Im}H) (\mathbf{h_A}-\mathbf{v_A}) (\mathbf{h_B}-\mathbf{v_B}) \cos(4 \bm{\theta_{H\text{A}}}+4 \bm{\theta_{H\text{B}}}-2 \bm{\theta_{Q\text{B}}}) \nonumber \\
&-4 (\text{Im}G-\text{Im}H) (\mathbf{h_A}-\mathbf{v_A}) (\mathbf{h_B}-\mathbf{v_B}) \cos(4 \bm{\theta_{H\text{A}}}+4 \bm{\theta_{H\text{B}}}-4 \bm{\theta_{Q\text{A}}}-2 \bm{\theta_{Q\text{B}}}) \nonumber \\ 
&+(A-B-C+D+2 \text{Re}G+2 \text{Re}H) (\mathbf{h_A}-\mathbf{v_A}) (\mathbf{h_B}-\mathbf{v_B}) \cos(4 (\bm{\theta_{H\text{A}}}+\bm{\theta_{H\text{B}}}-\bm{\theta_{Q\text{B}}})) \nonumber \\ 
&+(A-B-C+D-2 \text{Re}G-2 \text{Re}H) (\mathbf{h_A}-\mathbf{v_A}) (\mathbf{h_B}-\mathbf{v_B}) \cos(4 (\bm{\theta_{H\text{A}}}+\bm{\theta_{H\text{B}}}-\bm{\theta_{Q\text{A}}}-\bm{\theta_{Q\text{B}}})) \nonumber \\
&+(A-B-C+D-2 \text{Re}G-2 \text{Re}H) (\mathbf{h_A}-\mathbf{v_A}) (\mathbf{h_B}-\mathbf{v_B}) \cos(4 (\bm{\theta_{H\text{A}}}-\bm{\theta_{H\text{B}}}+\bm{\theta_{Q\text{B}}})) \nonumber \\
&+(A-B-C+D+2 \text{Re}G+2 \text{Re}H) (\mathbf{h_A}-\mathbf{v_A}) (\mathbf{h_B}-\mathbf{v_B}) \cos(4 (\bm{\theta_{H\text{A}}}-\bm{\theta_{H\text{B}}}-\bm{\theta_{Q\text{A}}}+\bm{\theta_{Q\text{B}}}))\nonumber \\
&-4 (\text{Im}G-\text{Im}H) (\mathbf{h_A}-\mathbf{v_A}) (\mathbf{h_B}-\mathbf{v_B}) \cos(4 \bm{\theta_{H\text{A}}}-4 \bm{\theta_{H\text{B}}}+2 \bm{\theta_{Q\text{B}}})\nonumber \\
&+4 (\text{Im}G-\text{Im}H) (\mathbf{h_A}-\mathbf{v_A}) (\mathbf{h_B}-\mathbf{v_B}) \cos(4 \bm{\theta_{H\text{A}}}-4 \bm{\theta_{H\text{B}}}-4 \bm{\theta_{Q\text{A}}}+2 \bm{\theta_{Q\text{B}}})\nonumber \\
&-8 (\text{Re}G-\text{Re}H) (\mathbf{h_A}-\mathbf{v_A}) (\mathbf{h_B}-\mathbf{v_B}) \cos(4 \bm{\theta_{H\text{A}}}-4 \bm{\theta_{H\text{B}}}-2 \bm{\theta_{Q\text{A}}}+2 \bm{\theta_{Q\text{B}}})\nonumber \\
&+4 (\text{Im}G+\text{Im}H) (\mathbf{h_A}-\mathbf{v_A}) (\mathbf{h_B}-\mathbf{v_B}) \cos(4 \bm{\theta_{H\text{A}}}-4 \bm{\theta_{H\text{B}}}-2 \bm{\theta_{Q\text{A}}}+4 \bm{\theta_{Q\text{B}}})\nonumber \\
&+8 (\text{Re}G-\text{Re}H) (\mathbf{h_A}-\mathbf{v_A}) (\mathbf{h_B}-\mathbf{v_B}) \cos(4 \bm{\theta_{H\text{A}}}+4 \bm{\theta_{H\text{B}}}-2 (\bm{\theta_{Q\text{A}}}+\bm{\theta_{Q\text{B}}}))\nonumber \\
&+8 (\text{Re}F+\text{Re}I) (\mathbf{h_A}-\mathbf{v_A}) (\mathbf{h_B}+\mathbf{v_B}) \sin(4 \bm{\theta_{H\text{A}}})\nonumber \\
&-2 (\text{Re}E-\text{Re}F+\text{Re}I-\text{Re}J) (\mathbf{h_A}-\mathbf{v_A}) (\mathbf{h_B}-\mathbf{v_B}) \sin(4 \bm{\theta_{H\text{A}}}-4 \bm{\theta_{H\text{B}}})\nonumber \\
&+8 (\text{Re}E+\text{Re}J) (\mathbf{h_A}+\mathbf{v_A}) (\mathbf{h_B}-\mathbf{v_B}) \sin(4 \bm{\theta_{H\text{B}}})\nonumber \\
&+2 (\text{Re}E+\text{Re}F-\text{Re}I-\text{Re}J) (\mathbf{h_A}-\mathbf{v_A}) (\mathbf{h_B}-\mathbf{v_B}) \sin(4 (\bm{\theta_{H\text{A}}}+\bm{\theta_{H\text{B}}}))\nonumber \\
&-8 (\text{Re}F+\text{Re}I) (\mathbf{h_A}-\mathbf{v_A}) (\mathbf{h_B}+\mathbf{v_B}) \sin(4 \bm{\theta_{H\text{A}}}-4 \bm{\theta_{Q\text{A}}})\nonumber \\
&-2 (\text{Re}E+\text{Re}F-\text{Re}I-\text{Re}J) (\mathbf{h_A}-\mathbf{v_A}) (\mathbf{h_B}-\mathbf{v_B}) \sin(4 \bm{\theta_{H\text{A}}}-4 \bm{\theta_{H\text{B}}}-4 \bm{\theta_{Q\text{A}}})\nonumber \\
&-16 (\text{Im}F+\text{Im}I) (\mathbf{h_A}-\mathbf{v_A}) (\mathbf{h_B}+\mathbf{v_B}) \sin(4 \bm{\theta_{H\text{A}}}-2 \bm{\theta_{Q\text{A}}})\nonumber \\
&-4 (\text{Im}F-\text{Im}I) (\mathbf{h_A}-\mathbf{v_A}) (\mathbf{h_B}-\mathbf{v_B}) \sin(4 \bm{\theta_{H\text{A}}}-4 \bm{\theta_{H\text{B}}}-2 \bm{\theta_{Q\text{A}}})\nonumber \\
&-4 (\text{Im}F-\text{Im}I) (\mathbf{h_A}-\mathbf{v_A}) (\mathbf{h_B}-\mathbf{v_B}) \sin(4 \bm{\theta_{H\text{A}}}+4 \bm{\theta_{H\text{B}}}-2 \bm{\theta_{Q\text{A}}})\nonumber \\
&+2 (\text{Re}E-\text{Re}F+\text{Re}I-\text{Re}J) (\mathbf{h_A}-\mathbf{v_A}) (\mathbf{h_B}-\mathbf{v_B}) \sin(4 (\bm{\theta_{H\text{A}}}+\bm{\theta_{H\text{B}}}-\bm{\theta_{Q\text{A}}}))\nonumber \\
&-8 (\text{Re}E+\text{Re}J) (\mathbf{h_A}+\mathbf{v_A}) (\mathbf{h_B}-\mathbf{v_B}) \sin(4 \bm{\theta_{H\text{B}}}-4 \bm{\theta_{Q\text{B}}})\nonumber \\
&-4 (\text{Im}F-\text{Im}I) (\mathbf{h_A}-\mathbf{v_A}) (\mathbf{h_B}-\mathbf{v_B}) \sin(4 \bm{\theta_{H\text{A}}}+4 \bm{\theta_{H\text{B}}}-2 \bm{\theta_{Q\text{A}}}-4 \bm{\theta_{Q\text{B}}})\nonumber \\
&-16 (\text{Im}E+\text{Im}J) (\mathbf{h_A}+\mathbf{v_A}) (\mathbf{h_B}-\mathbf{v_B}) \sin(4 \bm{\theta_{H\text{B}}}-2 \bm{\theta_{Q\text{B}}})\nonumber \\
&-4 (\text{Im}E-\text{Im}J) (\mathbf{h_A}-\mathbf{v_A}) (\mathbf{h_B}-\mathbf{v_B}) \sin(4 \bm{\theta_{H\text{A}}}+4 \bm{\theta_{H\text{B}}}-2 \bm{\theta_{Q\text{B}}})\nonumber \\
&-4 (\text{Im}E-\text{Im}J) (\mathbf{h_A}-\mathbf{v_A}) (\mathbf{h_B}-\mathbf{v_B}) \sin(4 \bm{\theta_{H\text{A}}}+4 \bm{\theta_{H\text{B}}}-4 \bm{\theta_{Q\text{A}}}-2 \bm{\theta_{Q\text{B}}})\nonumber \\
&-2 (\text{Re}E-\text{Re}F+\text{Re}I-\text{Re}J) (\mathbf{h_A}-\mathbf{v_A}) (\mathbf{h_B}-\mathbf{v_B}) \sin(4 (\bm{\theta_{H\text{A}}}+\bm{\theta_{H\text{B}}}-\bm{\theta_{Q\text{B}}}))\nonumber \\
&-2 (\text{Re}E+\text{Re}F-\text{Re}I-\text{Re}J) (\mathbf{h_A}-\mathbf{v_A}) (\mathbf{h_B}-\mathbf{v_B}) \sin(4 (\bm{\theta_{H\text{A}}}+\bm{\theta_{H\text{B}}}-\bm{\theta_{Q\text{A}}}-\bm{\theta_{Q\text{B}}}))\nonumber \\
&+2 (\text{Re}E+\text{Re}F-\text{Re}I-\text{Re}J) (\mathbf{h_A}-\mathbf{v_A}) (\mathbf{h_B}-\mathbf{v_B}) \sin(4 (\bm{\theta_{H\text{A}}}-\bm{\theta_{H\text{B}}}+\bm{\theta_{Q\text{B}}}))\nonumber \\
&+2 (\text{Re}E-\text{Re}F+\text{Re}I-\text{Re}J) (\mathbf{h_A}-\mathbf{v_A}) (\mathbf{h_B}-\mathbf{v_B}) \sin(4 (\bm{\theta_{H\text{A}}}-\bm{\theta_{H\text{B}}}-\bm{\theta_{Q\text{A}}}+\bm{\theta_{Q\text{B}}}))\nonumber \\
&+4 (\text{Im}E-\text{Im}J) (\mathbf{h_A}-\mathbf{v_A}) (\mathbf{h_B}-\mathbf{v_B}) \sin(4 \bm{\theta_{H\text{A}}}-4 \bm{\theta_{H\text{B}}}+2 \bm{\theta_{Q\text{B}}})\nonumber \\
&+4 (\text{Im}E-\text{Im}J) (\mathbf{h_A}-\mathbf{v_A}) (\mathbf{h_B}-\mathbf{v_B}) \sin(4 \bm{\theta_{H\text{A}}}-4 \bm{\theta_{H\text{B}}}-4 \bm{\theta_{Q\text{A}}}+2 \bm{\theta_{Q\text{B}}})\nonumber \\
&-4 (\text{Im}F-\text{Im}I) (\mathbf{h_A}-\mathbf{v_A}) (\mathbf{h_B}-\mathbf{v_B}) \sin(4 \bm{\theta_{H\text{A}}}-4 \bm{\theta_{H\text{B}}}-2 \bm{\theta_{Q\text{A}}}+4 \bm{\theta_{Q\text{B}}})\bigg),
\end{align}
and the bold values in $\bm{P}_{\psi,\phi}$ are vectors $\in\mathbb{R}^n$ for $n$ measurements.
It is important to note that we assume that all $\mathbf{h}$ and $\mathbf{v}$ within $\bm{P}_{\psi,\phi}\left(\bm{\theta_{Q\text{A}}},\bm{\theta_{H\text{A}}},\bm{\theta_{Q\text{B}}},\bm{\theta_{H\text{B}}}\right)$ can only be values of either 0 or 1. We also assume $\mathbf{h_A}\mathbf{v_A} = 0$ and $\mathbf{h_B}\mathbf{v_B} = 0$. These assumptions imply $\mathbf{h}^2 = \mathbf{h}$ and $\mathbf{v}^2 = \mathbf{v}$. As a consequence, numerous terms were set to 0 and allowed multiple variables to combine into the simplified expression above.

Similar to how we constructed a physical model of the system via a likelihood function with priors to ensure the state estimate is physical, we apply the same technique to the two-qubit tomography in the next section.

\subsection{Two-Qubit Python Code}

Using the same python libraries used previously, we load them with the code below.
\begin{lstlisting}[language=Python]
import arviz as az
import numpy as np
import pymc as pm
import pymc.math as pmm
import matplotlib.pyplot as plt
import aesara.tensor as at
\end{lstlisting}

Within the biphoton simulation presented in \texttt{main()}, we call the function \texttt{Prob(MmentsA,MmentsB,TrueState)} that takes measurement settings for systems A and B (using \texttt{MmentsA} and \texttt{MmentsB}) along with an identity matrix \texttt{TrueState} to calculate measurement probabilities. 

\begin{lstlisting}[style = mystyle]
def Prob(MmentsA,MmentsB,TrueState):
	# Density matrix has form
	#    A      ReE-iImE  ReF-iImF  ReG-iImG
	# ReE+iImE      B     ReH-iImH  ReI-iImI
	# ReF+iImF  ReH+iImH     C      ReJ-imJ
	# ReG+iImG  ReI+iImI  ReJ+imJ      D
	A = TrueState[0,0]
	B = TrueState[1,1]
	C = TrueState[2,2]
	D = TrueState[3,3]
	ReE = np.real(TrueState[1,0])
	ImE = np.imag(TrueState[1,0])
	ReF = np.real(TrueState[2,0])
	ImF = np.imag(TrueState[2,0])
	ReG = np.real(TrueState[3,0])
	ImG = np.imag(TrueState[3,0])
	ReH = np.real(TrueState[2,1])
	ImH = np.imag(TrueState[2,1])
	ReI = np.real(TrueState[3,1])
	ImI = np.imag(TrueState[3,1])
	ReJ = np.real(TrueState[3,2])
	ImJ = np.imag(TrueState[3,2])
    
	# extract measurement settings
	ThetaQA = MmentsA[:,1] 
	ThetaQB = MmentsB[:,1] 
	ThetaHA = MmentsA[:,0] 
	ThetaHB = MmentsB[:,0] 
	hA      = MmentsA[:,2] 
	hB      = MmentsB[:,2] 
	vA      = MmentsA[:,3] 
	vB      = MmentsB[:,3] 
    
	# measurement probability model
	prob = 1./32.*(8*A*hA*hB+8*B*hA*hB+8*C*hA*hB+8*D*hA*hB+8*A*hB*vA \
		+8*B*hB*vA+8*C*hB*vA+8*D*hB*vA+8*A*hA*vB+8*B*hA*vB+8*C*hA*vB \
		+8*D*hA*vB+8*A*vA*vB+8*B*vA*vB+8*C*vA*vB+8*D*vA*vB \
		+4*(A+B-C-D)*(hA-vA)*(hB+vB)*np.cos(4*ThetaHA) \
		+(A-B-C+D+2*ReG+2*ReH)*(hA-vA)*(hB-vB)*np.cos(4*ThetaHA-4*ThetaHB) \
		+4*(A-B+C-D)*(hA+vA)*(hB-vB)*np.cos(4*ThetaHB) \
		+(A-B-C+D-2*ReG-2*ReH)*(hA-vA)*(hB-vB)*np.cos(4*(ThetaHA+ThetaHB)) \
		+4*(A+B-C-D)*(hA-vA)*(hB+vB)*np.cos(4*ThetaHA-4*ThetaQA) \
		+(A-B-C+D-2*ReG-2*ReH)*(hA-vA)*(hB-vB)*np.cos(4*ThetaHA-4*ThetaHB-4*ThetaQA) \
		-4*(ImG+ImH)*(hA-vA)*(hB-vB)*np.cos(4*ThetaHA-4*ThetaHB-2*ThetaQA) \
		+4*(ImG+ImH)*(hA-vA)*(hB-vB)*np.cos(4*ThetaHA+4*ThetaHB-2*ThetaQA)\
		+(A-B-C+D+2*ReG+2*ReH)*(hA-vA)*(hB-vB)*np.cos(4*(ThetaHA+ThetaHB-ThetaQA)) \
		+4*(A-B+C-D)*(hA+vA)*(hB-vB)*np.cos(4*ThetaHB-4*ThetaQB) \
		-4*(ImG+ImH)*(hA-vA)*(hB-vB)*np.cos(4*ThetaHA+4*ThetaHB-2*ThetaQA-4*ThetaQB) \
		+4*(ImG-ImH)*(hA-vA)*(hB-vB)*np.cos(4*ThetaHA+4*ThetaHB-2*ThetaQB) \
		-4*(ImG-ImH)*(hA-vA)*(hB-vB)*np.cos(4*ThetaHA+4*ThetaHB-4*ThetaQA-2*ThetaQB) \
		+(A-B-C+D+2*ReG+2*ReH)*(hA-vA)*(hB-vB)*np.cos(4*(ThetaHA+ThetaHB-ThetaQB)) \
		+(A-B-C+D-2*ReG-2*ReH)*(hA-vA)*(hB-vB)*np.cos(4*(ThetaHA+ThetaHB-ThetaQA-ThetaQB))\
		+(A-B-C+D-2*ReG-2*ReH)*(hA-vA)*(hB-vB)*np.cos(4*(ThetaHA-ThetaHB+ThetaQB)) \
		+(A-B-C+D+2*ReG+2*ReH)*(hA-vA)*(hB-vB)*np.cos(4*(ThetaHA-ThetaHB-ThetaQA+ThetaQB)) \
		-4*(ImG-ImH)*(hA-vA)*(hB-vB)*np.cos(4*ThetaHA-4*ThetaHB+2*ThetaQB) \
		+4*(ImG-ImH)*(hA-vA)*(hB-vB)*np.cos(4*ThetaHA-4*ThetaHB-4*ThetaQA+2*ThetaQB) \
		-8*(ReG-ReH)*(hA-vA)*(hB-vB)*np.cos(4*ThetaHA-4*ThetaHB-2*ThetaQA+2*ThetaQB) \
		+4*(ImG+ImH)*(hA-vA)*(hB-vB)*np.cos(4*ThetaHA-4*ThetaHB-2*ThetaQA+4*ThetaQB) \
		+8*(ReG-ReH)*(hA-vA)*(hB-vB)*np.cos(4*ThetaHA+4*ThetaHB-2*(ThetaQA+ThetaQB)) \
		+8*(ReF+ReI)*(hA-vA)*(hB+vB)*np.sin(4*ThetaHA) \
		-2*(ReE-ReF+ReI-ReJ)*(hA-vA)*(hB-vB)*np.sin(4*ThetaHA-4*ThetaHB) \
		+8*(ReE+ReJ)*(hA+vA)*(hB-vB)*np.sin(4*ThetaHB) \
		+2*(ReE+ReF-ReI-ReJ)*(hA-vA)*(hB-vB)*np.sin(4*(ThetaHA+ThetaHB)) \
		-8*(ReF+ReI)*(hA-vA)*(hB+vB)*np.sin(4*ThetaHA-4*ThetaQA) \
		-2*(ReE+ReF-ReI-ReJ)*(hA-vA)*(hB-vB)*np.sin(4*ThetaHA-4*ThetaHB-4*ThetaQA) \
		-16*(ImF+ImI)*(hA-vA)*(hB+vB)*np.sin(4*ThetaHA-2*ThetaQA) \
		-4*(ImF-ImI)*(hA-vA)*(hB-vB)*np.sin(4*ThetaHA-4*ThetaHB-2*ThetaQA) \
		-4*(ImF-ImI)*(hA-vA)*(hB-vB)*np.sin(4*ThetaHA+4*ThetaHB-2*ThetaQA) \
		+2*(ReE-ReF+ReI-ReJ)*(hA-vA)*(hB-vB)*np.sin(4*(ThetaHA+ThetaHB-ThetaQA)) \
		-8*(ReE+ReJ)*(hA+vA)*(hB-vB)*np.sin(4*ThetaHB-4*ThetaQB) \
		-4*(ImF-ImI)*(hA-vA)*(hB-vB)*np.sin(4*ThetaHA+4*ThetaHB-2*ThetaQA-4*ThetaQB) \
		-16*(ImE+ImJ)*(hA+vA)*(hB-vB)*np.sin(4*ThetaHB-2*ThetaQB) \
		-4*(ImE-ImJ)*(hA-vA)*(hB-vB)*np.sin(4*ThetaHA+4*ThetaHB-2*ThetaQB) \
		-4*(ImE-ImJ)*(hA-vA)*(hB-vB)*np.sin(4*ThetaHA+4*ThetaHB-4*ThetaQA-2*ThetaQB) \
		-2*(ReE-ReF+ReI-ReJ)*(hA-vA)*(hB-vB)*np.sin(4*(ThetaHA+ThetaHB-ThetaQB)) \
		-2*(ReE+ReF-ReI-ReJ)*(hA-vA)*(hB-vB)*np.sin(4*(ThetaHA+ThetaHB-ThetaQA-ThetaQB)) \
		+2*(ReE+ReF-ReI-ReJ)*(hA-vA)*(hB-vB)*np.sin(4*(ThetaHA-ThetaHB+ThetaQB)) \
		+2*(ReE-ReF+ReI-ReJ)*(hA-vA)*(hB-vB)*np.sin(4*(ThetaHA-ThetaHB-ThetaQA+ThetaQB)) \
		+4*(ImE-ImJ)*(hA-vA)*(hB-vB)*np.sin(4*ThetaHA-4*ThetaHB+2*ThetaQB) \
		+4*(ImE-ImJ)*(hA-vA)*(hB-vB)*np.sin(4*ThetaHA-4*ThetaHB-4*ThetaQA+2*ThetaQB) \
		-4*(ImF-ImI)*(hA-vA)*(hB-vB)*np.sin(4*ThetaHA-4*ThetaHB-2*ThetaQA+4*ThetaQB))
    
	return prob
\end{lstlisting}

Using \texttt{main()} to simulate a measurement on a two-photon quantum state, we first define our true state to be a maximally-mixed Bell singlet state as
\begin{equation}
\rho = \frac{1}{2}\begin{bmatrix}
0 & 0 & 0 & 0 \\
0 & 1 &-1 & 0 \\
0 &-1 & 1 & 0 \\
0 & 0 & 0 & 0
\end{bmatrix}.
\end{equation}

We then use the half- quarter-waveplate formalism to simulate polarization projections with a PBS. In addition to adding systematic uncertainty to our measurements (where the error in waveplate angles is drawn from a normal distribution $\mathcal{N}(0, 2\pi/180)$), each PBS introduces crosstalk within each system (A and B). To simplify the model, we combine efficiencies after the PBS (\texttt{Mu} for transmitted components and \texttt{Nu} for reflected components) with the PBS transmission and reflection efficiencies for horizontal and vertical polarizations to construct subspace crosstalk matrices for systems A and B (\texttt{CrsstlkMatA} and \texttt{CrsstlkMatB}, respectively). From these, we construct the joint-space crosstalk matrix \texttt{CrsstlkMatAB}. With an input biphoton flux of 1000 biphotons per our measurement, we simulated the effects of crosstalk and Poisson noise at the detector. Finally, the function \texttt{BayesianFit(counts, params)} takes the resulting counts and measurement parameters into account to reconstruct our state and returns the trace for all parameters, the posterior-predictive checks for model validation, and the model itself.

\begin{lstlisting}[style = mystyle]
# simulate data collection and state reconstruction in main()
def main():
	# set the random-number generating seed to duplicate results
	RANDOM_SEED = 12345
	np.random.seed(RANDOM_SEED)

	# define a two-photon flux 
	TestFlux = 1000

	# Maximally-mixed Bell-singlet state
	TrueState = .5*np.array([[0.0, 0.0, 0.0,0.0],\
							 [0.0, 1.0,-1.0,0.0], \
							 [0.0,-1.0, 1.0,0.0], \
							 [0.0, 0.0, 0.0,0.0]])


	# density matrix has form
	#    A      ReE-iImE  ReF-iImF  ReG-iImG
	# ReE+iImE      B     ReH-iImH  ReI-iImI
	# ReF+iImF  ReH+iImH     C      ReJ-imJ
	# ReG+iImG  ReI+iImI  ReJ+imJ      D
	A = TrueState[0,0]
	B = TrueState[1,1]
	C = TrueState[2,2]
	D = TrueState[3,3]
	ReE = np.real(TrueState[1,0])
	ImE = np.imag(TrueState[1,0])
	ReF = np.real(TrueState[2,0])
	ImF = np.imag(TrueState[2,0])
	ReG = np.real(TrueState[3,0])
	ImG = np.imag(TrueState[3,0])
	ReH = np.real(TrueState[2,1])
	ImH = np.imag(TrueState[2,1])
	ReI = np.real(TrueState[3,1])
	ImI = np.imag(TrueState[3,1])
	ReJ = np.real(TrueState[3,2])
	ImJ = np.imag(TrueState[3,2])

	# waveplate standard deviations / uncertainty
	HWPA_std = 2*np.pi/180.
	HWPB_std = 2*np.pi/180.
	QWPA_std = 2*np.pi/180.
	QWPB_std = 2*np.pi/180.

	# fiber coupling efficiencies
	# Mu = PBS transmitted coupling efficiency
	# Nu = PBS reflected coupling efficiency

	MuNu_std = .02
	MuA = .6
	MuB = .7
	NuA = .6
	NuB = .8

	# Include PBS crosstalk
	# values reflect PBS can be lossy
	# assume we can only measure to an accuracy of 2%

	PBS_std = .02

	ThA = .98 # PBS transmission of horizontal polarization for system A
	RhA = .01 # PBS transmission of horizontal polarization for system A
	RvA = .97 # PBS transmission of vertical polarization for system A
	TvA = .01 # PBS transmission of vertical polarization for system A
	ThB = .96 # PBS transmission of horizontal polarization for system B
	RhB = .01 # PBS transmission of horizontal polarization for system B
	RvB = .97 # PBS transmission of vertical polarization for system B
	TvB = .01 # PBS transmission of vertical polarization for system B

	# measurements should follow [HWP, QWP, H, V] 
	# where QWP and HWP are fast axis angles (w.r.t. horizontal axis)
	# and H and V are either 0 or 1 depending on the projection
	Mments = np.array([  [0.,       0.     ,  1, 0], \
						 [0.,       0.     ,  0, 1], \
						 [np.pi/8,  np.pi/4,  1, 0], \
						 [np.pi/8,  np.pi/4,  0, 1], \
						 [0.,       np.pi/4,  1, 0], \
						 [0.,       np.pi/4,  0, 1] ])

	# Will be generating 36 measurements
	MmentsA = np.kron(Mments,np.ones([6,1]))
	MmentsB = np.kron(np.ones([6,1]),Mments)

	# assume angles defined with a standard deviation of 2*pi/180
	Theta_std = 2*np.pi/180

	# systematic deviations in all measurements
	ThetaQA = MmentsA[:,1] + np.random.normal(0,Theta_std)
	ThetaQB = MmentsB[:,1] + np.random.normal(0,Theta_std)
	ThetaHA = MmentsA[:,0] + np.random.normal(0,Theta_std)
	ThetaHB = MmentsB[:,0] + np.random.normal(0,Theta_std)
	hA      = MmentsA[:,2]
	hB      = MmentsB[:,2]
	vA      = MmentsA[:,3]
	vB      = MmentsB[:,3]

	# probabilities of measurement output (before sorting with PBS)
	prob = Prob(MmentsA,MmentsB,TrueState)

	# construct crosstalk matrix for system A
	PBS_A_crsstlk = np.zeros([2,2])
	PBS_A_crsstlk[0,0] = MuA*ThA
	PBS_A_crsstlk[0,1] = MuA*TvA
	PBS_A_crsstlk[1,0] = NuA*RhA
	PBS_A_crsstlk[1,1] = NuA*RvA
	# add systematic error in system A's crosstalk components
	PBS_A_crsstlk = np.abs(PBS_A_crsstlk+np.random.normal(0,np.sqrt(PBS_std**2 + MuNu_std**2),(2,2)))
	# probabilites cannot exceed 1
	PBS_A_crsstlk[PBS_A_crsstlk > 1] = 1

	# construct crosstalk matrix for system B
	PBS_B_crsstlk = np.zeros([2,2])
	PBS_B_crsstlk[0,0] = MuB*ThB
	PBS_B_crsstlk[0,1] = MuB*TvB
	PBS_B_crsstlk[1,0] = NuB*RhB
	PBS_B_crsstlk[1,1] = NuB*RvB
	# add systematic error in system A's crosstalk components
	PBS_B_crsstlk = np.abs(PBS_B_crsstlk+np.random.normal(0,np.sqrt(PBS_std**2 + MuNu_std**2),(2,2)))
	# probabilities cannot exceed 1
	PBS_B_crsstlk[PBS_B_crsstlk > 1] = 1

	# construct joint-space crosstalk matrix for system AB
	CrsstlkMatA = np.kron(np.eye(3), PBS_A_crsstlk) 
	CrsstlkMatB = np.kron(np.eye(3), PBS_B_crsstlk)
	CrsstlkMatAB = np.kron(CrsstlkMatA,CrsstlkMatB)

	# apply crosstalk to flux of a perfect system
	n = np.dot(CrsstlkMatAB,prob*TestFlux)

	# apply Poisson noise
	counts = np.random.poisson(n)

	# group settings and uncertainties in a dictionary
	Uncerts = {"ThA": ThA,
			   "ThB": ThB,
			   "TvA": TvA,
			   "TvB": TvB,
			   "RhA": RhA,
			   "RhB": RhB,
			   "RvA": RvA,
			   "RvB": RvB,
			   "MuA": MuA,
			   "NuA": NuA,
			   "MuB": MuB,
			   "NuB": NuB,
			   "PBS_std": PBS_std,
			   "MuNu_std": MuNu_std,
			   "Theta_std":Theta_std}
	# group measurements and uncertainty into a dictionary settings for cleaner code 
	params = {"MmentsA": MmentsA, "MmentsB": MmentsB, "Uncerts":Uncerts}
	
	# construct state estimate
	[trace,ppc,poisson_model] = BayesianFit(counts, params)
	
if __name__ == "__main__":
	main()
\end{lstlisting}

Within the function, \texttt{BayesianFit()}, we use the function \texttt{GetTotFlux()} to perform a least-squares fit using our estimated crosstalk matrix's inverse and detector counts to approximate the input biphoton flux. Thus, we obtain 9 flux estimates where each flux is used in 4 projections for a total of 36 measurements.

\begin{lstlisting}[style = mystyle]
def GetTotFlux(counts,params):
	Uncerts = params["Uncerts"]

	# (Phi + Phi^T)*(Psi + Psi^T) = 1
	# measurement order:
	# [H V D A L R] * [H V D A L R] = \
	#		H[H V D A L R] V[H V D A L R] D[H V D A L R] A[H V D A L R] L[H V D A L R] R[H V D A L R]
	#(h+v)*(h+v) = hh + hv + vh + vv  --> 0,1,6,7
	#(h+v)*(d+a) = hd + ha + vd + va  --> 2,3,8,9
	#(h+v)*(l+r) = hl + hr + vl + vr  --> 4,5,10,11
	#(d+a)*(h+v) = dh + dv + ah + av  --> 12,13,18,19
	#(d+a)*(d+a) = dd + da + ad + aa  --> 14,15,20,21
	#(d+a)*(l+r) = dl + dr + al + ar  --> 16,17,22,23
	#(l+r)*(h+v) = lh + lv + rh + rv  --> 24,25,30,31
	#(l+r)*(d+A) = ld + la + rd + ra  --> 26,27,32,33
	#(l+R)*(l+r) = ll + lr + rl + rr  --> 28,29,34,35

	ThA = Uncerts["ThA"]
	ThB = Uncerts["ThB"]
	TvA = Uncerts["TvA"]
	TvB = Uncerts["TvB"]
	RhA = Uncerts["RhA"]
	RhB = Uncerts["RhB"]
	RvA = Uncerts["RvA"]
	RvB = Uncerts["RvB"]
	MuA = Uncerts["MuA"]
	NuA = Uncerts["NuA"]
	MuB = Uncerts["MuB"]
	NuB = Uncerts["NuB"]
 
	# crosstalk and coupling efficiencies muddy our estimate of the total flux
	#  ( [ muA*ThA  muA*TvA ] \otimes [ muB*ThB  muA*TvB ] ) \dot ( [hA] \otimes [hB] )
	#  ( [ nuA*RhA  nuA*RvA ]         [ nuB*RhB  nuA*RvB ] )      ( [vA]         [VB] )

	crsstlkPBS_A = np.array([[MuA*ThA, MuA*TvA],[NuA*RhA, NuA*RvA]])
	crsstlkPBS_B = np.array([[MuB*ThB, MuB*TvB],[NuB*RhB, NuB*RvB]])
	crsstlkMat_A = np.kron(np.eye(3),crsstlkPBS_A)
	crsstlkMat_B = np.kron(np.eye(3),crsstlkPBS_B)
	crsstlkPBS_AB = np.kron(crsstlkMat_A,crsstlkMat_B)
	
	# multiply our counts by the pseudo-inverse of crsstlkPBS_AB 
	pinvCrsstlkAB = np.linalg.pinv(crsstlkPBS_AB)
	fluxEst = np.dot(pinvCrsstlkAB,counts)
	# to get the total flux, need to sum the 4 relevant projections
	TotFlux = np.zeros(9)
	TotFlux[0] = np.sum(fluxEst[[0,1,6,7]])
	TotFlux[1] = np.sum(fluxEst[[2,3,8,9]])
	TotFlux[2] = np.sum(fluxEst[[4,5,10,11]])
	TotFlux[3] = np.sum(fluxEst[[12,13,18,19]])
	TotFlux[4] = np.sum(fluxEst[[14,15,20,21]])
	TotFlux[5] = np.sum(fluxEst[[16,17,22,23]])
	TotFlux[6] = np.sum(fluxEst[[24,25,30,31]])
	TotFlux[7] = np.sum(fluxEst[[26,27,32,33]])
	TotFlux[8] = np.sum(fluxEst[[28,29,34,35]])
	TotFlux = np.round(TotFlux)
	
	return TotFlux
\end{lstlisting}

The function \texttt{BayesianFit()} performs the Baysian analysis with PyMC. Similar to what was done in the single-qubit state reconstruction, we first define the priors for the biphoton flux, the density matrix parameters ($t_i$), the waveplate angles, and the PBS transmission and reflection components. Again, we use the fact that the normal distribution has the following property $\mathcal{N}(\bar{x},\sigma_x) = \bar{x}+\sigma_x \mathcal{N}(0,1)$ for all normally distributed priors. 

Combining these prior distributions to form deterministic variables, we model the probability of a detector count using \texttt{prob} and our crosstalk matrix \texttt{CrsstlkMatAB}. The variable \texttt{n\_obs} serves as our likelihood function for observing a detector count (being a Gaussian approximation to a Poisson distribution such that $\mathcal{P}(n)\approx\mathcal{N}(n,\sqrt{n})$). To reduce the likelihood of the posterior distribution misbehaving when the number of observed counts goes to 0, we use a slight modification with $\mathcal{P}(\bm{n})\approx\mathcal{N}(\bm{n},\sqrt{\max\bm{n}})$ where $\bm{n}\in\mathbb{R}^{36}$ is a vector of photon counts. 

To sample the posterior distribution, we instruct PyMC to use 4 CPU cores and to tune with 1000 samples per core followed by 1500 draws per core with the code snipped below. Again, the parameter \texttt{target\_accept} is set close to 1 to tell the sampler to make smaller step sizes for difficult posterior distributions.
\begin{lstlisting}[style = mystyle]
with poisson_model:
		trace = pm.sample(draws = 1500, tune = 1000, cores=4, target_accept = 0.98)
\end{lstlisting}
Finally, we use the obtained traces and our model parameters to simulate input data with the code 
\begin{lstlisting}[style = mystyle]
ppc = pm.sample_posterior_predictive(trace, model=poisson_model, var_names = ["n_obs"])
\end{lstlisting} 
to check how consistent our state estimate is with the model and input data.

\begin{lstlisting}[style = mystyle]
def BayesianFit(counts, params): 
	# unpack measurements and parameters
	MmentsA = params["MmentsA"]
	MmentsB = params["MmentsB"]
	Uncerts = params["Uncerts"]

	hA = MmentsA[:,2]
	hB = MmentsB[:,2]
	vA = MmentsA[:,3]
	vB = MmentsB[:,3]

	ThA = Uncerts["ThA"]
	ThB = Uncerts["ThB"]
	TvA = Uncerts["TvA"]
	TvB = Uncerts["TvB"]
	RhA = Uncerts["RhA"]
	RhB = Uncerts["RhB"]
	RvA = Uncerts["RvA"]
	RvB = Uncerts["RvB"]
	MuA = Uncerts["MuA"]
	NuA = Uncerts["NuA"]
	MuB = Uncerts["MuB"]
	NuB = Uncerts["NuB"]
	PBS_std = Uncerts["PBS_std"]
	MuNu_std = Uncerts["MuNu_std"]
	Theta_std = Uncerts["Theta_std"]

	# get least-squares estimate of biphoton input flux using PBS crosstalk matrix
	TotFlux = GetTotFlux(counts,params)

	poisson_model = pm.Model()
	with poisson_model:

		# Gaussian approximation to Poisson distribution
		# Split mean and standard deviation using N(x,sigma) = x + sigma*N(0,1)
		# 9 flux estimates used for 36 measurements
		stdFlux = pm.Normal("stdFlux",np.zeros(9),sigma=1,size = 9)
		fluxDist = pmm.abs_(TotFlux.flatten() + stdFlux*np.sqrt(np.max(TotFlux)))
		
		# combine 9 flux estimates for 36 measurements appropriately
		temp_a = fluxDist[[0,0,1,1,2,2]] 
		temp_b = fluxDist[[3,3,4,4,5,5]] 
		temp_c = fluxDist[[6,6,7,7,8,8]] 
		flux = at.concatenate([temp_a,temp_a,temp_b,temp_b,temp_c,temp_c])
		
		# priors for density matrix calculation
		t0 = pm.Uniform("t0", lower = -1.0, upper = 1.0)
		t1 = pm.Uniform("t1", lower = -1.0, upper = 1.0)
		t2 = pm.Uniform("t2", lower = -1.0, upper = 1.0)
		t3 = pm.Uniform("t3", lower = -1.0, upper = 1.0)
		t4 = pm.Uniform("t4", lower = -1.0, upper = 1.0)
		t5 = pm.Uniform("t5", lower = -1.0, upper = 1.0)
		t6 = pm.Uniform("t6", lower = -1.0, upper = 1.0)
		t7 = pm.Uniform("t7", lower = -1.0, upper = 1.0)
		t8 = pm.Uniform("t8", lower = -1.0, upper = 1.0)
		t9 = pm.Uniform("t9", lower = -1.0, upper = 1.0)
		t10 = pm.Uniform("t10", lower = -1.0, upper = 1.0)
		t11 = pm.Uniform("t11", lower = -1.0, upper = 1.0)
		t12 = pm.Uniform("t12", lower = -1.0, upper = 1.0)
		t13 = pm.Uniform("t13", lower = -1.0, upper = 1.0)
		t14 = pm.Uniform("t14", lower = -1.0, upper = 1.0)
		t15 = pm.Uniform("t15", lower = -1.0, upper = 1.0)

		zWP = pm.Normal("zWP",np.zeros(4),sigma = 1,size = (4))
		ThetaQA =  MmentsA[:,1] + Theta_std*zWP[0]
		ThetaHA =  MmentsA[:,0] + Theta_std*zWP[1]
		ThetaQB =  MmentsB[:,1] + Theta_std*zWP[2]
		ThetaHB =  MmentsB[:,0] + Theta_std*zWP[3]
		
		z1 = pm.Normal("z1",np.zeros(8),sigma = 1,size = (8))
		combPBS_std = np.sqrt(PBS_std**2+MuNu_std**2)
		ThA_Dist = pmm.abs_(ThA*MuA + combPBS_std*z1[0])  
		TvA_Dist = pmm.abs_(TvA*MuA + combPBS_std*z1[1]) 
		RhA_Dist = pmm.abs_(RhA*NuA + combPBS_std*z1[2])  
		RvA_Dist = pmm.abs_(RvA*NuA + combPBS_std*z1[3]) 
		ThB_Dist = pmm.abs_(ThB*MuB + combPBS_std*z1[4])  
		TvB_Dist = pmm.abs_(TvB*MuB + combPBS_std*z1[5]) 
		RhB_Dist = pmm.abs_(RhB*NuB + combPBS_std*z1[6]) 
		RvB_Dist = pmm.abs_(RvB*NuB + combPBS_std*z1[7]) 
		
		# combine t_i components
		# Deterministic variables have a recorded trance but do not add randomness to model
		tr = pm.Deterministic("tr",pmm.sqr(t0)+pmm.sqr(t1)+pmm.sqr(t2)+pmm.sqr(t3)+pmm.sqr(t4) \
							  +pmm.sqr(t5)+pmm.sqr(t6)+pmm.sqr(t7)+pmm.sqr(t8)+pmm.sqr(t9) \
							  +pmm.sqr(t10)+pmm.sqr(t11)+pmm.sqr(t12)+pmm.sqr(t13)+pmm.sqr(t14) \
							  +pmm.sqr(t15) )
		A = pm.Deterministic( "A", (pmm.sqr(t0)+pmm.sqr(t1)+pmm.sqr(t2)+pmm.sqr(t4)+pmm.sqr(t5) \
									+pmm.sqr(t9)+pmm.sqr(t10))/tr )
		B = pm.Deterministic( "B", (pmm.sqr(t3)+pmm.sqr(t6)+pmm.sqr(t7)+pmm.sqr(t11)+pmm.sqr(t12))/tr )
		C = pm.Deterministic( "C", (pmm.sqr(t8)+pmm.sqr(t13)+pmm.sqr(t14))/tr )
		D = pm.Deterministic( "D",  pmm.sqr(t15)/tr )
		ReE = pm.Deterministic( "ReE", (t10*t12 + t1*t3 + t4*t6 + t5*t7 + t11*t9)/tr )
		ImE = pm.Deterministic( "ImE", (t10*t11 + t2*t3 + t5*t6 - t4*t7 - t12*t9)/tr )
		ReF = pm.Deterministic( "ReF", (t10*t14 + t4*t8 + t13*t9)/tr )
		ImF = pm.Deterministic( "ImF", (t10*t13 + t5*t8 - t14*t9)/tr )
		ReG = pm.Deterministic( "ReG", t15*t9/tr  )
		ImG = pm.Deterministic( "ImG", t15*t10/tr )
		ReH = pm.Deterministic( "ReH", (t11*t13 + t12*t14 + t6*t8)/tr )
		ImH = pm.Deterministic( "ImH", (t12*t13 - t11*t14 + t7*t8)/tr )
		ReI = pm.Deterministic( "ReI", (t11*t15)/tr )
		ImI = pm.Deterministic( "ImI", (t12*t15)/tr )
		ReJ = pm.Deterministic( "ReJ", (t13*t15)/tr )
		ImJ = pm.Deterministic( "ImJ", (t14*t15)/tr )
		
		# probability of output before PBS crosstalk distorts measurement
		prob = 1./32.*(8*A*hA*hB+8*B*hA*hB+8*C*hA*hB+8*D*hA*hB+8*A*hB*vA \
			+8*B*hB*vA+8*C*hB*vA+8*D*hB*vA+8*A*hA*vB+8*B*hA*vB+8*C*hA*vB+8*D*hA*vB+8*A*vA*vB+8*B*vA*vB \
			+8*C*vA*vB+8*D*vA*vB+4*(A+B-C-D)*(hA-vA)*(hB+vB)*pmm.cos(4*ThetaHA) \
			+(A-B-C+D+2*ReG+2*ReH)*(hA-vA)*(hB-vB)*pmm.cos(4*ThetaHA-4*ThetaHB) \
			+4*(A-B+C-D)*(hA+vA)*(hB-vB)*pmm.cos(4*ThetaHB) \
			+(A-B-C+D-2*ReG-2*ReH)*(hA-vA)*(hB-vB)*pmm.cos(4*(ThetaHA+ThetaHB)) \
			+4*(A+B-C-D)*(hA-vA)*(hB+vB)*pmm.cos(4*ThetaHA-4*ThetaQA) \
			+(A-B-C+D-2*ReG-2*ReH)*(hA-vA)*(hB-vB)*pmm.cos(4*ThetaHA-4*ThetaHB-4*ThetaQA) \
			-4*(ImG+ImH)*(hA-vA)*(hB-vB)*pmm.cos(4*ThetaHA-4*ThetaHB-2*ThetaQA) \
			+4*(ImG+ImH)*(hA-vA)*(hB-vB)*pmm.cos(4*ThetaHA+4*ThetaHB-2*ThetaQA) \
			+(A-B-C+D+2*ReG+2*ReH)*(hA-vA)*(hB-vB)*pmm.cos(4*(ThetaHA+ThetaHB-ThetaQA)) \
			+4*(A-B+C-D)*(hA+vA)*(hB-vB)*pmm.cos(4*ThetaHB-4*ThetaQB) \
			-4*(ImG+ImH)*(hA-vA)*(hB-vB)*pmm.cos(4*ThetaHA+4*ThetaHB-2*ThetaQA-4*ThetaQB) \
			+4*(ImG-ImH)*(hA-vA)*(hB-vB)*pmm.cos(4*ThetaHA+4*ThetaHB-2*ThetaQB) \
			-4*(ImG-ImH)*(hA-vA)*(hB-vB)*pmm.cos(4*ThetaHA+4*ThetaHB-4*ThetaQA-2*ThetaQB) \
			+(A-B-C+D+2*ReG+2*ReH)*(hA-vA)*(hB-vB)*pmm.cos(4*(ThetaHA+ThetaHB-ThetaQB)) \
			+(A-B-C+D-2*ReG-2*ReH)*(hA-vA)*(hB-vB)*pmm.cos(4*(ThetaHA+ThetaHB-ThetaQA-ThetaQB)) \
			+(A-B-C+D-2*ReG-2*ReH)*(hA-vA)*(hB-vB)*pmm.cos(4*(ThetaHA-ThetaHB+ThetaQB)) \
			+(A-B-C+D+2*ReG+2*ReH)*(hA-vA)*(hB-vB)*pmm.cos(4*(ThetaHA-ThetaHB-ThetaQA+ThetaQB)) \
			-4*(ImG-ImH)*(hA-vA)*(hB-vB)*pmm.cos(4*ThetaHA-4*ThetaHB+2*ThetaQB) \
			+4*(ImG-ImH)*(hA-vA)*(hB-vB)*pmm.cos(4*ThetaHA-4*ThetaHB-4*ThetaQA+2*ThetaQB) \
			-8*(ReG-ReH)*(hA-vA)*(hB-vB)*pmm.cos(4*ThetaHA-4*ThetaHB-2*ThetaQA+2*ThetaQB) \
			+4*(ImG+ImH)*(hA-vA)*(hB-vB)*pmm.cos(4*ThetaHA-4*ThetaHB-2*ThetaQA+4*ThetaQB) \
			+8*(ReG-ReH)*(hA-vA)*(hB-vB)*pmm.cos(4*ThetaHA+4*ThetaHB-2*(ThetaQA+ThetaQB)) \
			+8*(ReF+ReI)*(hA-vA)*(hB+vB)*pmm.sin(4*ThetaHA) \
			-2*(ReE-ReF+ReI-ReJ)*(hA-vA)*(hB-vB)*pmm.sin(4*ThetaHA-4*ThetaHB) \
			+8*(ReE+ReJ)*(hA+vA)*(hB-vB)*pmm.sin(4*ThetaHB) \
			+2*(ReE+ReF-ReI-ReJ)*(hA-vA)*(hB-vB)*pmm.sin(4*(ThetaHA+ThetaHB)) \
			-8*(ReF+ReI)*(hA-vA)*(hB+vB)*pmm.sin(4*ThetaHA-4*ThetaQA) \
			-2*(ReE+ReF-ReI-ReJ)*(hA-vA)*(hB-vB)*pmm.sin(4*ThetaHA-4*ThetaHB-4*ThetaQA) \
			-16*(ImF+ImI)*(hA-vA)*(hB+vB)*pmm.sin(4*ThetaHA-2*ThetaQA) \
			-4*(ImF-ImI)*(hA-vA)*(hB-vB)*pmm.sin(4*ThetaHA-4*ThetaHB-2*ThetaQA) \
			-4*(ImF-ImI)*(hA-vA)*(hB-vB)*pmm.sin(4*ThetaHA+4*ThetaHB-2*ThetaQA) \
			+2*(ReE-ReF+ReI-ReJ)*(hA-vA)*(hB-vB)*pmm.sin(4*(ThetaHA+ThetaHB-ThetaQA)) \
			-8*(ReE+ReJ)*(hA+vA)*(hB-vB)*pmm.sin(4*ThetaHB-4*ThetaQB) \
			-4*(ImF-ImI)*(hA-vA)*(hB-vB)*pmm.sin(4*ThetaHA+4*ThetaHB-2*ThetaQA-4*ThetaQB) \
			-16*(ImE+ImJ)*(hA+vA)*(hB-vB)*pmm.sin(4*ThetaHB-2*ThetaQB) \
			-4*(ImE-ImJ)*(hA-vA)*(hB-vB)*pmm.sin(4*ThetaHA+4*ThetaHB-2*ThetaQB) \
			-4*(ImE-ImJ)*(hA-vA)*(hB-vB)*pmm.sin(4*ThetaHA+4*ThetaHB-4*ThetaQA-2*ThetaQB) \
			-2*(ReE-ReF+ReI-ReJ)*(hA-vA)*(hB-vB)*pmm.sin(4*(ThetaHA+ThetaHB-ThetaQB)) \
			-2*(ReE+ReF-ReI-ReJ)*(hA-vA)*(hB-vB)*pmm.sin(4*(ThetaHA+ThetaHB-ThetaQA-ThetaQB)) \
			+2*(ReE+ReF-ReI-ReJ)*(hA-vA)*(hB-vB)*pmm.sin(4*(ThetaHA-ThetaHB+ThetaQB)) \
			+2*(ReE-ReF+ReI-ReJ)*(hA-vA)*(hB-vB)*pmm.sin(4*(ThetaHA-ThetaHB-ThetaQA+ThetaQB)) \
			+4*(ImE-ImJ)*(hA-vA)*(hB-vB)*pmm.sin(4*ThetaHA-4*ThetaHB+2*ThetaQB) \
			+4*(ImE-ImJ)*(hA-vA)*(hB-vB)*pmm.sin(4*ThetaHA-4*ThetaHB-4*ThetaQA+2*ThetaQB) \
			-4*(ImF-ImI)*(hA-vA)*(hB-vB)*pmm.sin(4*ThetaHA-4*ThetaHB-2*ThetaQA+4*ThetaQB))
		
		# construct joint-space crosstalk matrix
		CrsstlkMatA = at.slinalg.kron(at.eye(3), \
							at.stack([ThA_Dist,TvA_Dist,RhA_Dist,RvA_Dist]).reshape((2,2)))
		CrsstlkMatB = at.slinalg.kron(at.eye(3), \
							at.stack([ThB_Dist,TvB_Dist,RhB_Dist,RvB_Dist]).reshape((2,2)))
		CrsstlkMatAB = at.slinalg.kron(CrsstlkMatA,CrsstlkMatB)
		
		# apply crosstalk to noiseless counts
		NoiselessCounts = prob*flux
		Ncounts = pmm.dot(CrsstlkMatAB,NoiselessCounts.reshape(36,1))
		
		# likelihood distribution for number of observed counts
		n_obs = pm.TruncatedNormal("n_obs" , mu=Ncounts.flatten(), \ 
								sigma=np.sqrt(np.max(counts)),lower=0,observed = counts.flatten())

	with poisson_model:
		trace = pm.sample(draws = 1500, tune = 1000, cores=4, target_accept = 0.98)

	# posterior predictive checks
	ppc = pm.sample_posterior_predictive(trace, model=poisson_model, var_names = ["n_obs"])

	# plot trace variables and print summary
	az.plot_trace(trace, compact=True); plt.show()
	print(az.summary(trace,round_to=2))

	return [trace,ppc,poisson_model]
\end{lstlisting}

While the joint-space Stokes vector (with elements $S_{i,j}$ for $\{i,j\}\in \{0,1,2,3\}$) is not calculated above, it can easily be included in the model by adding the following code to the PyMC model.
\begin{lstlisting}[style = mystyle]
		# traces for two-photon Stokes parameters can be obtained by adding following to model  
		S00 = pm.Deterministic( "S00", A + B + C + D ); S01 = pm.Deterministic( "S01", 2*(ReE + ReJ) )
		S02 = pm.Deterministic( "S02", 2*(ImE + ImJ) ); S03 = pm.Deterministic( "S03", A - B + C - D )
		S10 = pm.Deterministic( "S10", 2*(ReF + ReI) ); S11 = pm.Deterministic( "S11", 2*(ReG + ReH) )
		S12 = pm.Deterministic( "S12", 2*(ImG - ImH) ); S13 = pm.Deterministic( "S13", 2*(ReF - ReI) )
		S20 = pm.Deterministic( "S20", 2*(ImF + ImI) ); S21 = pm.Deterministic( "S21", 2*(ImG + ImH) )
		S22 = pm.Deterministic( "S22", 2*(ReH - ReG) ); S23 = pm.Deterministic( "S23", 2*(ImF - ImI) )
		S30 = pm.Deterministic( "S30", A + B - C - D ); S31 = pm.Deterministic( "S31", 2*(ReE - ReJ) )
		S32 = pm.Deterministic( "S32", 2*(ImE - ImJ) ); S33 = pm.Deterministic( "S33", A - B - C + D )
\end{lstlisting}

The two-qubit model above uses systematic errors in the measurement parameters provided in \texttt{main()}. These errors were simulated by adding random terms drawn from a normal distribution having a mean of 0 and a well-defined standard deviation. However, we cannot be sure that these random variables were drawn within one standard deviation of the normal-distribution's mean. Thus, the model is more likely to exhibit inconsistencies in the form of broad multimodal distributions for the final density matrix elements. In those cases, the prior mean values should be allowed to vary as the provided measurement-setting mean values may be drastically different from what we are telling the model. Fortunately in an actual experiment, the measurement settings are likely to be known with greater precision (within one standard deviation of a well-defined mean).

\end{document}